\documentclass[aps,prx,reprint,amsmath,amssymb,amsfonts,notitlepage]{revtex4-1}
\AtBeginDocument{\usepackage{booktabs}}        

\usepackage{microtype}
\usepackage{wasysym}
\usepackage{times}
\usepackage[varg]{txfonts}
\usepackage{mathrsfs}
\usepackage{bm}

\usepackage{color}
\usepackage{graphicx}
\usepackage{hyperref}
\usepackage{natbib}
\usepackage{overpic}
\usepackage{braket}

\usepackage[export]{adjustbox}

\newcommand{\avg}[1]{\langle #1 \rangle}

\newcommand{\cc}[1]{{#1}^{*}}
\newcommand{\cb}[1]{\bar{#1}}

\newcommand{\tsup}[1]{\textsuperscript{#1}}
\newcommand{\tsub}[1]{\textsubscript{#1}}

\newcommand{\meV}{\ {\rm meV}}

\newcommand{\K}{\ {\rm K}}

\newcommand{\hc}{{\rm h.c.}}
\newcommand{\subref}[2]{\ref{#1}\hyperref[#1]{#2}}

\renewcommand{\vec}[1]{{\boldsymbol{#1}}}
\newcommand{\mat}[1]{\vec{#1}}
\newcommand{\trp}[1]{{#1}^{\intercal}}
\newcommand{\vhat}[1]{\vec{\hat{#1}}}
\newcommand{\muB}{\mu_{\rm B}}

\newcommand{\tss}[1]{\textsubscript{#1}}

\newcommand{\abo}[2]{#1\tss{2}#2\tss{2}O\tss{7}}

\newcommand{\dto}{\abo{Dy}{Ti}}
\newcommand{\hto}{\abo{Ho}{Ti}}
\newcommand{\eto}{\abo{Er}{Ti}}
\newcommand{\yto}{\abo{Yb}{Ti}}
\newcommand{\tto}{\abo{Tb}{Ti}}

\newcommand{\rth}[1]{{#1}\tsup{3+}}
\newcommand{\hhh}{[\frac{1}{2}\frac{1}{2}\frac{1}{2}]}

\definecolor{cred}{RGB}{228,26,28}
\definecolor{cblue}{RGB}{8,48,107}
\definecolor{cgreen}{RGB}{77,175,74}
\definecolor{cgray}{RGB}{150,150,150}
\definecolor{clgray}{RGB}{200,200,200}
\definecolor{cpurple}{RGB}{152,78,163}
\definecolor{corange}{RGB}{255,127,0}
\definecolor{cgold}{RGB}{230,171,2}

\definecolor{ccc}{RGB}{87,16,110}
\usepackage[nolist,nohyperlinks]{acronym}
\newacro{DM}[DM]{{Dzyaloshinskii-Moriya}}
\newacro{AIAO}[AIAO]{all-in/all-out}
\newacro{PC}[PC]{Palmer-Chalker}
\newacro{QSI}[QSI]{quantum spin ice}
\newacro{QSL}[QSL]{quantum spin liquid}
\newacro{SI}[SI]{spin ice}
\newacro{SFM}[SFM]{splayed ferromagnet}
 
\newcommand{\captitle}[1]{#1.}

\hypersetup{colorlinks=true,linkcolor=ccc,citecolor=ccc,urlcolor=ccc} 

\begin{document}

\title{Frustrated quantum rare-earth pyrochlores}

\author{Jeffrey G. Rau}
\affiliation{Max-Planck-Institut f\"ur Physik komplexer Systeme, 01187 Dresden, Germany}
\author{Michel J. P. Gingras}
\affiliation{Department of Physics and Astronomy, University of Waterloo, Ontario, N2L 3G1, Canada}
\affiliation{Canadian Institute for Advanced Research, MaRS Centre, West Tower 661 University Ave., Suite 505, Toronto, ON, M5G 1M1, Canada}

\begin{abstract}
In this review we provide an introduction to the physics of a series of frustrated quantum rare-earth pyrochlores. We first give a background on the microscopic single- and two-ion physics of these materials, discussing the origins and properties of their exchange interactions and their minimal low-energy effective models, before outlining what is known about their classical and quantum phases. We then make use of this
understanding to discuss four important material examples: \eto{}, \yto{}, \tto{} and \abo{Pr}{Zr}, covering in some detail what is known experimentally and theoretically for each, and summarize some key questions that remain open. Finally, we offer an outlook on some alternative material platforms for realizing similar physics and discuss what we see as prospects for future investigations on these quantum rare-earth pyrochlores.

\end{abstract}

\date{\today}

\maketitle


\tableofcontents

\section{Introduction}
Condensed matter systems perhaps serve as the widest and most diverse playground for studying the physics of classical and quantum many-body phenomena. In many cases, the interactions between the microscopic constituents of the system cooperate, or only compete weakly, to drive the pertinent degrees of freedom into some ordered state. Examples include simple crystals and metals, liquid crystals, conventional superconductors and many magnetically ordered systems such as ferro- and anti-ferromagnets.  

Alternatively, instead of cooperating, interactions can instead strongly compete, with no clear choice of the ultimate ground state of the system. One generically refers to a system with multiple interactions that have mutually incompatible tendencies to be \emph{frustrated}. Systems that are highly frustrated can have many degenerate or near-degenerate states at low-energy, opening a route toward realizing unconventional phases, as well as potentially leading to  exotic low energy excitations.

Many paradigmatic and well-studied examples of highly frustrated condensed matter systems have arisen in the field of magnetism, in the context both of classical and quantum spin systems. Much of this effort has focused on \emph{geometrically} frustrated magnets~\cite{ramirez2001geometrical,lacroix2013introduction}, where the interactions between the spins are uniform and anti-ferromagnetic. Here, the frustration arises solely due to the spatial arrangement of the spins preventing the desired mutually anti-parallel alignment with all neighboring spins. Examples are typically drawn from Heisenberg, Ising and XY-like spin systems where the spin lattice is built from triangular or tetrahedral units; this includes triangular, kagome and pyrochlore lattices (among many others) which are naturally geometrically frustrated.

If the frustration is sufficiently high, the magnetic ordering usually expected at low temperatures can be averted entirely. One key motivation for the study of such systems is that, in some cases, the ultimate ground state can be a \emph{\ac{QSL}}~\cite{savary2016quantum}. These are intrinsically quantum states of matter, defined by a \emph{topological} order~\cite{wen2004quantum} of sorts which characterizes the long-range entanglement present in the ground state wave-function. This topological order is also accompanied by \emph{fractionalized} excitations that can have exotic mutual statistics as well as an unusual response to typical experimental probes. Even if a \ac{QSL} is not ultimately stabilized, proximity to such state may have strong effects on the low energy properties of the system, controlling much of its behavior. Going further, even in frustrated systems that are not proximate to a \ac{QSL}, interesting physics can be at play, for example in the development of unconventional long-range orders.

Alternative types of \emph{non}-geometric frustration have attracted increasing attention in recent years. The simplest examples can be constructed straightforwardly from unfrustrated systems; for example, starting with an anti-ferromagnet on a bipartite lattice with nearest-neighbor exchange, one can add further neighbor exchanges that disrupt the two-sublattice N\'eel ordering.  A different kind of non-geometric frustration has been realized recently in highly anisotropic magnets, induced by large atomic spin-orbit coupling~\cite{rau2016spin}. In these systems the relative importance of different anisotropic exchange interactions generates the frustration, potentially realizing an entirely different kind of limit than what can be found in geometrically frustrated systems. Examples of magnets frustrated by anisotropy are drawn from heavy transition metal magnets as well as from magnets built from lanthanide (rare-earth) and actinide series ions. A prominent set of such materials are the so-called ``Kitaev magnets"~\cite{winter2017models} realized in Na\tsub{2}IrO\tsub{3}, $(\alpha,\beta,\gamma)$-Li\tsub{2}IrO\tsub{3} and $\alpha$-RuCl\tsub{3} whose low-energy physics is close to Kitaev's celebrated honeycomb model~\cite{kitaev2006anyons}.

\begin{figure}
\centering
\begin{overpic}[width=0.9\columnwidth]{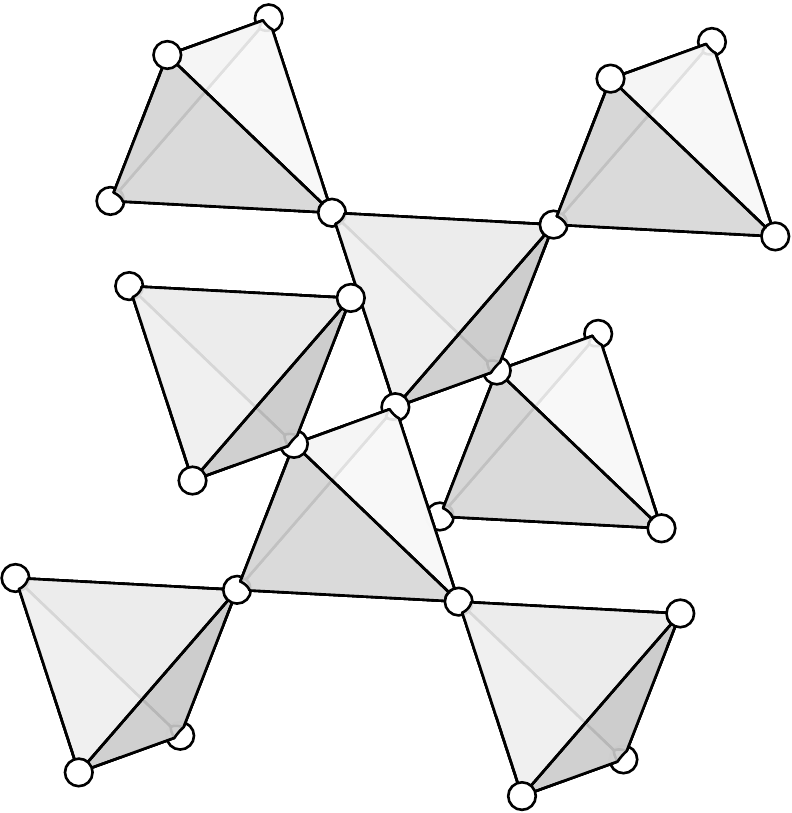}
\put(85,35){\rth{R}}
\put(0,0){(a)}
\end{overpic}
\begin{overpic}[width=0.4\columnwidth]{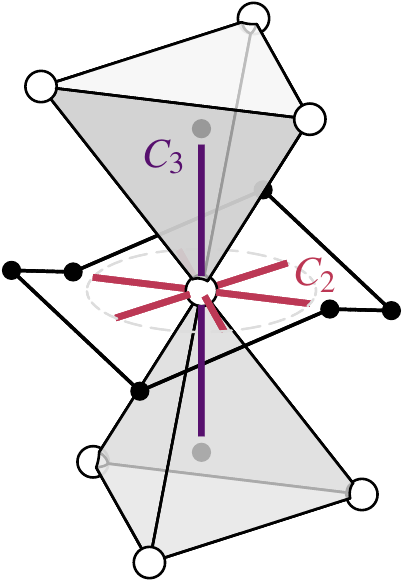}
\put(0,0){(b)}
\put(55,38){O\tsup{2-}}
\end{overpic}
\hspace{1cm}
\begin{overpic}[width=0.25\columnwidth]{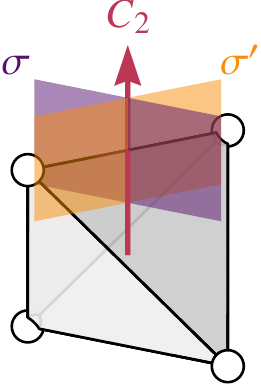}
\put(-10,0){(c)}
\end{overpic}
 \caption{\label{fig:lattice}
 (a) Crystal structure of the magnetic ions in the \abo{R}{M} rare-earth pyrochlores. (b) Local environment of a rare-earth ion showing the surrounding oxygen ions. The local XY plane of the ion, as well as the $C_2$ and $C_3$ rotation axes of the $D_{3d}$ site symmetry group are illustrated. (c) Illustration of the symmetries of the nearest-neighbor bond of the pyrochlore lattice, showing the two reflections ($\sigma$ and $\sigma'$) as well as a two-fold rotation ($C_2$).
  }
\end{figure}

In this review, we give an introduction to the physics of a family of frustrated highly anisotropic rare-earth magnets on the pyrochlore lattice [see Fig.~\subref{fig:lattice}{(a)}]. These are of the form \abo{R}{M}, where R is trivalent rare-earth ion and M a non-magnetic tetravalent transition metal ion~\cite{gardner2010rmp}. The best known of these compounds are the dipolar spin ices \abo{Dy}{Ti} and \abo{Ho}{Ti} which are almost entirely classical~\cite{rau2015magnitude} highly frustrated magnets. Our focus here is not on these classical systems~\cite{gingras2011spin}, but on the \emph{quantum} rare-earth pyrochlores of this series. Due to the  large spin-orbit coupling, along with the localized nature of the rare-earth ions, one naturally finds highly anisotropic exchange interactions which, in many cases, are strongly frustrated. After a review of the microscopic physics at play,  we delve into the study of four specific compounds: \eto{}, \yto{}, \tto{} and \abo{Pr}{Zr}. Each of these represents an archetypal example from a group of related materials, covering a range of behavior, from the weakly frustrated \eto{}, to the highly frustrated \yto{} and \tto{}, to randomly disordered \abo{Pr}{Zr}. None of these systems conform strongly to the spin ice limit~\cite{gingras2011spin}, or even the geometrically frustrated Heisenberg limit~\cite{canals1998pyrochlore}, but nevertheless exhibit new and interesting behavior. Finally, for each compound we provide an outlook to the future, as well as discussing some alternative material platforms for realizing similar physics.

\section{Microscopic background}
\label{sec:intro}
\subsection{Single-ion physics}
\label{sec:single-ion}
In rare-earth magnets, the single-ion physics strongly dominates over the two-ion exchange interactions and thus must be dealt with first (see Ref.~[\onlinecite{wybourne1965}] for details). This single-ion physics is tractable, displaying a reasonably clear hierarchy of energy scales, with Coulomb interactions dominating over spin-orbit coupling which itself dominates over the effects of the crystalline environment. The free-ion ground state is found by first minimizing the Coulomb and spin-orbit energies, in accordance with Hund's rules, yielding a ground state manifold with definite total angular momentum, $J$.  Next, the effect of the crystal field in solid removes most of this remaining $(2J+1)$-fold degeneracy~\cite{newman2007crystal}, with the charged ions of the crystal imposing electric fields~\footnote{A full accounting of all the sources of crystal field effects is complicated, with pure electrostatics being only one contribution. We refer the reader to the discussion in Ref.~[\onlinecite{newman2007crystal}] for a more complete discussion.} that breaks the rotational symmetry and splits the $J$-manifold.

For the \abo{A}{B} pyrochlore systems we are most interested in, the $D_{3d}$ site symmetry~\cite{gardner2010rmp} includes a $C_3$ rotation, three $C_2$ rotations (in the plane perpendicular to the three-fold axis) and their combinations with inversion [see Fig.~\subref{fig:lattice}{(b)}]. In such an environment, the crystal field potential, $\mathcal{V}(\vec{J})$, can be written as
\begin{align}
 \mathcal{V}(\vec{J}) &= 
 B_{20} O_{20}(\vec{J})+
 B_{40} O_{40}(\vec{J})+
 B_{43} O_{43}(\vec{J})\nonumber \\&+
 B_{60} O_{60}(\vec{J})+
 B_{63} O_{63}(\vec{J})+
 B_{66} O_{66}(\vec{J}),
\end{align}
where the operators $O_{KQ}(\vec{J})$ are Stevens operator equivalents~\cite{stevens1952matrix}, polynomials of order-$K$ in the total angular momentum $\vec{J}$. The precise details of this Hamiltonian and the associated splittings vary from material to material and from ion to ion. Typically, these energy scales are large relative to the two-ion interactions, but small relative to the energy required to reach the other $J$ manifolds. We can further restrict ourselves to the lowest-lying states of the free ion ground state $J$-manifold as determined by the crystalline electric field interaction.

For materials of interest, these low lying states take the form of a doublet. We will denote the states of this doublet as $\ket{\pm}$ and define pseudo-spin operators, $\vec{S}$, as
\begin{align}
 S^z &\equiv \frac{\ket{+}\bra{+} - \ket{-}\bra{-}}{2}, &
 S^{\pm} &\equiv \ket{\pm}\bra{\mp}.
 \label{eq:pseudospin}
\end{align}
For Kramers ions with an odd number of electrons (half-integer $J$), all of the crystal field levels are doubly degenerate due to Kramers' theorem. For non-Kramers ions (integer $J$) that have an even number of $4f$ electrons, any two-fold degeneracy must be enforced by the crystal symmetry.

Each of these crystal field doublets can be classified in terms of the irreducible representations of the site symmetry group $D_{3d}$ and its double group ~\cite{bradley2010mathematical}. Altogether, there are three distinct types, two Kramers doublets and one non-Kramers doublet. The most familiar is the $\Gamma_4$ doublet; we will refer to this as an ``effective spin-1/2'' as, in all aspects of symmetry, it behaves identically to an $S=1/2$ spin. The second type of doublet is built from the two one-dimensional irreducible representations $\Gamma_5$ and $\Gamma_6$. Since these two representations do not mix under the spatial symmetries, but only through time-reversal, this doublet behaves very differently than an effective spin-1/2~\cite{huang2014do}. This $\Gamma_5 \oplus \Gamma_6$ doublet has been dubbed a ``dipolar-octupolar doublet'', as $S^z$ corresponds to a component of the magnetic dipole moment, while the $S^{\pm}$ components correspond to parts of a magnetic octupole moment, changing the angular momentum in units of three~\footnote{Note that for $D_{3d}$ symmetry, the magnetic dipoles and octupoles are not necessarily distinct, with $S^z$ and $S^x$ transforming identically.}.

Finally, we have the ``non-Kramers doublet'' which transforms in the $E_g$ irreducible representation. This is similar in spirit to the dipolar-octupolar case: while the $S^z$ component transforms like a magnetic dipole, the $S^{\pm}$ components transform like \emph{electric} quadrupoles~\cite{lee2012generic}. This key feature, that $S^{\pm}$ is time-reversal even, makes this doublet qualitatively distinct from the two Kramers cases. We stress that the $D_{3d}$ site symmetry is necessary to preserve this degeneracy; without it, local electric field gradients can couple to the quadrupole moment (appearing as on-site terms like $S^{\pm}$) and split the doublet (see Sec.~\ref{sec:pr}).

These symmetry properties have consequences for the components of the $g$-factor of the ion which defines the magnetic moment $\vec{\mu}_i$ at site $i$ 
\begin{equation}
 \label{eq:moment}
 \vec{\mu}_i = -g_J \muB P \vec{J}_iP \equiv -\muB\left[
 g_z \vhat{z}^{}_i S^z_i + g_{\pm} \left(\vhat{x}^{}_i S^x_i + \vhat{y}^{}_i S^y_i\right)\right] .
\end{equation}
Here, $g_J$ is the Land\'e $g$-factor, $\muB$ the Bohr magneton and $P$ is a projector into the ground crystal field doublet. The vectors $\vhat{x}_i$, $\vhat{y}_i$, $\vhat{z}_i$ define a local frame~\cite{ross2011quantum} for the rare-earth ion, with $\vhat{z}_i$ along the $C_3$ axis, and $\vhat{y}_i$ along a $C_2$ axis [see Fig.~\subref{fig:lattice}{(b)}]. For the effective spin-1/2, one  can have strong Ising moments with $g_z \gg g_{\pm}$, strongly XY moments with $g_{\pm} \gg g_z$, or something in between $g_z \sim g_{\pm}$ depending on the details of the crystal field. However, for the dipolar-octupolar and non-Kramers doublets, since the transverse components $S^{\pm}$ transform as octupoles and quadrupoles (respectively) one has that $g_{\pm} = 0$ exactly \footnote{While this is plainly true for the non-Kramers case, we must note that for the dipolar-octupolar doublet, strictly, the statement is that one can always find a basis that for the doublet such that $g_{\pm}=0$.}, and thus they carry strictly Ising-like magnetic moments. We summarize the properties of these three types of doublets in Table~\ref{tab:doublets}.

\begin{table*}[ht]
 \begin{tabular}{lcccccccc}
 \toprule
 Doublet & Irrep. & $g_z$ & $g_{\pm}$ & Time. rev. & $C_3$ & $C_2$
  & States & Examples 
  \\
  \midrule
  Effective spin-1/2 &$\Gamma_4$ & $\neq 0$ & $\neq 0$ & 
  $\vec{S} \rightarrow -\vec{S}$ &
  $\begin{aligned}
   S^z &\rightarrow +S^z\\
   S^{\pm} &\rightarrow e^{\pm \frac{2\pi i}{3}}S^{\pm}  
  \end{aligned}$                   
  &
  $\begin{aligned}
   S^z &\rightarrow -S^z\\
   S^{\pm} &\rightarrow +S^{\mp}  
  \end{aligned}$                   
  & $\ket{\pm \tfrac{1}{2}}$, $\ket{\pm\tfrac{5}{2}}$,$\cdots$ &
  \begin{tabular}{c}
   \eto{}, \\
   \yto{} 
  \end{tabular}
  \\
  Dipolar-octupolar &$\Gamma_5 \oplus \Gamma_6$ & $\neq 0$ & $0$ &
  $\vec{S} \rightarrow -\vec{S}$  &
  $\vec{S} \rightarrow +\vec{S}$  & 
  $\begin{aligned}
   S^z &\rightarrow -S^z\\
   S^{\pm} &\rightarrow +S^{\mp}  
  \end{aligned}$                   
  & $\ket{\pm \tfrac{3}{2}}$, $\ket{\pm\tfrac{9}{2}}$,$\cdots$ &
  \begin{tabular}{c}
   \dto{}
  \end{tabular}
  \\
  \midrule 
  Non-Kramers & $E_g$ & $\neq 0$ & $0$ &
  $\begin{aligned}
   S^z &\rightarrow -S^z\\
   S^{\pm} &\rightarrow +S^{\mp}  
  \end{aligned}$ &
  $\begin{aligned}
   S^z &\rightarrow S^z\\
   S^{\pm} &\rightarrow e^{\pm \frac{2\pi i}{3}} S^{\pm}  
  \end{aligned}$ &
  $\begin{aligned}
   S^z &\rightarrow -S^z\\
   S^{\pm} &\rightarrow +S^{\mp}  
  \end{aligned}$
  & $\ket{\pm 1}$, $\ket{\pm 4}$, $\ket{\mp 5}$, $\cdots$ &
  \begin{tabular}{c}
   \hto{}, \\
   \tto{} 
  \end{tabular}
  \\  
 \bottomrule
 \end{tabular}
 \caption{\label{tab:doublets}
  \captitle{Types of crystal field doublets in rare-earth pyrochlores} We give the conventional name and notation for the irreducible representation of $D_{3d}$ to which it belongs~\cite{bradley2010mathematical} as well as the action of the elements of $D_{3d}$ on the pseudo-spin operators, $\vec{S}$, of each doublet. In addition, we give examples of $J$-manifold states that can be combined to create each doublet, as well as providing some material examples.
 }
\end{table*}

\subsection{Two-ion physics}
\label{sec:two-ion}
The interactions between rare-earth ions are considerably more complex than the single-ion physics discussed in the previous section~\cite{onoda2011se,rau2015magnitude,iwahara2015se}. Given this complexity, we will comment on these interactions in broad and somewhat phenomenological terms. In this regard, it is useful to move back up (in energy scale) from a description in terms of the crystal field doublets to a description in terms of the full free-ion $J$-manifolds, before descending back to the pseudo-spins discussed in Sec.~\ref{sec:single-ion}. While the description in terms of the $J$-manifolds is quite involved, we will see that the final description in terms of the pseudo-spins is relatively compact and simple.

Generically, pair-wise interactions between the angular momenta $\vec{J}_i$ can be written as
\begin{equation}
 \label{eq:multipolar}
 \frac{1}{2}\sum_{ij} \sum_{KQ} \sum_{K'Q'} \mathcal{M}^{KQ,K'Q'}_{ij} \mathscr{O}_{KQ}(\vec{J}_i) \mathscr{O}_{K'Q'}(\vec{J}_j) 
 + \sum_{i} \mathcal{V}(\vec{J}_i),
\end{equation}
where the $\mathscr{O}_{KQ}(\vec{J})$ are multipole operators, polynomials of order $K$ in the $\vec{J}$ operator (with $|Q| \leq K$). The second piece is the on-site crystal field potential, ${\mathcal V}(\vec{J})$, discussed in Sec.~\ref{sec:single-ion}. Explicitly, one can chose a basis such that the matrix elements of these multipole operators are proportional to Clebsch-Gordan coefficients (see e.g. Ref.~[\onlinecite{rau2016vcff}]) with $\bra{J,M}{\mathscr{O}_{KQ}(\vec{J})}\ket{J,M'} \propto \braket{J,M;K,Q|J,M'}$, though we will not need this level of detail here. 

Simple bilinear exchanges such as ${\vec{J}_i\!\cdot\!\vec{J}_j}$, used widely in the literature~\cite{jensen1991rare}, are interactions between rank-1 multipoles with $K=K'=1$. Generically, one does \emph{not} expect the multipolar exchanges to take such a na\"ive form; interactions between multipoles of rank greater than one are expected to be as, if not more, significant than these rank-1 terms. This includes interactions such as quadrupole-quadrupole (rank-2), octupole-octupole (rank-3) as well as cross terms such as dipole-octupole (rank-1, rank-3) and so forth for higher ranks.~\footnote{Note that, as for the single-ion physics, the spin-only moments of \rth{Gd} or Eu\tsup{2+} are special cases~\cite{gardner2010rmp} that do not readily conform to the expectations discussed here.} 

For an ion with angular momentum $J$, these multipoles can have rank up to $2J$, so there are in principle many, many interactions encoded in $\mathcal{M}^{KQ,K'Q'}_{ij}$ to consider per bond. However, there are some strong constraints arising from the microscopic mechanisms that generate these interactions. Examples include electro- and magneto-static interactions, spin-phonon interactions, as well as direct- and super-exchange. For our purposes, the most important of these are the super-exchange interactions and, in some cases, the magnetic dipole-dipole interactions~\cite{gardner2010rmp}. It is important to note that for essentially all the mechanisms discussed above, the interactions are strongly suppressed for ranks $K > 7$. This is due to the nearly free ion nature of the rare-earth ions limiting the maximal total angular momentum transferred by the $4f$ electron involved in each step of the super-exchange process to $7/2$~\cite{rau2015magnitude,iwahara2015se}. To a good approximation, we can thus generally restrict $K,K' \leq 7$ in Eq. (\ref{eq:multipolar}).

For some systems, this bound, when combined with the single-ion physics, can prove highly restrictive. For example, in \dto{} or \hto{}, the crystal field ground doublet is primarily of the form $\sim \ket{J,\pm J}$ with $J=15/2$ and $J=8$ (respectively). The rank-15 and rank-16 operators needed to transition between the $\ket{J,\pm J}$ states are thus strongly suppressed by \emph{any} of the two-ion interaction mechanisms discussed above, and the exchange interactions are essentially rendered classical~\cite{rau2015magnitude}. In the opposite extreme, for example for Yb\tsup{3+}, one has $J=7/2$ where \emph{only} rank $\leq 7$ operators are needed to mix the states of crystal field ground doublet, regardless of the ground doublet composition. One can thus have cases where the single- and two-ion anisotropy are intimately linked, while for other ions or crystal environments, they are essentially unrelated.

Admittedly, the complexity of these multipolar models is somewhat disheartening given the enormous number of free parameters encoded by the multipolar exchanges, even after taking into account the rank-7 bound and relevant lattice symmetries. However, all is not lost: for most cases of interest, the separation between the exchange and the crystal field energy scales allows some significant simplification. At the coarsest level, one may carry out first-order degenerate perturbation theory, taking the crystal field energy scale, $\Lambda$, to be much larger than the multipolar exchange scale $\mathcal{M}$, obtaining a model only in terms of the pseudo-spins [Eq.~(\ref{eq:pseudospin})]. This amounts to simply projecting Eq.~(\ref{eq:multipolar}) into the subspace of the ground doublets,
\begin{align}
 \label{eq:projection}
  &\frac{1}{2}\sum_{ij} \sum_{KQ} \sum_{K'Q'} \mathcal{M}^{KQ,K'Q'}_{ij} \mathscr{O}_{KQ}(\vec{J}_i) \mathscr{O}_{K'Q'}(\vec{J}_j) 
 + \sum_{i} \mathcal{V}(\vec{J}_i) \nonumber \\
 &\xrightarrow{P(\cdots)P} \frac{1}{2} \sum_{ij} \trp{\vec{S}}_i \mat{\mathcal{J}}_{ij} \vec{S}_j + {\rm const.},
\end{align}
where $\mat{\mathcal{J}}_{ij}$ is an exchange matrix between the pseudo-spins, $\vec{S}_i$. Thankfully, the exchange matrices in this smaller subspace are more tightly constrained by symmetry than the full multipolar exchanges entering into Eq.~(\ref{eq:multipolar}). The symmetries relevant for a nearest-neighbor bond of the pyrochlore lattice include a two-fold rotation and two reflections, as shown in Fig.~\subref{fig:lattice}{(c)}. One can show that for both types of Kramers ground doublets, there are four symmetry allowed nearest-neighbor exchanges~\cite{ross2011quantum}, while for a non-Kramers doublet there are only three~\cite{onoda2011se,lee2012generic}. Without any constraints from the multipole ranks or the compositions of the crystal field doublets, and absent other information, one would anticipate generic behavior, with all of these exchanges of the same order of magnitude, and such is expected for all the compounds of interest in this review.

\begin{widetext}
The symmetry-allowed nearest-neighbor model appropriate for the effective spin-1/2 doublet is the most complex, taking the form~\cite{curnoe2007,ross2011quantum}
\begin{align}
 \label{eq:effective-spin}
 \sum_{\avg{ij}} \big[ 
  J_{zz} S^{z}_i S^z_j 
  - J_{\pm}\left(S^+_iS^-_j + S^-_i S^+_j\right) 
  + J_{\pm\pm} \left(\gamma_{ij} S^+_i S^+_j + \cc{\gamma}_{ij} S^-_i S^-_j \right)
  + J_{z\pm} \left(\zeta_{ij} \left[S^z_i S^+_j + S^+_i S^z_j\right] + 
 \cc{\zeta}_{ij} \left[S^z_i S^-_j + S^-_i S^z_j \big]
 \right)
  \right],
\end{align}
where $\avg{ij}$ denotes the nearest-neighbors of the pyrochlore lattice. The allowed exchanges include an Ising coupling, $J_{zz}$, and XY-like exchange, $J_{\pm}$, as well as $J_{\pm\pm}$ and $J_{z\pm}$ couplings which carry bond-dependent phase factors $\zeta_{ij} = -\cc{\gamma}_{ij}$~\cite{ross2011quantum}, and thus do not have as simple of a geometric interpretation. In a global frame for the pseudo-spins, these four exchanges can be recast as a Heisenberg, Kitaev, pseudo-dipolar and \ac{DM} exchange~\cite{yan2017exchange,rau2018frustration}. We note that the sign of $J_{z\pm}$ is somewhat arbitrary as it can be changed by a local $C_2$ pseudo-spin rotation about the $\vhat{z}_i$ axis. While simple in this local basis, when expressed in a global frame, one finds equivalences between superficially very different exchange parameters~\cite{rau2018frustration}.

The model for ions with a non-Kramers doublet is identical to the effective spin-1/2 case, save for time-reversal symmetry forcing $J_{z\pm}=0$; one thus has~\cite{curnoe2007,lee2012generic}
\begin{equation}
 \label{eq:non-kramers}
 \vspace{-0.1cm}
 \sum_{\avg{ij}} \left[
  J_{zz} S^{z}_i S^z_j 
  - J_{\pm}\left(S^+_iS^-_j + S^-_i S^+_j\right) 
  + J_{\pm\pm} \left(\gamma_{ij} S^+_i S^+_j + \cc{\gamma}_{ij} S^-_i S^-_j\right)
  \right].
\end{equation}
This affords the system an accidental symmetry; due to the decoupling of $S^z$ and $S^{\pm}$, a $C_2$ pseudo-spin rotation about the $\vhat{z}_i$ is now a symmetry of the model, not merely a duality. We stress that this is an accidental symmetry, present only for two-spin interactions, which can be lifted by multi-spin interactions. We also note the sign of $J_{\pm\pm}$ can be flipped by a pseudo-spin rotation by $\pi/2$ about $\vhat{z}_i$, leaving the other couplings invariant, so one can take $J_{\pm\pm} \geq 0$ without loss of generality.

Last, we consider the dipolar-octupolar case. Due to the trivial action of the $C_3$ rotation on these states, one finds the same basic form as the effective spin-1/2 case, except that the phase factors are absent, with $\gamma_{ij} = \zeta_{ij} = 1$. This can be recast as~\cite{huang2014do}
\begin{equation}
 \label{eq:dipolar-octupolar}
 \sum_{\avg{ij}} \left[
  J_{xx} S^x_i S^x_j+
  J_{yy} S^y_i S^y_j+
  J_{zz} S^{z}_i S^z_j+
  J_{xz} \left(S^x_i S^z_j + S^z_i S^x_j\right)\right],
\end{equation}
where we have defined $J_{xx} \equiv 2(J_{\pm\pm}-J_{\pm})$, $J_{yy} \equiv -2(J_{\pm\pm}+J_{\pm})$ and $J_{xz}\equiv 2 J_{z\pm}$. Unlike the effective spin-1/2 and non-Kramers cases, there is no bond-dependence in the exchange interactions. We note that the $J_{xz}$ exchange can be removed by a redefinition of the pseudo-spin axes~\cite{huang2014do}, though one must be mindful that this transformation must be also applied to the definition of the magnetic moment, Eq.~(\ref{eq:moment}), and to any further neighbor exchanges. 
\end{widetext}
\subsection{Virtual crystal field corrections}
\label{sec:vcff}
To close this section, we discuss another route to generating exchange interactions between the pseudo-spins, distinct from the microscopic mechanisms mentioned earlier, which involves corrections due to the finite crystal field energy scale. Recall that the projection of the multipolar interactions to the pseudo-spin model [Eq.~(\ref{eq:projection})] results from first order degenerate perturbation theory in the small parameter $\mathcal{M}/\Lambda$, where $\mathcal{M}$ is a typical multipolar exchange scale [Eq.~(\ref{eq:multipolar})] and $\Lambda$ is the gap to the first excited crystal field level. Going beyond first order introduces additional effective interactions between the pseudo-spins which depend on the details of the multipolar exchanges and the crystal field potential. We refer to these as \emph{virtual crystal field corrections}~\cite{molavian2007dyn,molavian2009towards}. 

For all types of doublets, the second order correction generates renormalization of the nearest-neighbor exchanges discussed above, as well as new second- and third-neighbor exchanges, with a scale set roughly by $\sim \mathcal{M}^2/\Lambda$. In addition, the non-Kramers case also admits a \emph{three-spin} interaction term, also appearing at second-order, of the form~\cite{molavian2009towards}
\begin{equation}
 \label{eq:three-spin}
 \sum_{\avg{ijk}} \left[\mathcal{K}_{ijk} S^z_i S^+_j S^z_k +\hc\right],
\end{equation}
where $\avg{ij}$ and $\avg{jk}$ are nearest-neighbor bonds and $i \neq k$. This is forbidden for both types of Kramers doublets as it breaks time-reversal symmetry. Note that this interaction explicitly breaks the accidental $C_2$ symmetry present in the non-Kramers model with only two-spin interactions [Eq.~(\ref{eq:non-kramers})]. Additional three-spin terms that do not involve $S^z$ are also allowed by symmetry, but are not generated at second order by virtual crystal field corrections. We note that these kinds of perturbative corrections also affect the \emph{observables} of the system, which can also acquire corrections at order $\mathcal{M}/\Lambda$~\cite{molavian2009towards}.

\section{Phases}
\label{sec:phases}

\begin{figure*}
\centering
\hspace{1cm}
\includegraphics[width=\textwidth]{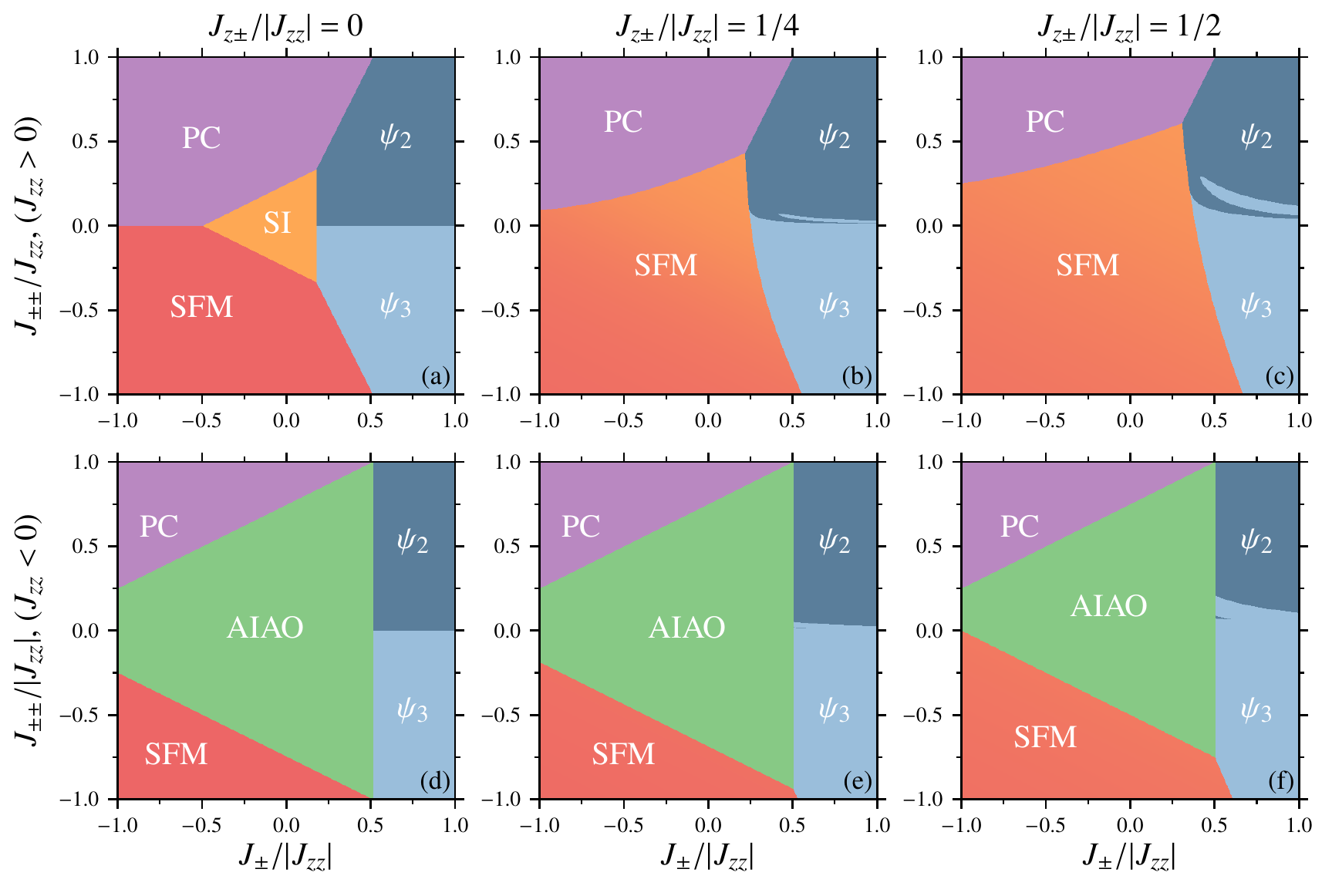}
  \includegraphics[width=0.9\textwidth]{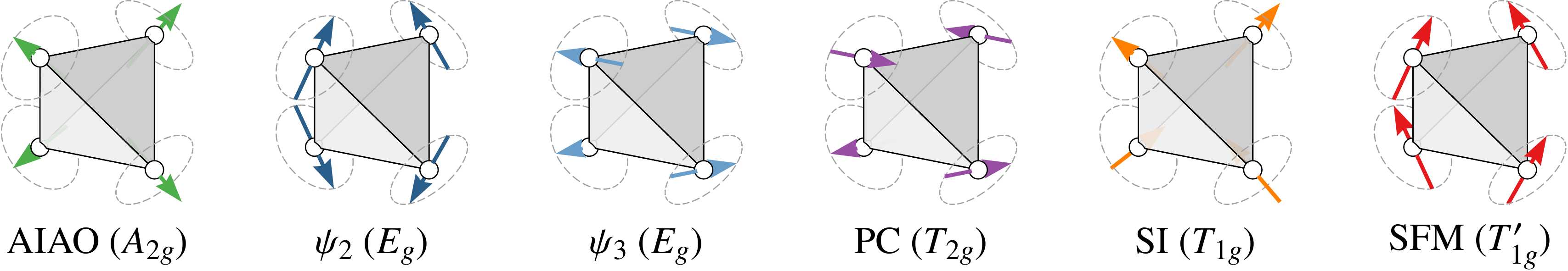}
 \caption{\label{fig:phases}
  (a-f) Classical phase diagrams of the effective spin-1/2 and non-Kramers doublet models [Eqs.~(\ref{eq:effective-spin},\ref{eq:non-kramers})]. We show phases for (a-c) $J_{zz}>0$ and (d-f) $J_{zz}<0$ with (a,d) $J_{z \pm} = 0$, (b,e) $J_{z\pm} = |J_{zz}|/4$ and (c,f) $J_{z\pm} = |J_{zz}|/2$.  The phase diagrams for $J_{z\pm} < 0$ are related to these by a local $C_2$ rotation about $\vhat{z}_i$. In the $E_g$ ($\Gamma_5$) phase, we show the state chosen semi-classically at order $O(1/S)$, which is always $\psi_2$ or $\psi_3$. The splay angle, $\theta_s$ in the splayed ferromagnet (SFM) phase is indicated by a color gradient, interpolating between orange which represents the $T_{1g}$ state (ordered spin ice) and red indicating the $T_{1g}'$ state where the pseudo-spins lie in the local XY plane. Example configurations of pseudo-spins for each of these orders (see Table~\ref{tab:orders}) are shown.
  }
\end{figure*}

The phase diagrams of the minimal, nearest-neighbor models for effective spin-1/2 [Eq.~(\ref{eq:effective-spin})], non-Kramers doublet [Eq.~(\ref{eq:non-kramers})] and dipolar-octupolar doublets [Eq.~(\ref{eq:dipolar-octupolar})] have been been studied in a number of works. In this section, we survey what is known about the phases that appear in such models at zero temperature.

\subsection{Classical phases and ordered states}
\label{sec:phases:classical}
A broad outline of the phases that occur can be exposed through an analysis of the classical ordered ground states, i.e. simple product states of the pseudo-spins. A rather comprehensive discussion for the effective spin-1/2 case can be found in Ref.~[\onlinecite{yan2017exchange}], from which one can infer many of the results for the non-Kramers case~\cite{lee2012generic}. The classical dipolar-octupolar phase diagram has been studied in Ref.~[\onlinecite{huang2014do}]. In all cases, one generically finds the classical ordered phases do not enlarge the primitive unit cell~\cite{yan2017exchange} and thus can be classified by how they transform under the point group of the pyrochlore lattice~\cite{bradley2010mathematical}.

The classification of these phases for the effective spin-1/2 and non-Kramers case are essentially identical, since they only differ under the action of time-reversal. Altogether, we identify five distinct types of ordered states~\cite{yan2017exchange}: $A_{2g}$, $E_g$, $T_{1g}$, $T_{1g}'$ and $T_{2g}$~\cite{bradley2010mathematical}, summarized in Table~\ref{tab:orders}.  For the effective spin-1/2 case, all of these are magnetic orders, while for the non-Kramers case, the $E_g$, $T_{1g}'$ and $T_{2g}$ states correspond to quadrupolar orders that do not break time-reversal. Both the $T_{1g}$ and $T_{1g}'$ orders transform identically under spatial symmetries, and thus can be mixed freely for the effective spin-1/2 case -- we refer to any such state as a \ac{SFM}. The mixing between the $T_{1g}$ and $T_{1g}'$ is typically characterized by the ``splay angle" $\theta_s$, which varies smoothly with the exchanges, and measures the angle of the pseudo-spins from the appropriate global $[100]$ cubic axis~\cite{yan2017exchange}. The classical phase diagram as a function of $J_{\pm}/|J_{zz}|$, $J_{\pm\pm}/|J_{zz}|$ for both signs of $J_{zz}$ and several values of $J_{z\pm}/|J_{zz}|$ is shown in Fig.~\ref{fig:phases}, accompanied by illustrations of each of the ordered states.

As much of the classical phase diagram consists of robust ground states (with only symmetry-enforced degeneracies), the effects of quantum fluctuations would na\"ively be expected to be mostly benign. However, there are exceptions, several of which are realized by the materials examples we will come to later. One important example is the $E_g$ phase, the so-called  ``$\Gamma_5$ manifold", where there is an accidental $U(1)$ classical degeneracy~\cite{savary2012obd,zhitomirsky2012obd,wong2013obd,yan2017exchange}. Explicitly, one can rotate the pseudo-spins continuously in the local XY plane about their local $[111]$ axes without changing the classical energy. Two high-symmetry basis states of this manifold are the $\psi_2$ and $\psi_3$ states (illustrated in Fig.~\ref{fig:phases}). In the phase diagrams shown in Fig.~\ref{fig:phases}, we have indicated which of the $\Gamma_5$ states (always $\psi_2$ or $\psi_3$) is selected by the leading quantum corrections to the classical energy (see Sec.~\ref{sec:eto} for a more detailed discussion). Additional $U(1)$ degeneracies can be found at various phase-boundary between the classical phases, such as between the $E_g$ and $T_{1g}$ or $T_{2g}$ orders (see Ref.~\cite{yan2017exchange} for a more complete discussion).

\begin{table*}[ht]
 \begin{tabular}{ccc}
  \toprule
  Name(s) & Order parameter & Examples \\
  \midrule
  \begin{tabular}{c}
   $A_{2g}$ ($\Gamma_3$), \\ All-in/All-out
  \end{tabular} & \footnotesize $M_{A_{2g}} \equiv S^z_1+S^z_2+S^z_3+S^z_4$ & None \\ \midrule
   $E_g$ $(\Gamma_5)$
      &\footnotesize $\vec{M}_{E_g} \equiv \left(\begin{array}{c}
      S^x_1+S^x_2+S^x_3+S^x_4\\      
      S^y_1+S^y_2+S^y_3+S^y_4
     \end{array}\right) $
      & \begin{tabular}{c}
      \abo{Er}{Ti}, \\ \abo{Er}{Ge}, \\ \abo{Yb}{Ge}
     \end{tabular}\\ \midrule
  \begin{tabular}{c}
   $T_{2g}$ ($\Gamma_7$), \\ Palmer-\\Chalker
  \end{tabular} & \footnotesize $\vec{M}_{T_{2g}} \equiv \left(\begin{array}{c}
           S^y_1 + S^y_2 -S^y_3 -S^y_4\\
           \left(-\frac{\sqrt{3}}{2} S^x_1-\frac{1}{2}S^y_1\right)-
           \left(-\frac{\sqrt{3}}{2} S^x_2-\frac{1}{2}S^y_2\right)+
           \left(-\frac{\sqrt{3}}{2} S^x_3-\frac{1}{2}S^y_3\right)-
           \left(-\frac{\sqrt{3}}{2} S^x_4-\frac{1}{2}S^y_4\right)
           \\
           \left(\frac{\sqrt{3}}{2} S^x_1-\frac{1}{2}S^y_1\right)-
           \left(\frac{\sqrt{3}}{2} S^x_2-\frac{1}{2}S^y_2\right)-
           \left(\frac{\sqrt{3}}{2} S^x_3-\frac{1}{2}S^y_3\right)+
           \left(\frac{\sqrt{3}}{2} S^x_4-\frac{1}{2}S^y_4\right)           
   \end{array}\right)$ & \begin{tabular}{c}
               \abo{Er}{Sn}, \\ \abo{Er}{Pt}
  \end{tabular} \\ \midrule
  \begin{tabular}{c}
   $T_{1g}$ ($\Gamma_9$), \\ Ordered- \\ spin ice   
  \end{tabular} &\footnotesize $\vec{M}_{T_{1g}} \equiv \left(\begin{array}{c}
 S^z_1+S^z_2-S^z_3-S^z_4\\
 S^z_1-S^z_2+S^z_3-S^z_4\\
 S^z_1-S^z_2-S^z_3+S^z_4   
          \end{array}\right)$ & \begin{tabular}{c}\abo{Tb}{Sn}
         \end{tabular}\\ 
\begin{tabular}{c}
 $T_{1g}'$ ($\Gamma_9$), \\ Splayed- \\ Ferromagnet
\end{tabular}  & \footnotesize $\vec{M}_{T_{1g}'} \equiv \left(
        \begin{array}{c}
         S^x_1 + S^x_2 -S^x_3 -S^x_4\\
         \left(-\frac{1}{2}S^x_1+\frac{\sqrt{3}}{2} S^y_1\right)-
         \left(-\frac{1}{2}S^x_2+\frac{\sqrt{3}}{2} S^y_2\right)+
         \left(-\frac{1}{2}S^x_3+\frac{\sqrt{3}}{2} S^y_3\right)-
         \left(-\frac{1}{2}S^x_4+\frac{\sqrt{3}}{2} S^y_4\right)\\
         \left(-\frac{1}{2}S^x_1-\frac{\sqrt{3}}{2} S^y_1\right)-
         \left(-\frac{1}{2}S^x_2-\frac{\sqrt{3}}{2} S^y_2\right)-
         \left(-\frac{1}{2}S^x_3-\frac{\sqrt{3}}{2} S^y_3\right)+
         \left(-\frac{1}{2}S^x_4-\frac{\sqrt{3}}{2} S^y_4\right)         
         \\
        \end{array}
  \right)
       $       & \begin{tabular}{c}
        \abo{Yb}{Ti},\\ \abo{Yb}{Sn}
                \end{tabular}
  \\ \bottomrule
\end{tabular}
\caption{
 \label{tab:orders} Classical order parameters for each of the five types of $\vec{Q}=0$ phases for effective spin-1/2 or non-Kramers doublets on the pyrochlore lattice. Some material examples are given. There are two different $T_{1g}$ types of order that can mix freely for the effective spin-1/2 case, while for the non-Kramers case they differ under the action of time-reversal. Pseudo-spin configurations for each ordering type are shown in Fig.~\ref{fig:phases}.
}
\end{table*}
Another exception arises for the non-Kramers case [see Fig.~\subref{fig:phases}{(a,d)}] for $|J_{\pm}|,|J_{\pm\pm}| \ll J_{zz}$ the lowest energy $\vec{Q}=0$ state is $T_{1g}$, a so-called ``ordered" \ac{SI} state~\cite{lee2012generic}, a specific case of the general \ac{SFM} states. In fact, for this case, \emph{all} \ac{SI} states are classical ground states; we refer to all such states in this region in Fig.~\subref{fig:phases}{(a)} generically as spin ice (SI). We will discuss briefly how this degeneracy is lifted by quantum effects in Sec.~\ref{sec:phases:quantum}. Note that this classical degeneracy does not survive in the effective spin-1/2 case where generically $J_{z\pm}\neq 0$, with the ground state being a mix of $T_{1g}$ and $T_{1g}'$.

The classification of ordered states for the dipolar-octupolar doublet~\cite{huang2014do} is somewhat different, as shown in Table~\ref{tab:dipolar-octupolar}. Since both $S^x$ and $S^z$ transform in the same way as magnetic dipoles, one has several versions of the all-in/all-out and ordered spin ice states found in the effective spin-1/2 and non-Kramers cases. However, since the $S^y$ pseudo-spin operator is invariant under all spatial symmetries, there are two qualitatively new kinds of order: a kind of octupolar ordered spin ice, as well as an octupolar variant of the all-in/all-out order ($A_{1g}$) which has the intriguing property that it  \emph{only} breaks time-reversal symmetry. For a discussion of the classical phase diagram of the dipolar-octupolar model [Eq.~(\ref{eq:dipolar-octupolar})], we refer the reader to Ref.~[\onlinecite{huang2014do}].

\subsection{Quantum phases}
\label{sec:phases:quantum}
Extensive classical degeneracies in the exchange models of Eqs.~(\ref{eq:effective-spin}-\ref{eq:dipolar-octupolar}) are mostly confined to phase boundaries and special isolated points in the space of exchange parameters. The best known and studied of these is that of the spin ice manifold~\cite{gingras2011spin} which appears when $J_{zz} > 0$ and $J_{\pm} = J_{\pm\pm} = J_{z\pm} = 0$ (analogously for the dipolar-octupolar case). The classical thermal physics of this point displays a number of fascinating phenomena~\cite{gingras2011spin}, from the emergence of a version of magnetostatics (a ``Coulomb'' phase) to the appearance of effective magnetic monopoles as elementary excitations. As discussed in Sec.~\ref{sec:phases:classical}, this also holds over a finite region of phase space for the non-Kramers case [see Fig.~\subref{fig:phases}{(a)}]. When the effects of quantum fluctuations are included ($J_{zz}\gg J_{\pm}, J_{\pm\pm} \gg J_{z\pm}$), a \ac{QSL} is known to be induced, the so-called \ac{QSI} phase, described by a $U(1)$ gauge theory with an emergent photon excitation (see Ref.~[\onlinecite{gingras2014quantum}] for a review). This phase is stable to all perturbations and thus occupies a finite region of phase space about the classical spin ice point~\cite{hermele2004pyro}. Other exotic phases have been proposed to be nearby the classical \ac{SI} point, such as the Coulomb ferromagnet of Ref.~[\onlinecite{savary2012coulombic}], though clear evidence of such a phase in the nearest-neighbor models of Eq.~(\ref{eq:effective-spin}-\ref{eq:dipolar-octupolar}) has yet to be found~\cite{mcclarty2015chain}.

Finally, we note that there exist other kinds of extensively degenerate manifolds in the classical phase diagram. The most prominent of these is that of the anti-ferromagnetic Heisenberg point (for example, near $J_{zz} = -2J_{\pm}$, $J_{\pm\pm} = J_{z\pm}=0$). Classically, this model is known to host a spin liquid~\cite{moessner1998pyro} phase with Coulomb correlations, similar to that found in classical \ac{SI}. Much of the physics of the Heisenberg model in the quantum limit is unknown, though a variety of exotic states have been proposed~\cite{canals1998pyrochlore,henley2006pyroobd,burnell2009}. A more unusual classical degenerate manifold, found in a highly anisotropic regime, was recently studied in Ref.~[\onlinecite{benton2016spin}] and has a description in terms of a higher-rank gauge theory. How such a manifold responds to the inclusion of quantum fluctuations is unknown, but is a topic of current interest.

\begin{table}[ht]
 \begin{tabular}{ccc} 
\toprule
  Name(s) & Order parameter & Examples \\
  \midrule
  \begin{tabular}{c}
   $A_{1g}$ ($\Gamma_1$), \\
   Octupolar all-in/all-out
  \end{tabular} & \footnotesize $M_{A_{1g}} \equiv S^y_1+S^y_2+S^y_3+S^y_4$ & None \\ \midrule
  \begin{tabular}{c}
   $A_{2g}$ ($\Gamma_3$)
  \end{tabular} & \footnotesize $M_{A_{2g}} \equiv S^z_1+S^z_2+S^z_3+S^z_4$ & \abo{Nd}{Zr}, \\
  \begin{tabular}{c}
   All-in/All-out
  \end{tabular} & \footnotesize $M_{A_{2g}'} \equiv S^x_1+S^x_2+S^x_3+S^x_4$ & \abo{Nd}{Hf} \\ \midrule
  \begin{tabular}{c}
   $T_{1g}$ ($\Gamma_9$), \\ Ordered spin ice   
  \end{tabular} &\footnotesize $\vec{M}_{T_{1g}} \equiv \left(\begin{array}{c}
                                 S^z_1+S^z_2-S^z_3-S^z_4\\
                                 S^z_1-S^z_2+S^z_3-S^z_4\\
                                 S^z_1-S^z_2-S^z_3+S^z_4   
                                \end{array}\right)$ & None \\
  \begin{tabular}{c}
  \end{tabular} &\footnotesize $\vec{M}_{T_{1g}'} \equiv \left(\begin{array}{c}
                                 S^x_1+S^x_2-S^x_3-S^x_4\\
                                 S^x_1-S^x_2+S^x_3-S^x_4\\
                                 S^x_1-S^x_2-S^x_3+S^x_4   
                                 \end{array}\right)$ & \\ \midrule
  \begin{tabular}{c}
   $T_{2g}$ ($\Gamma_7$), \\ Ordered- \\ octupolar spin ice
  \end{tabular} &\footnotesize $\vec{M}_{T_{2g}} \equiv \left(\begin{array}{c}
                                 S^y_1+S^y_2-S^y_3-S^y_4\\
                                 S^y_1-S^y_2+S^y_3-S^y_4\\
                                 S^y_1-S^y_2-S^y_3+S^y_4   
          \end{array}\right)$ & None \\ \bottomrule
 \end{tabular}
\caption{
 \label{tab:dipolar-octupolar} Classical order parameters for each of the six types of $\vec{Q}=0$ phases for dipolar-octupolar doublets~\cite{huang2014do} on the pyrochlore lattice in the local basis. The two types of all-in/all-out order can mix freely, as can the two types of ordered spin ice states.
}
\end{table}
\section{Order by disorder in Erbium pyrochlores}
\label{sec:eto}

With this background in hand, we consider specific material examples from the \abo{A}{B} family of pyrochlores that realize different aspects of this physics.  The first of the quantum pyrochlores that we discuss is \eto{}. This compound is the most conventional of this family of materials, but still harbors many surprises and rich physics at low energy. We will introduce the physics of this material somewhat ahistorically, but will highlight the experimental and theoretical milestones that moved our understanding forward as we encounter them. Altogether, \eto{} represents a beautiful example of a frustrated anisotropic quantum magnet that can be, and has been, understood in fine detail both experimentally and theoretically.

At the atomic level, \rth{Er} has a $J=15/2$ free ion ground state manifold, with the crystal field selecting an effective spin-1/2 ground state. Since it is reasonably well separated~\cite{champion2003obd} from the excited crystal field levels ($\Lambda \sim 6\meV$), the nearest-neighbor effective spin-1/2 model of Eq.~(\ref{eq:effective-spin}) should be a reasonable description. Based on XY-like moments~\cite{huiskamp-1969-physica,champion2003obd} and the negative Curie-Weiss temperature, early studies~\cite{champion2003obd} assumed that the physics was that of a Heisenberg anti-ferromagnet~\cite{champion2003obd}, but with the moments pinned to the local XY plane. In the local basis, this maps to exchange parameters $J_{\pm} > 0$, $J_{\pm\pm} = 2J_{\pm}$ and $J_{zz}=J_{z\pm} = 0$. Classically, such a model includes the $\Gamma_5$ manifold as a ground state (see Sec.~\ref{sec:phases}), though its full ground state manifold is larger and more complicated~\cite{champion2004soft,mcclarty2014obdxy}. More realistic values of the anisotropic exchange parameters have since been determined by fitting the spectra observed in inelastic neutron scattering at high fields. For example, Ref.~[\onlinecite{savary2012obd}] finds good agreement with the parameters
\begin{align}
 \label{eq:savary-eto}
 J_{zz}
 &= -2.5 \cdot 10^{-2} \meV,
 &
 J_{\pm}
 &= +6.5 \cdot 10^{-2} \meV, \nonumber \\
 J_{\pm\pm}
 &= +4.2\cdot 10^{-2} \meV, &
 J_{z\pm}
 &= -0.88 \cdot 10^{-2} \meV.
\end{align}
where the $g$-factors were determined to be $g_z = 2.45$ and $g_{\pm} = 5.97$. While qualitatively similar to the na\"ive expectations for an easy plane XY anti-ferromagnet, with $J_{\pm},J_{\pm\pm} \gtrsim |J_{zz}|, |J_{z\pm}|$, there are some significant differences, with no clear separation of scales between the four exchanges. Other studies~\cite{petit2014obd} have found similar values for the exchanges. In addition, the predictions of these parameters [Eq.~(\ref{eq:savary-eto})] for thermodynamic quantities at high-temperature are consistent with experiments~\cite{oitmaa2013obd}.

Experimentally, \eto{} orders anti-ferromagnetically at $T_N \sim 1.2\K$ via a second-order phase transition~\cite{champion2003obd}. Early neutron diffraction measurements indicated that the ordered phase was drawn from the $\Gamma_5$ manifold, consistent with early expectations as well as the parameters of Eq.~(\ref{eq:savary-eto}). Later experiments~\cite{poole2007magnetic} identified the ground state as being $\psi_2$ (see Sec.~\ref{sec:phases:classical}), establishing that the degeneracy expected classically is indeed lifted below $T_N$.  Exploring and explaining the mechanism of this degeneracy lifting has been the focus of much of the theoretical and experimental works on \eto{}.

\subsection{Order-by-disorder and ground state selection}

\begin{figure}
    \centering
    \includegraphics[width=\columnwidth,valign=t]{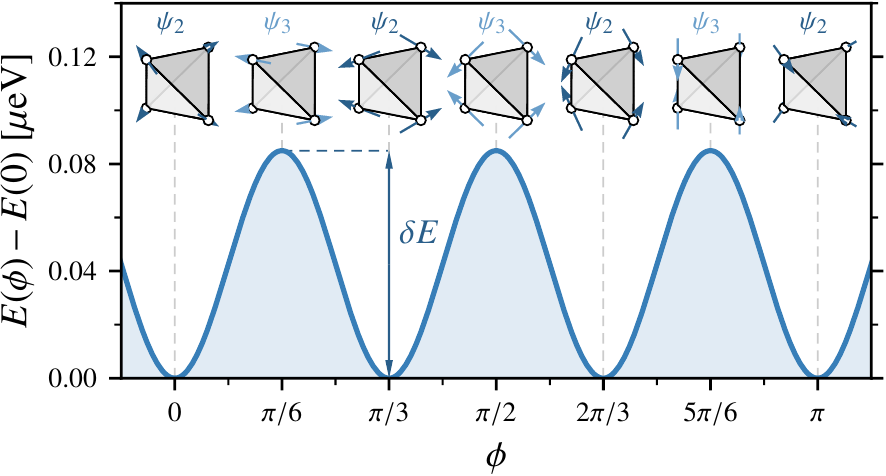}
    \caption{\label{fig:eto-zpm}
    Quantum zero point energy, $E(\phi)$, as a function of $\Gamma_5$ state (inset) indexed by the angle $\phi$, computed at $O(1/S)$ in spin-wave theory using the parameters of Ref.~[\onlinecite{savary2012obd}].
    }
\end{figure}

Much of the interest in the physics of \eto{} stems from the accidental classical degeneracy (i.e. not due to a symmetry) of these $\Gamma_5$ states. How this is resolved in practice has been dubbed ``order-by-disorder''~\cite{villain1980order,shender1982antiferromagnetic,henley1989}. Interpreted broadly, this refers to the lifting of an accidental degeneracy that appears in some artificial limit (e.g. classical spins, zero temperature, no disorder, etc) upon moving away from this limit. 

One of the earliest and most commonly discussed flavor of order-by-disorder is order-by-\emph{thermal}-disorder~\cite{villain1980order}. In this case, one has a degeneracy in the classical energy at zero temperature, but not in the free energy at any finite temperature. In other words, while the energies of the states are the same, the landscape of nearby excited states is different, and thus the entropies are different. We will discuss several other kinds of order-by-disorder, but they all share this same essential character. Several types relevant for \eto{} are:
\begin{enumerate}
\item Order-by-\emph{thermal}-disorder~\cite{zhitomirsky2012obd,zhitomirsky2014obd,mcclarty2014obdxy,oitmaa2013obd,javanparast2015}: degeneracy lifted through finite temperature.
 Two distinct types: near zero temperature ($T \rightarrow 0^+$) and near criticality ($T \rightarrow T_N - 0^+$).
\item Order-by-\emph{quantum}-disorder~\cite{savary2012obd,zhitomirsky2012obd,mcclarty2014obdxy}: degeneracy lifted through quantum zero point spin fluctuations in the semi-classical limit ($1/S \rightarrow 0$).
\item Order-by-\emph{structural} disorder~\cite{andreanov2015obd,maryasin2014obd}: degeneracy lifted by the dilution of the magnetic Er$^{3+}$ ions or the introduction of random exchange disorder.
\item Order-by-\emph{virtual-crystal-field-fluctuations}~\cite{mcclarty2009energetic,rau2016vcff}: degeneracy lifted by
 higher-order multi-spin interactions generated by virtual crystal field corrections (see Sec.~\ref{sec:vcff}).
\end{enumerate}

The selection of the ground state in \eto{} is a particular good starting point to study order-by-disorder since the accidental classical degeneracy is present for any symmetry allowed two-spin interactions of arbitrary range~\cite{savary2012obd}. This implies that any selection must proceed through some fluctuation effect (e.g. quantum, thermal, etc), otherwise proceeding energetically via multi-spin interactions. This can be seen straightforwardly in a Landau-Ginzburg description: due to the high symmetry of the pyrochlore lattice, the effective free-energy for the $\Gamma_5$ order parameter~\footnote{Allowing for a canting of the moments away from the local XY plane endows the Landau-Ginzburg free-energy with fourth-order terms~\cite{wong2013obd,javanparast2015} such as $\sim m^3 m_z \cos(3\phi)$ where $m$ is magnitude of the XY part, and $m_z$ the out-of-plane part. Such terms can be removed by solving for the equilibrium value of $m_z \propto m^3 \cos(3\phi)$, thus obtaining a free-energy of the form given in Eq.~(\ref{eq:landau-eto})} takes the form
\begin{equation}
 \label{eq:landau-eto}
 F[\vec{m}] \sim A_2 m^2 + A_4 m^4 + A_6 m^6 - B_6 m^6 \cos(6\phi) + \cdots,
\end{equation}
where $\vec{m} \equiv m(\cos{\phi}\vhat{x} + \sin{\phi}\vhat{y})$ describes the pseudo-spin configuration in the local basis. The vector $\vec{m}$ transforms in the $E_g$ representation of the point group of the pyrochlore lattice, with the angle $\phi$ tuning between the different states of the $\Gamma_5$ manifold (as shown in Fig.~\ref{fig:eto-zpm}). This immediately implies that, at the classical level, multi-spin interactions are needed to break the degeneracy since the classical energy directly maps to something of the form of Eq.~(\ref{eq:landau-eto}) when evaluated for a $\Gamma_5$ state, but with only the parts quadratic in $m$. Assuming the terms higher than sixth order are small, the selection effect is encoded in the sign of $B_6$, with $B_6 > 0$ selecting $\psi_2$ and $B_6<0$ selecting $\psi_3$, with an overall energy difference of $\delta E = 2 B_6$. 

This selection energy from order-by-quantum-disorder can be explicitly computed at $O(1/S)$ in the semi-classical, $S \rightarrow \infty$ limit~\cite{savary2012obd,zhitomirsky2012obd}. At leading order, one can describe the small fluctuations about the ordered state as a set of independent bosonic magnon modes. Each of these modes contributes to the ground state energy through its zero-point motion, distinguishing the classically degenerate $\Gamma_5$ states. One finds that this zero-point energy selects $\psi_2$ at $O(1/S)$, as is found experimentally~\cite{poole2007magnetic}. The dependence of this zero-point energy on the $\Gamma_5$ angle, $\phi$, is illustrated in Fig.~\ref{fig:eto-zpm}, where one finds a very small energy difference (per pseudo-spin) of $\delta E \sim 0.086 \mu{\rm eV}$ between the $\psi_2$ and $\psi_3$ states. In reality, several order-by-disorder effects should be operational in \eto{}: since the effects of quantum fluctuations or any direct multi-spin terms are not tunable in any reasonable way, all should be present~\cite{mcclarty2009energetic,rau2016vcff}. 

Determining the order-by-disorder mechanism is thus a quantitative question, and one may ultimately only be able to identify a mechanism as being dominant over all others. With these theoretical expectations for order-by-disorder in \eto{}  having been outlined, we next consider some of the implications of order-by-disorder for experiments in this material.

\subsection{Pseudo-Goldstone mode}

\begin{figure}
    \centering
    \includegraphics[width=0.7\columnwidth,valign=t]{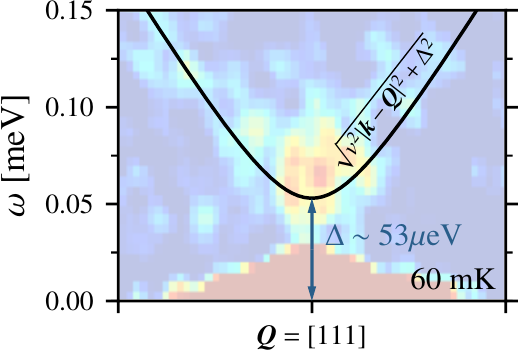}
    \caption{\label{fig:eto-pg}
    Inelastic neutron scattering intensity at low-energy near $\vec{Q} = [111]$ showing the pseudo-Goldstone mode with small gap of $\Delta \sim 53\ \mu {\rm eV}$. Adapted from \citet{ross2014obd}.
    }
\end{figure}

A key feature of order-by-disorder scenarios is the presence of a nearly gapless mode in the spectrum. This ``pseudo-Goldstone mode''~\cite{champion2003obd} is a remnant of the accidental classical degeneracy, with the cost to produce long-wavelength deformations within this manifold being small, but made non-zero due to order-by-disorder. The presence of such a mode in \eto{} is a necessary consequence of order-by-disorder and thus provides a quantitative goalpost for benchmarking any theory of the order-by-disorder mechanism. 

More explicitly, one expects classically that at zero temperature the $\Gamma_5$ ordered phase will be characterized by a gapless magnon, $\sim v|\vec{q}-\vec{Q}|$, emanating from the magnetic Bragg peak at $\vec{Q} \sim [111]$. Including the effects of any type of order-by-disorder will induce a gap $\Delta$ and give a spectrum of the form $\sim \sqrt{v^2|\vec{q}-\vec{Q}|^2 + \Delta^2}$, as illustrated in Fig.~\ref{fig:eto-pg}. One roughly expects this gap to be related to the magnitude of the selection energy as $\Delta^2 \sim J \delta E$, where $\delta E = 2B_6$ is the energy difference between $\psi_2$ and $\psi_3$ states and $J$ is a typical exchange scale~\cite{jensen1991rare}. If the effects of order-by-disorder are weak, in the sense that the appropriate classical limit is nearly reached, the gapping of this pseudo-Goldstone mode, $\Delta$, may be much smaller than the dominant energy scales of the problem. For example, if one estimates the gap using the selection energy, $\delta E$, found at $1/S$ in spin-wave theory one finds $\Delta \sim 0.03 \meV$~\cite{savary2012obd}, consistent with what is found using interacting spin-wave theory~\cite{rau2018pseudo}. This is significantly smaller than the bandwidth of the magnon excitations in \eto{}, which span roughly $\sim 0.5 \meV$~\cite{ruff2008spinwaves,savary2012obd}.  

Evidence for such a pseudo-Goldstone mode in \eto{} was first detected experimentally in the approximately $\sim T^3$ specific heat at low temperature~\cite{huiskamp-1969-physica,champion2003obd,derotier2012eto,sosin2010eto}. More directly, early inelastic neutron scattering experiments~\cite{ruff2008spinwaves} observed an approximately linearly dispersing mode near $\vec{Q} \sim [111]$, but could not resolve the gap, bounding it to be $\Delta \lesssim 0.1\meV$. More recent detailed inelastic neutron scattering studies~\cite{ross2014obd,petit2014obd,lhotel2017gap} focusing on this mode have resolved the gap and found it so be $\Delta \sim 0.04-0.05 \meV$, an order of magnitude smaller than the na\"ive energy scale (see Fig.~\ref{fig:eto-pg}). The presence of this mode in \eto{}, with its strongly suppressed energy scale, is a clear experimental signal that one or more of the aforementioned order-by-disorder mechanisms is at work in \eto{}.

\subsection{Current and future directions}

As we have seen in the previous sections, the presence of order-by-disorder has been reasonably well-established in \eto{} through a variety of direct experimental observations and theoretical arguments. Much of the more recent interest in \eto{} has focused on the effects of perturbations to the selection at low-temperature, such as through the application of magnetic fields or through the introduction of non-magnetic ions, such as \rth{Y}, to dilute the magnetic \rth{Er} sites.

The effect of a magnetic field on \eto{} has a number of interesting regimes, such as quantum criticality in a $[1\cb{1}0]$ field~\cite{ruff2008spinwaves}, as well as rich physics at low-fields where aspects of the order-by-disorder come into play. This low-field behavior has been explored in detail in recent theoretical~\cite{maryasin2016clock} and experimental~\cite{gaudet2017clock,lhotel2017gap} studies. Due to the small energy scales, the competition between the order-by-disorder selection and the selection by the applied field occurs at small magnetic fields, of order $\lesssim 0.5\ {\rm T}$. This selection is highly direction dependent, with qualitatively different behavior for $\psi_2$ and $\psi_3$ ordering.  Experimentally, most of these transitions have been observed in \eto{}, and are qualitatively consistent with theoretical expectations~\cite{maryasin2016clock,gaudet2017clock}. Given the inherent tunability of an applied field, coupled with the diverse set of behaviors predicted, the exploration of this low-field behavior serves as an instructive setting for addressing quantitative questions about order-by-disorder in \eto{}.

The question of order-by-structural disorder has also been studied in some detail recently. Early theoretical calculations predicted that site dilution or exchange disorder would favor the selection of a $\psi_3$ ground state~\cite{maryasin2014obd,andreanov2015obd}. At larger disorder strengths, recent studies~\cite{andrade2018clusterglass} find a ``cluster spin glass'' phase, consisting of a frozen mix $\psi_2$ and $\psi_3$ clusters, past a dilution of $\sim 35\%$ or so. When the competition with the selection effects in the clean limit is considered, one expects a transition between a $\psi_2$ and $\psi_3$ at some critical dilution or disorder strength which can be estimated to be roughly $10\%-30\%$ for \eto{}, based on the size of measured pseudo-Goldstone gap. One then expects the clean $\psi_2$ phase to evolve into such a cluster glass phase, with the appearance of a $\psi_3$ phase at intermediate dilution depending on the precise value of the order-by-disorder selection energy scale in the clean limit. Experimentally, this has been realized by the synthesis of solid solutions of the form Er\tsub{2-$2\delta$}Y$_{2\delta}$Ti$_2$O$_7$ with $\delta=0, 0.1$ and $0.2$~\cite{niven2014magnetic,gaudet2016dilution}. Analysis of elastic and inelastic neutron scattering data has been interpreted as being consistent with the $\delta = 0.1$ sample being in the $\psi_2$ phase (as in the clean limit) and the $\delta = 0.2$ sample being in the cluster glass phase. Whether there is an intermediate $\psi_3$ phase between these two dilution strengths~\cite{andrade2018clusterglass} remains an open question.

The physics discussed in this section has focused on \eto{}, though there are several rare-earth pyrochlore materials that share much of the same physics. This includes \abo{Er}{Ge}, which has been suggested to exhibit a $\psi_3$ ground state~\cite{dun2015ergeo} (rather than a $\psi_2$ state) and \abo{Yb}{Ge}~\cite{hallas2016ybgeo}, which is also thought to host a $\psi_3$ ground state (as we will discuss in Sec.~\ref{sec:yto}). Further detailed investigations of \eto{}, as well as \abo{Er}{Ge} and \abo{Yb}{Ge}, should prove valuable in improving our global understanding of order-by-disorder physics.

\section{Exotic order in Ytterbium pyrochlores}
\label{sec:yto}
We next consider the \abo{Yb}{Ti} pyrochlore which has received significant attention as a potential candidate for realizing \ac{QSI}~\cite{ross2011quantum}. Despite these early expectations, \yto{} is, in some sense, more conventional than \eto{}, appearing to order into a simple ferromagnet~\cite{yasui2003ferromagnetic,chang2012higgs,yaouanc2016novel,gaudet2016gapless,kermarrec2017ground,viv2017excitations}, without any residual classical degeneracy. However, this ferromagnetic state unexpectedly shows highly unusual dynamical behavior, yielding a puzzle as to what underpins the physics in this material. 

The atomic physics of \rth{Yb} is simple, described by a single-hole in the $4f$ manifold, which has a $J=7/2$ free-ion ground state manifold. Due to the large crystal field energy scale of $\sim 75\meV$~\cite{gaudet2015cef}, we can restrict to only the ground effective spin-1/2 doublet, and thus the model of Eq.~(\ref{eq:effective-spin}) should be a very good description. Attempts at determining the appropriate exchange parameters for \abo{Yb}{Ti} have yielded several quite different models~\cite{ross2011quantum,petit2015dynamics,thompson2017}, depending on exactly what data was analyzed and what sample was studied. The most systematic of these studies is the recent work of Ref.~[\onlinecite{thompson2017}] where inelastic neutron scattering spectra in large $[001]$ magnetic fields was analyzed, finding
\begin{align}
 \label{eq:coldea-yto}
 J_{zz}
 &= +2.6 \cdot 10^{-2} \meV,
 &
 J_{\pm}
 &= +7.4 \cdot 10^{-2} \meV, \nonumber \\
 J_{\pm\pm}
 &= +4.8\cdot 10^{-2} \meV, &
 J_{z\pm}
 &= -15.9 \cdot 10^{-2} \meV.
\end{align}
One sees that the $J_{z\pm}$ coupling is dominant, with these parameters far from the \ac{QSI} regime put forth in early studies. The $g$-factors were also determined simultaneously, finding $g_z \sim 2.14$ and $g_{\pm} \sim 4.17$. A similar set of exchanges parameters was also reported in Ref.~[\onlinecite{petit2015dynamics}].
 
Experimentally, though there is some sample dependence, the current consensus appears to be that \abo{Yb}{Ti} orders into a \ac{SFM} state through a first-order transition at $T_c \sim 0.25 \K$~\cite{yasui2003ferromagnetic,chang2012higgs,yaouanc2016novel,gaudet2016gapless,kermarrec2017ground,viv2017excitations}, with the magnetic moments roughly along the cubic $[100]$ axes, with small splay angle~\footnote{Some aspects of this transition are contentious, given that the magnetic Bragg peaks are difficult to resolve atop the structural Bragg peaks.}. This is consistent with classical expectations~\cite{thompson2017} for the exchange parameters of Eq.~(\ref{eq:coldea-yto}). The details of this transition, such as the height of the specific heat peak, the precise $T_c$ or the splay angle, depend on the specific sample considered~\cite{yasui2003ferromagnetic,chang2012higgs,yaouanc2016novel,gaudet2016gapless,kermarrec2017ground,viv2017excitations}, suggesting a sensitivity to disorder or internal stress~\cite{kermarrec2017ground}.

\subsection{Spin dynamics in the ordered state}

\begin{figure}[t]
    \centering
    \includegraphics[width=0.9\columnwidth,valign=t]{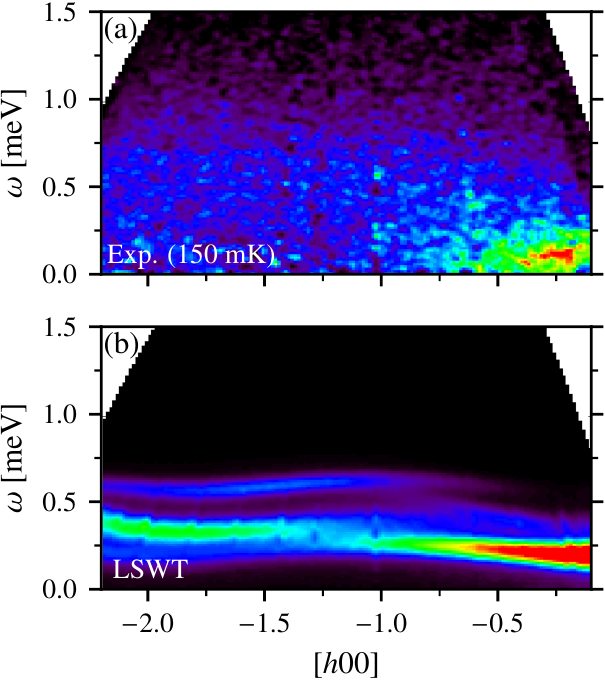}
    \caption{\label{fig:yto-swt}
    (a) Experimental excitation spectrum for \yto{} at $150$ mK and zero field inside the SFM phase~\cite{thompson2017} showing a broad continuum of excitations. (b) Prediction of linear spin-wave theory using the parameters~\cite{thompson2017} of Eq.~(\ref{eq:coldea-yto}) at zero-field showing well-defined magnon modes and a clear gap at zero wave-vector. (a,b) Adapted from \citet{thompson2017}.
    }
\end{figure}

The key mystery in \abo{Yb}{Ti}, and more broadly in \abo{Yb}{Ge} and \abo{Yb}{Sn}, lies in its zero-field spin dynamics. The na\"ive theoretical expectations for the dynamics in the \ac{SFM} state are entirely unremarkable. Due to the large anisotropy in the exchange interactions, there should be no gapless or nearly gapless modes and thus one should just expect a gapped magnon, perhaps with a minimum near $\vec{Q}=0$, the ordering wave-vector. Using the exchange parameters for \abo{Yb}{Ti} one would predict a one-magnon gap to be roughly $\sim 0.2 \meV$~\cite{thompson2017}, as illustrated in Fig.~\subref{fig:yto-swt}{(b)}. This is well within experimental bounds, and does not require the significant experimental effort put forth to observe the small order-by-disorder gap in \eto{}.

However, what is actually found in \yto{}, in both single-crystals~\cite{petit2015dynamics,thompson2017} and powder samples~\cite{hallas2016universal}, is in strong disagreement with these expectations. The observed spectrum is nearly gapless with no well-defined spin-wave modes. The spectrum has been described as a ``ferromagnetic continuum''~\cite{petit2015dynamics,hallas2016universal,gaudet2016gapless,thompson2017}, with a broad intensity distribution emanating from $\vec{Q}=0$, as shown in Fig.~\subref{fig:yto-swt}{(a)}. A similar continuum is observed in powder samples of \abo{Yb}{Sn} which has the same \ac{SFM} ground state~\cite{hallas2016universal}.

Even more striking is how this excitation spectrum evolves as a function of magnetic field. In the work of Ref.~[\onlinecite{thompson2017}], a field along $[001]$ was considered, as to preserve the symmetries of the \ac{SFM} ground state. It was found that this continuum evolves smoothly from the sharp spin-waves expected, and indeed found, at sufficiently large field, without any apparent phase transition. As one lowers the field, the one- and two-magnon excitations merge completely below $B \sim 1\ {\rm T}$~\cite{pan2014low,thompson2017}, signaling a strong departure from na\"ive semi-classical expectations. There are a number of exotic phases one could propose to explain the zero-field continuum; for example, the excitations could be a spinon particle-hole continuum of a gapless $U(1)$ spin-liquid~\cite{wang2016qsl} (but modified to carry a small ferromagnetic moment -- a kind of fractionalized \ac{SFM} state). Such proposals are, however, superficially at odds with the zero-field state being smoothly connected to the trivial high-field state.

The mystery deepens when considering the sister material \abo{Yb}{Ge}~\cite{hallas2014chem,hallas2016ybgeo}. This compound \emph{does not} order in a \ac{SFM} state, but into the $\Gamma_5$ manifold~\cite{hallas2014chem} in which \eto{} orders. Given the exchange regime found in \abo{Yb}{Ti}, one might theoretically expect that order-by-quantum-disorder in \abo{Yb}{Ge} will select a $\psi_3$ ground state~\cite{jaubert2015multiphase}. One then expects similar phenomenology to \eto{}: novel order-by-disorder physics atop what is essentially a simple classical ordered state~\cite{jaubert2015multiphase}. However, the spin dynamics of \abo{Yb}{Ge} show a similar ferromagnetic continuum as seen in \abo{Yb}{Ti}, despite its \emph{anti}-ferromagnetic $\Gamma_5$ ground state~\cite{hallas2016universal}. This seems somewhat contradictory; from the field dependence in \abo{Yb}{Ti}, the continuum of excitations would seem to be a property of the \ac{SFM} ground state, evolving smoothly into the trivial polarized paramagnet, yet it also appears in the \emph{anti}-ferromagnetic ground state of \abo{Yb}{Ge}.

What is driving this physics has not yet been resolved; it is not clear if the explanation is (in some sense) conventional, or if it necessitates structural disorder in some essential way or is truly exotic, such as involving proximity to a \ac{QSL} of some kind. In the next section, we outline some more conventional ideas that may underlie some of this physics.

\subsection{Splayed ferromagnet and multi-phase competition}

\begin{figure}
    \centering
    \includegraphics[width=0.9\columnwidth,valign=t]{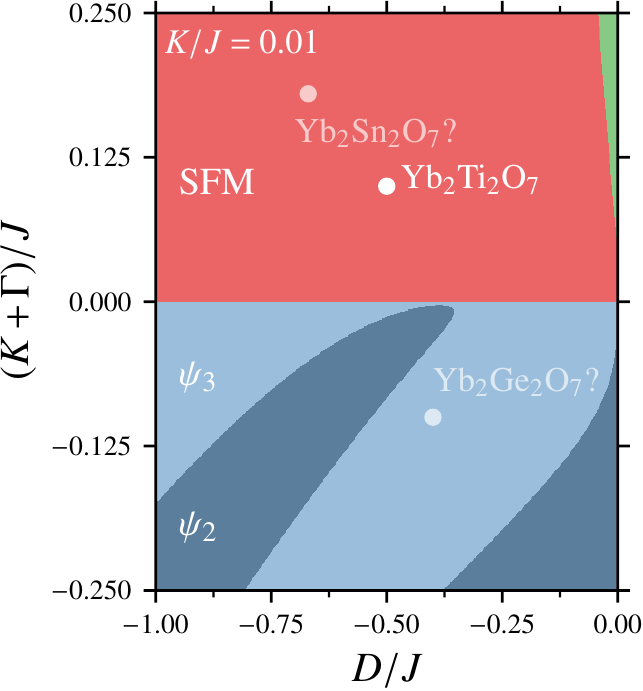}
    \caption{\label{fig:yto-phases}
    Phase diagram of model Eq.~(\ref{eq:effective-spin}) in the (dual) global basis (see Sec.~\ref{sec:yto}) showing the phase boundary between the splayed-ferromagnet (SFM) and
    the $\Gamma_5$ manifold relevant for \yto{}. Phases are identified using the same color scheme as in Fig.~\ref{fig:phases}. Hypothetical positions of the related \abo{Yb}{Sn} (SFM) and \abo{Yb}{Ge} ($\Gamma_5$) compounds are indicated. 
    }
\end{figure}
To get a better understanding of the \abo{Yb}{M} family (M = Ti, Ge, Sn) we focus on what is understood at the classical and semi-classical level, though it can only present a partial picture of what is going on in \yto{}. One distinctive feature of the exchange parameters of Eq.~(\ref{eq:coldea-yto}) is that they sit near a (classical) phase boundary between the \ac{SFM} state and the $\Gamma_5$ manifold of states~\cite{jaubert2015multiphase,petit2015dynamics}. This feature is also shared with several of the earlier proposals for the exchange interactions in \yto{}~\cite{ytorods,ross2011quantum,petit2015dynamics}. 

Some of the aspects of the classical phase boundary can be more clearly grasped by changing to a global frame for the exchanges. As mentioned in Sec.~\ref{sec:phases}, the four $(J_{zz},J_{\pm},J_{\pm\pm},J_{z\pm})$ exchanges of Eq.~(\ref{eq:effective-spin}) can be mapped~\cite{yan2017exchange} to four exchanges $(J,K,\Gamma,D)$ where $J$ is a Heisenberg exchange, $K$ and $\Gamma$ are Kitaev and pseudo-dipolar exchanges and $D$ is a \ac{DM} interaction. Before performing this mapping, we one can first apply the ``duality" discussed in Sec.~\ref{sec:two-ion}, rotating by $\pi$ about $\vhat{z}_i$ to flip the sign of $J_{z\pm}$~\cite{rau2018frustration}. For the parameters of Eq.~(\ref{eq:coldea-yto}), after reversing the sign of $J_{z\pm}$, one finds that $J$ and $D$ are dominant, with $J >0$ and $D<0$ and $K,\Gamma \ll J,|D|$. Classically, the model with $K = \Gamma = 0$ has been studied in Refs.~[\onlinecite{elhajal2005,canals2008,chern2010pyrochlore}], and it describes the relevant phase boundary between the \ac{SFM} and $\Gamma_5$ phases. As illustrated in Fig.~\ref{fig:yto-phases}, one finds that $K+\Gamma >0$ favors a \ac{SFM} state, while $K + \Gamma <0$ favors the $\Gamma_5$ states. Just as in \eto{}, several kinds of order-by-disorder~\cite{jaubert2015multiphase,petit2015dynamics} are expected to play an important role in the physics of the \ac{SFM}-$\Gamma_5$ phase boundary. First and foremost is the stabilization of a (primarily) $\psi_3$ state by order-by-quantum disorder at zero-temperature~\cite{jaubert2015multiphase}. This stabilization energy is not present on the \ac{SFM} side, and thus we expect quantum fluctuations to enlarge the $\Gamma_5$ region at the expense of the \ac{SFM} region~\cite{jaubert2015multiphase}.

The proximity of these compounds to this \ac{SFM}-$\Gamma_5$ phase boundary may play an important role in explaining the sensitivity of these compounds to structural disorder or (possibly) to the origin of the unusual dynamics. For example, it may alter some of the na\"ive expectations for the excitation spectrum of the \ac{SFM} phase that we outlined earlier. Other possibilities, within such a framework, may be more extrinsic and involve both the interplay between the $\Gamma_5$-\ac{SFM} competition and structural disorder. As we have seen in \eto{}, disorder can reveal a quite rich phenomenology for such systems with strongly competing ground states.

\subsection{Current and future directions}
One key experimental question is the role of structural disorder in these
all compounds. Does it play a fundamental role in the unusual dynamics? Or would this physics remain in a perfect, clean sample? Some of these questions could be answered by experiments that look at the dynamics in (nominally) better quality samples, such as those that were reported in Ref.~[\onlinecite{arpino2017}]. A complementary view is provided by experiments on \yto{} under applied pressure~\cite{kermarrec2017ground}. These experiments find that the application of moderate hydrostatic pressure can stabilize the \ac{SFM} state~\cite{kermarrec2017ground}, removing some of the ambiguities that have plagued its detection in samples under ambient conditions. This could indicate that most of the \yto{} samples under study could have some distribution of residual strains that are affecting the low-energy physics. While the application of pressure sharpens the appearance of the static order, it remains to be seen how the unusual dynamics is modified. An alternate interpretation is that pressure is pushing \yto{} away from the $\Gamma_5$-\ac{SFM} boundary~\cite{jaubert2015multiphase} and thus rendering its physics more classical, and less frustrated. Measurements of the dynamics of \yto{} under pressure would be very useful to resolve these questions.

More phenomenologically, one may try to gain some understanding from the ``universality'' of the appearance of the continuum of excitations~\cite{hallas2016universal}, given it appears in all three compounds, with both \ac{SFM} and $\Gamma_5$ ground states. One might speculate that there is a ``parent'' state from which the dynamics is natural and originates,  with the \ac{SFM} and $\Gamma_5$ states regarded as secondary instabilities. Within such an interpretation, the broad peak in specific heat at higher-temperatures, observed in all these compounds, could be viewed as signaling the entrance into a manifold of states representative of an  (unknown) parent state, with the ordering transitions being secondary instabilities. Studies of the recently synthesized \abo{Yb}{Pt}~\cite{cai2016platinum}, which has not been as well characterized, would be useful in the regard, though the somewhat different ionic physics of Pt\tsup{4+} could complicate a direct comparison.

We end this section with some of the key questions that remain to be answered for \yto{} and the related \abo{Yb}{Ge} and \abo{Yb}{Sn} materials:
\begin{enumerate}
\item Is there a parent state from which the unusual dynamics derives, and what is this parent state?
\item What role does the proximity to the $\Gamma_5$-\ac{SFM} boundary play?
\item To what extent do extrinsic effects, such as structural disorder, play an important role?
\end{enumerate}

\section{Spin liquid in Terbium pyrochlores}
\label{sec:tto}
The compound \tto{} has resisted attempts at understanding since its initial synthesis~\cite{huiskamp-1969-physica}. Unlike \eto{} and \yto{}, \tto{} appears to not  magnetically order and thus has long been considered a \ac{QSL} candidate. In this section, we offer a perspective on the current experimental status of \tto{} and sketch the challenges that have plagued theoretical efforts in explaining its behavior. The plain fact that \tto{} has resisted such attempts for so long is a testament to the depth and breadth of the physics exhibited by this compound. The numerous mysteries and puzzles to be addressed, only a few of which we touch on here, represent a definite opportunity, for both theory and experiment, to move forward our understanding of systems with complex local degrees of freedom and many competing orders.

At the free-ion level, \rth{Tb}, has a $J=6$ free ion ground state manifold which is further split by the crystal field, yielding a non-Kramers doublet~\cite{gingras2000iontto,mirebeau2007ttotsno,zhang2014ttocef,princep2015ttocef}. Unlike the \rth{Er} and \rth{Yb} cases, the first excited state, another non-Kramers doublet, is only separated by the relatively small energy $\Lambda \sim 1.4\meV$. This renders the nature of the two-ion interactions is more subtle. If the manifolds associated with the two low-lying levels do not cross (consistent with the experimental spectra~\cite{gardner2004spin,mirebeau2007ttotsno,guitteny2013ttoexcitations,fennell2014magnetoelastic2}), then a description in terms of the ground doublets should be possible. The question then falls to the size of the virtual crystal corrections discussed in Sec.~\ref{sec:vcff}. If significant, the physics of Eq.~(\ref{eq:non-kramers}) would then need to be augmented with include second- and third-neighbor (anisotropic) exchanges as well as three-spin terms in vein of Eq.~(\ref{eq:three-spin})~\cite{molavian2007dyn,molavian2009towards}.

While the crystal field scale $\Lambda \sim 1.4\meV$ is small relative the crystal field scale in other rare-earth magnets, it is still about an order of magnitude larger than typical exchange scales in rare-earths, for example as found in \eto{} or \yto{}. Taking $\mathcal{M} \sim 0.2 \meV$, one might expect the first correction to be of order $\mathcal{M}^2/ \Lambda \sim 0.02\meV \ll \mathcal{M}$, neglecting any matrix element effects. While likely not the whole story, to move forward in our discussion of \tto{}, we will thus adopt a somewhat pragmatic view, cautiously invoking the pseudo-spin model appropriate for a non-Kramers doublet, [Eq.~(\ref{eq:non-kramers})], as a first description of \tto{} and treating these virtual crystal field corrections as secondary effects.

Given the Ising-like magnetic moments, early efforts focused on scenarios based on a (argued) proximity to classical spin ice~\cite{gingras2000iontto,kao2003understanding,enjalran2004theory}. Such an interpretation faces several challenges, such as the wrong sign of Curie-Weiss temperature and the presence of paramagnetic scattering at $[002]$, which is forbidden for Ising-like moments~\cite{kao2003understanding,enjalran2004theory}. Some of these issues can be partially resolved by considering a simple model of (rank-1) Heisenberg exchange and dipolar interactions in the full $J$-manifold, as to include admixing with the higher crystal field states~\cite{molavian2007dyn,molavian2009towards}, effectively allowing for a finite transverse $g$-factor (see Sec.~\ref{sec:vcff}). This led to a picture where \tto{} sits near the boundary with the \ac{AIAO} states ($J_{zz}=0$) due to virtual crystal corrections inducing a strong renormalization of $J_{zz}$~\cite{molavian2007dyn}.

Some recent works~\cite{petit2012tto,guitteny2013ttoexcitations,takatsu2016quadrupole} have attempted to estimate the three $(J_{zz},J_{\pm},J_{\pm\pm})$ parameters from comparisons to experimental data, but the results are less conclusive than for \eto{} or even \yto{}. One representative example is provided by Ref.~[\onlinecite{takatsu2016quadrupole}] which proposes
\begin{align}
 J_{zz} &= 0.34 \meV , & J_{\pm} &= 0 \meV , & J_{\pm\pm} &= 0.146 \meV.    
\end{align}
These parameters put the system close to \ac{SI} and two quadrupolar ordered phases, analogues of the $\Gamma_5$ manifold discussed for \eto{} and \yto{}, and of the $T_{2g}$, ``Palmer-Chalker" state (see Fig.~\ref{fig:phases} and Table~\ref{tab:orders}). More controlled approaches to determining these parameters, such as the high field studies undertaken for \eto{} and \yto{}, are complicated by the presence of the low-lying first excited crystal field doublet at $\Lambda \sim 1.4\meV$.

\subsection{Spin liquid and correlations}
\begin{figure}[t]
    \centering
    \includegraphics[width=0.6\columnwidth,valign=t]{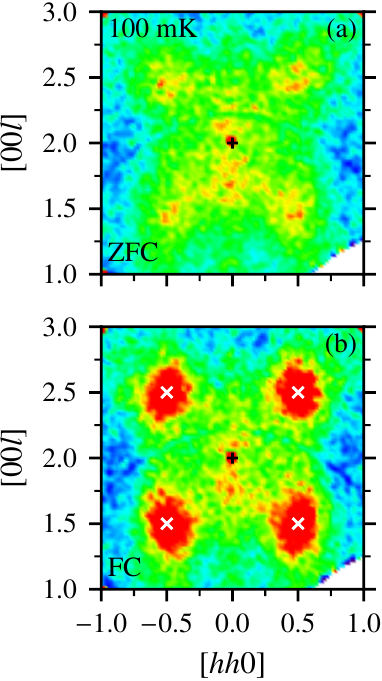}
    \caption{\label{fig:tto-hhh}
    (a,b) Broad scattering at (a) $[002]$ when zero-field cooled and (b) sharper quasi-Bragg peaks wave-vectors equivalent to $\hhh$ when field-cooled in slighty off stoichiometric Tb$_{2+x}$Ti$_{2-x}$O$_{7+y}$ with $x \sim -0.001$ at $T = 100 $mK. Adapted from \citet{kermarrec2015hhh3}.
    }
\end{figure}

The experimental study of \tto{} has a long and complicated history. A key question to be answered is the nature of the paramagnetic state that persists to low temperatures.
Early studies of the low-energy properties of \tto{} focused on this lack of ordering~\cite{gardner1999first,gardner2001tto}. This can be seen in the specific heat and the susceptibility, which show only broad features down to the lowest measured temperatures. This absence of ordering led to \tto{} being put forth as a \ac{QSL} candidate. 

However, the lack of clear ordering does not necessarily imply a \ac{QSL} ground state, more telling evidence can be found in the magnetic correlations. The development of a \emph{correlated} paramagnet in \tto{} can be seen as one goes below $\sim 20\K$~\cite{gardner2001tto}. This is most manifest in (nominally) elastic neutron scattering measurements where one observes an increase in diffuse elastic intensity around the wave-vector $[002]$ at low temperature~\cite{gardner2001tto}, as shown in Fig.~\subref{fig:tto-hhh}{(a)}. More recent experiments~\cite{fritsch2013hhh,fritsch2014hhh2} have observed that this feature sharpens into quasi-Bragg peaks near wave-vectors equivalent to $\hhh$, as shown in Fig.~\subref{fig:tto-hhh}{(b)}. In addition to these peaks, the surrounding diffuse intensity forms a distinctive ``butterfly'' shape~\cite{fennell2012powerlaw}. The appearance of these features is dependent on the experimental history: the $\hhh$ peaks appear under field cooled conditions~\cite{fritsch2014hhh2}, but not when zero-field cooled. This suggest some freezing at low temperatures, though not necessarily total, accompanied by nontrivial static magnetic correlations with characteristic wave-vector $\hhh$.

We now turn to dynamical probes. In addition to the presence of static correlations, it has been established through a variety of methods that there are dynamic magnetic correlations down to low temperatures~\cite{kanada1999neutron,gardner2003dynamictto}. This can be seen directly in the inelastic neutron scattering spectrum which shows broad low-energy excitations. More recent experiments, such as those of Refs.~\cite{guitteny2013ttoexcitations,takatsu2012ins,kadowaki2015composite,takatsu2016quadrupole}, see a small gap of $\sim 0.1 \meV$ with excitations extending in energy up to $\sim~0.3 \meV$ or so. The presence of such dynamics has been corroborated by $\mu$-SR measurements~\cite{gardner2003dynamictto}, as well as AC susceptibility measurements~\cite{gardner2003dynamictto,hamaguchi2004tto}. How the glassy behavior of the static $\hhh$ feature fits into such a picture is unclear.

The lack of ordering, the presence of both static and dynamic correlations at low temperatures suggests that \tto{} realizes some kind of spin liquid state, or possibly some kind of partially frozen glassy state. The nature of this state remains unclear, though new insights have been recently uncovered from the introduction of minute amounts of disorder into \tto{}.

\subsection{Stoichiometry and quadrupolar order}

\begin{figure}[t]
    \centering
    \includegraphics[width=0.9\columnwidth,valign=t]{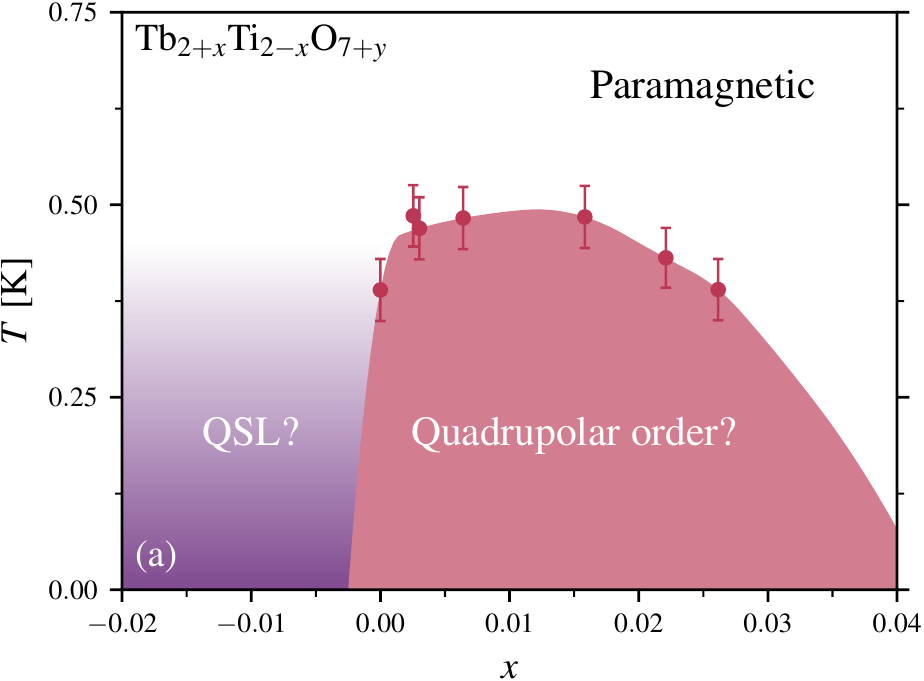}
    \caption{\label{fig:tto-sto}
    Phase diagram of Tb$_{2+x}$Ti$_{2-x}$O$_{7+y}$ as a function of the composition parameter $x$, showing the disordered and ordered phases appearing for $x \lesssim 0$ and $x \gtrsim 0$. Adapted from \citet{wakita2016quantum}. 
    }
\end{figure}
A relatively recent milestone~\cite{taniguichi2002longrangeorder} is the controlled synthesis of off-stoichiometric samples of \tto{}, with formula Tb$_{2-x}$Ti$_{2+x}$O$_{7+y}$. Such samples show an extreme sensitivity to the value of the composition parameter $x$, even in a small window of $1-2\%$ deviations from stoichiometry. It was found~\cite{taniguichi2002longrangeorder} that the lack of ordering observed previously only occurs for $-0.02 \lesssim x \lesssim 0$, with the $ 0\lesssim x \lesssim 0.03$ samples showing an ordering transition in the specific heat at $\sim 0.5 \K$, as illustrated in Fig.~\ref{fig:tto-sto}. The ordered state for $x \gtrsim 0$ samples appears to be non-magnetic, suggesting a quadrupolar ordering, such as the quadrupolar $E_g$ ($\Gamma_5$), $T_{1g}'$ or $T_{2g}$ (Palmer-Chalker) states discussed in Sec.~\ref{sec:phases}~\cite{kadowaki2015composite,takatsu2016quadrupole}. Such a quadrupolar phase has been argued to be qualitatively consistent the behavior of the phase diagram as a function of magnetic field~\cite{takatsu2016quadrupole}.  The sensitivity to very small deviations in stoichiometry also provides a plausible explanation for some of sample dependence reported in the literature for \tto{}.

Much of the phenomenology established in the early studies has been re-visited in light of this question of stoichiometry. The signatures of the (static) spin correlations seen in (nominally) stoichiometric samples are also present in off-stoichiometry samples~\cite{kermarrec2015hhh3}, independent of the presence of the transition at $\sim 0.5 \K$~\cite{taniguichi2002longrangeorder}. The dynamic correlations observed in inelastic neutron scattering have also been studied in the off-stoichiometric samples. As in the (nominally) stoichiometric samples, one finds an excitation gapped by $\sim 0.1 \meV$, but which sharpens significantly as one goes to larger values of $x$~\cite{kadowaki2018continuum}. This overall behavior has been interpreted as the proximity of \tto{} to the boundary of a \ac{QSI} phase and a quadrupolar ordered phase~\cite{takatsu2016quadrupole}, with these excitations interpreted as the pseudo-spin waves associated with the quadrupolar order~\cite{kadowaki2015composite}. However, direct detection of this putative quadrupolar order is still lacking.

\subsection{Discussion and outlook}
\label{sec:tto:discussion}

There are a number of major open questions concerning \tto{}. Some are theoretical questions that we have broached earlier, such as the applicability of the pseudo-spin model. If the virtual crystal field corrections are sufficiently large, this mapping may break down completely, suggesting it may be necessary to include \emph{both} low-lying doublets on equal footing. How the presence of such a ``pseudo-orbital" degree of freedom  would affect the physics remains to be seen. There are also several open experimental questions, such as the nature of the ordered state seen for $x \gtrsim 0$ samples and of the disordered state seen for $x \lesssim 0$. Further, there is the overall question of \emph{why} \tto{} is so sensitive deviations from stoichiometry, e.g. some manifestation of an extreme sensitivity to disorder. Studies of related compounds such as \abo{Tb}{Sn}, which shows magnetic ordering into an ice state~\cite{mirebeau2005ordered}, or \abo{Tb}{Hf}, which is thought to have strong structural disorder~\cite{sibille2017coulomb}, may shed some light on these questions.

Another important line of inquiry to which we have not done justice is that of magneto-elastic excitations~\cite{ruff2010magnetoelastic,fennell2014magnetoelastic2,constable2017vibronic} in \tto{}. This thread goes back to early experiments that studied its high temperature magneto-elastic properties~\cite{belov1983giant,mamsurova1986low}. While likely primarily  a single-ion effect, their size and temperature dependence highlights the large electric quadrupole moment carried by the \rth{Tb} electronic states. Due to the non-Kramers structure of the ground and first excited doublets, this large quadrupolar moment can carry over to the low-energy physics. Recent inelastic neutron scattering studies have found that these kinds of magneto-elastic effects persist in the low-energy spectrum. This can be directly seen in the hybridization of the phonon modes with low-lying crystal field levels at $\Lambda \sim 1.4 \meV$. Like the static correlations, this hybridization appears independent of the precise stoichiometry~\cite{ruminy2016magnetoelastic1}. What role such magnetoelastic effects play in the low-energy physics of \tto{} remains an interesting topic for further study.

We also note that there are a number of other experimental features of \tto{} that we did not touch upon here. This includes the rich behaviors observed under applied magnetic fields~\cite{ttodoublelayer,ttofield2,yin2013field,fritsch2014hhh2,takatsu2016quadrupole}, as well as the highly unusual features that have been observed in the thermal conductivity~\cite{hirschberger2015large}. With many fundamental questions remaining to be answered, the Tb$_2$M$_2$O$_7$ (M=Ti, Sn, Ge) family, more broadly, still presents promising opportunities to expand our understanding of highly frustrated magnetic systems, especially those with significant magneto-elastic couplings and pseudo-orbital degrees of freedom.

\section{Disorder and spin liquids in Praseodymium pyrochlores}
\label{sec:pr}
The final quantum pyrochlore we discuss is \abo{Pr}{Zr}, which has recently attracted attention as a \ac{QSI} candidate, both theoretically~\cite{savary2017disorder,benton2017quantum} and experimentally~\cite{kimura2013quantum,petit2016przro1,bonville2016przro2}.

\begin{figure}
    \centering
    \includegraphics[width=0.8\columnwidth,valign=t]{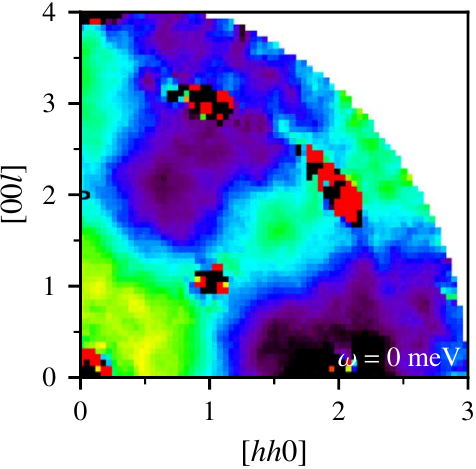}    
    \caption{\label{fig:przro-qe} Quasi-elastic neutron scattering intensity of \abo{Pr}{Zr} at $T = 0.1\K$ showing spin-ice-like features. Adapted from \citet{kimura2013quantum}.
    }
\end{figure}

The free-ion ground state of \rth{Pr} is a $J=4$ manifold, with a non-Kramers doublet selected by the crystal field in \abo{Pr}{Zr}~\cite{kimura2013quantum,bonville2016przro2}. Due to the large crystal field energy scale, $\sim 10 \meV$~\cite{kimura2013quantum,bonville2016przro2}, one can likely ignore virtual crystal field corrections, and thus the two-ion interactions take the form of Eq.~(\ref{eq:non-kramers}) with three independent couplings.  Early work~\cite{onoda2011se,kimura2013quantum}, estimated that this compound was near the \ac{QSI} limit, with $J_{zz} >0$ and $|J_{\pm}|, |J_{\pm\pm}| \ll J_{zz}$; more recent studies~\cite{petit2016przro1,martin2017disorder} have suggested larger values of $J_{\pm}$, with some uncertainty in $J_{zz}$. This was experimentally inferred from the lack of ordering~\cite{kimura2013quantum,bonville2016przro2}, the broad hump in the specific heat~\cite{kimura2013quantum,bonville2016przro2} and the spin-ice-like pattern~\cite{kimura2013quantum} in the (quasi-elastic) structure factor [see Fig.~\ref{fig:przro-qe}]. Some theoretical calculations of the exchange interactions have also suggested that such a regime was not unreasonable for the \abo{Pr}{M} family~\cite{onoda2011se}.

\subsection{Structural disorder}

The key feature missing from this description, that distinguishes \abo{Pr}{Zr} from the other pyrochlores we have discussed so far, is the presence of strong disorder.

\begin{figure}
    \centering
    \includegraphics[width=0.8\columnwidth,valign=t]{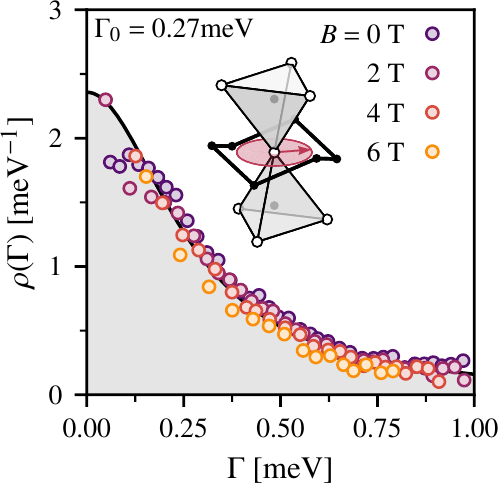}
    \caption{\label{fig:przro-dis} Collapse of the inelastic neutron scattering spectrum of \abo{Pr}{Zr} at $T = 1.4 \K$ for several magnetic fields onto a half-Lorentzian distribution $\rho(\Gamma) = 2\Gamma_0 (\Gamma^2 +\Gamma_0^2)^{-1}/\pi$ with $\Gamma_0 = 0.27\meV$. Only energies larger then the Zeeman energy are included.  Adapted from \citet{wen2017disorder}. (inset) Illustration of off-centering displacement of the \rth{Pr} ion that can give rise to a finite on-site transverse field.
    }
\end{figure}

The presence of structural disorder in \abo{Pr}{Zr} was first exposed in studies of its
inelastic neutron scattering spectrum~\cite{wen2017disorder}. At reasonably low temperatures, one finds a continuum of scattering that extends from low energies and out to several meV, as shown in Fig.~\ref{fig:przro-dis}. Accounting for detailed balance, one finds that this continuum persists essentially unchanged to high temperatures~\cite{wen2017disorder}. This strongly suggests that this represents a distribution of quenched structural disorder that strongly affects the magnetic physics of the \rth{Pr} ions.

Recall that the effect of structural disorder on non-Kramers doublets is more dramatic than for the effective spin-1/2 or dipolar-octupolar doublets. For Kramers ions, since the degeneracy of their doublet is protected by time-reversal symmetry, structural distortions can only directly introduce disorder in the two-ion interactions. In contrast, for non-Kramers doublets, if the $D_{3d}$ site symmetry is broken, for example by disorder or strain, the $E_g$ doublet is no longer protected and splits. Within the manifold of pseudo-spin states, such a perturbation maps to a transverse field
\begin{equation}
 \sum_{i} \left[ {\Gamma}^{}_i S^+_i + \cc{\Gamma}_i S^-_i  \right],
\end{equation}
where $\Gamma_i \equiv |\Gamma_i| e^{i\theta_i}$ depends on the microscopic details of the disorder. These local (transverse) ``fields'' must also appear in concert with any exchange disorder which is also inevitably induced by structural disorder; the relative of importance of such bond disorder remains to be seen.

If we assume that this transverse-field disorder is dominant, then one can extract a probability distribution for $|\Gamma_i|$ from the continuum observed experimentally. The analysis of Ref.~[\onlinecite{wen2017disorder}] finds a ``half"-Lorentzian distribution
\begin{equation}
 \rho(\Gamma) = \frac{2}{\pi}\left(\frac{\Gamma_0}{\Gamma^2 + \Gamma_0^2}\right),
\end{equation}
with $\Gamma_0 \sim 0.27 \meV$ providing a good description [see Fig.~\ref{fig:przro-dis}]. These random fields represent a significant perturbation to typical rare-earth exchange scales and thus should be critical in understanding the physics of \abo{Pr}{Zr}. This distribution of transverse-fields is also consistent with the strong reduction and smearing of the hyperfine contribution to the specific heat observed in early studies~\cite{kimura2013quantum}; the large transverse fields pin the pseudo-spins in the local XY (quadrupolar) directions which couple only weakly with the nuclear spins~\cite{martin2017disorder}.

The spatial distribution and origin of this disorder is less clear, though there have been some tantalizing experimental hints. First, it was found in Ref.~[\onlinecite{wen2017disorder}] that the $[200]$ Bragg peak, forbidden in a clean sample, is (weakly) visible in \abo{Pr}{Zr}. This forbidden peak also has an anomalously large width, suggesting a correlation length of $\xi \sim 23\AA$ (two unit cells) for the structural disorder~\cite{wen2017disorder}. Second, some allowed Bragg peaks, such as $[220]$ remain resolution limited. A simple picture that is consistent with these features is that the structural disorder is ``off-centering'' of the \rth{Pr} ions, with the rare-earth ions displaced in the plane perpendicular to the local $\vhat{z}_i$ axis, as illustrated in the inset of Fig.~\ref{fig:przro-dis}. Rough point-charge estimates of the transverse fields induced by such displacements give a length scale of $\sim 0.1\AA$ to get splittings comparable to what is seen experimentally. Such small off-centering disorder is also consistent with the anisotropic uncertainties reported in refinements of the crystal structure of \abo{Pr}{Zr}~\cite{koohpayeh2014synthesis}. Similar off-centering has also been studied in the related (non-magnetic) pyrochlore compounds  \abo{La}{Zr}~\cite{tabira1999structured,tabira2001annular} and \abo{Bi}{Ti}~\cite{shoemaker2011biruo}. A more detailed analysis of the diffuse part of elastic neutron scattering is broadly consistent with this interpretation~\cite{martin2017disorder}, finding evidence for a distribution of random strains in \abo{Pr}{Zr}.

A reasonable description of \abo{Pr}{Zr} must then include these strong random transverse fields. What kind of ground state this leads to depends on the details of the exchanges as well as the relative importance of exchange disorder. Evidence of static spin correlations in the diffuse neutron scattering, as shown in Fig.~\ref{fig:przro-qe}, suggest that the exchanges may favor ice-like states~\cite{kimura2013quantum,petit2016przro1,bonville2016przro2,martin2017disorder} -- implying a purely single-ion picture is not sufficient to fully understand \abo{Pr}{Zr}. For strong exchange and weak exchange disorder, one might then expect a disordered variant of \ac{QSI}~\cite{savary2017disorder,benton2017quantum}, though the visibility of such phase could be limited to extremely low temperatures. For weak exchange, one would likely find a frozen quadrupolar ``glass'', with the pseudo-spins pinned along the direction preferred by the local disorder configuration. Since the distribution, $\rho(\Gamma)$, appears to extend to small $\Gamma$, even in such a case there should be regions where the disorder is weak and the exchange dominates. Most interesting perhaps is the intermediate regime, where the exchange interactions and the disorder are on equal footing. How the competition between these two (na\"ively) opposing tendencies is actually borne out in \abo{Pr}{Zr} remains an interesting and open question.

\subsection{Discussion and outlook}
There remain several questions to resolve for \abo{Pr}{Zr}, as well as new and intriguing experimental observations that demand explanation and understanding. Indeed, the study of \abo{Pr}{Zr}, with its intrinsic disorder, may provide some bearing toward understanding the sample dependence found in \yto{} or the extreme sensitivity to off-stoichiometry found in \tto{}.

First, there is the very recent work of Ref.~[\onlinecite{tokiwa2018discovery}] which finds an anomalous upturn in the thermal conductivity of \abo{Pr}{Zr} at $\sim 0.5 \K$. This increase is interpreted as a feature of a \ac{QSI} ground state, signaling the onset of thermal transport by the emergent photon excitation~\cite{hermele2004pyro}. In light of the evidence for significant on-site disorder, this interpretation likely requires additional justification, for example, via a demonstration that disorder is less important in the samples in which the thermal conductivity was measured. However, the appearance of the peak is still unusual in itself: within a picture of strong transverse field disorder, how can this increase be explained? One possibility involves a renormalization of the phonon velocity due to interaction with the pseudo-spins; this could be significant since they carry quadrupolar moments and thus could have significant magneto-elastic couplings. The hyperfine interactions could also play a role, as they onset at a comparable temperature; while they cannot contribute directly to thermal transport, they could in principle affect the pseudo-spins and thus the phonons through spin-phonon interactions.

A more promising route toward a \ac{QSI} state in these compounds may be through the sister compound \abo{Pr}{Hf}~\cite{anand2016prhfo1,sibille2016prhfo}. In this compound, very similar phenomenology is observed: no ordering, a broad specific heat peak, and a spin-ice-like pattern in neutron scattering. However, there is no reported evidence of the same kind of structural disorder seen in \abo{Pr}{Zr}. An analysis in Ref.~[\onlinecite{sibille2017experimental}] suggests that a modulation of the neutron scattering intensity is consistent with the expectations for \ac{QSI}, though there are potentially alternate explanations for such a feature. However, it remains that there is an absence of direct evidence for disorder as found in \abo{Pr}{Zr} as well as the presence of a finite energy excitation that suggests non-trivial spin dynamics, suggesting the physics could be rather different from that of \abo{Pr}{Zr}. While there are other kinds of disorder that could be invisible to the kind of analysis used for \abo{Pr}{Zr} (e.g. perhaps random displacement of the \rth{Pr} ions \emph{along} $\vhat{z}_i$), this compound represents a promising avenue for future studies.

\section{Outlook}
Beyond the four specific material examples that we have focused on, we would like to close this review by commenting on some other systems, some similar, some quite different, that could realize some of the same physics -- and potentially be helpful in understanding the broader aspects of frustrated magnetism in such highly anisotropic compounds.

The first of these systems are another class of rare-earth magnets with a pyrochlore lattice, the spinels~\cite{lau2005} of the form A\tsub{2}RX\tsub{4} (where A = Mg, Cd and X =  Se, S). These have been synthesized for several different rare-earths such as \rth{Ho}, \rth{Er} and \rth{Yb} in powder form. Early studies have shown much the same diversity seen in the \abo{R}{M} pyrochlores, with examples of classical dipolar spin ice~\cite{erspinel1,erspinel2,erspinel3}, as well as potential candidates for order-by-quantum-disorder as found in \eto{}~\cite{higo2017spinels1,ybspinel2}. More speculatively, it is possible that the Yb-spinels~\cite{higo2017spinels1,ybspinel2} may exhibit similar physics to \abo{Yb}{Ge} and its cousins~\cite{rau2018frustration} -- if so, they could shed new light on the underlying mystery of the \abo{Yb}{M} family. Synthesis of single crystals as well as detailed studies of the excitations of these rare-earth pyrochlore spinels are thus needed.

A family of materials more far afield are the recently synthesized transition metal fluoride pyrochlores~\cite{fluoridesynthesis1,fluoridesynthesis4,fluoridesynthesis3}. These transition metal magnets are expected to be more isotropic than their rare-earth counterparts, though there has been some evidence for XY-like features in their excitation spectra~\cite{fluoridexy1,ross2017fluoride}. One key feature of these compounds is the presence of structural (charge) disorder on the non-magnetic pyrochlore sites of the crystal lattice. A recent study on one such compound~\cite{plumb2017continuum} shows that, in spite of the disorder, it is a good realization of a classical pyrochlore Heisenberg antiferromagnet over a broad range of temperatures~\cite{conlon2009}. While the presence of the structural disorder complicates the low-energy physics, it also affords new opportunities for study of frustrated disordered systems, similar in spirit to \abo{Pr}{Zr}. Indeed, a natural possibility could be that the charge disorder is frustrated in itself, satisfying an ``ice-rule", as was originally discussed by Anderson~\cite{anderson}. 

More speculatively, one may ask whether any of this kind of physics is realizable in transition metal magnets with strong spin-orbit coupling~\cite{witczak2014correlated,rau2016spin}. Many of the experimental challenges in studying rare-earth magnets stem from their intrinsically low energy scales, requiring experiments at very low temperature and very high resolution in spectroscopic probes. Many of these issues would be significantly alleviated in transition metal realizations, where the exchange scale could be orders of magnitude larger. Indeed, this could even open avenues to new kinds of experimental probes, such as resonant elastic and inelastic X-ray scattering, that potentially also directly measure less conventional, non-magnetic excitations. As it stands, the currently known heavy transition metal pyrochlores (such as the pyrochlore iridates) are not strongly insulating and do not show much diversity in their magnetic physics, nearly all of them ordering in an \ac{AIAO} state~\cite{witczak2014correlated}. Perhaps if the \ac{DM} interaction in these or related compounds could be driven negative (indirect)~\cite{elhajal2005,canals2008,chern2010pyrochlore}, some of the physics discussed in Sec.~\ref{sec:yto} could be realized.

In this review, we have covered the microscopic and theoretical background necessary for understanding the quantum rare-earth pyrochlores, as well as discussed four important material examples in detail. Many of these materials raise foundational questions that remain unanswered, such as the nature of their ground state and its low-lying excitations. For \yto{} and \tto{} in particular, the answers to these questions have been elusive for two decades.  Our primary aim was to provide an overview of the rich physics that can be explored in these systems and outline the key questions and perspectives we feel will be useful in moving towards resolving these puzzles. Given the recent serious breakthroughs that have occurred for each of these compounds, we are hopeful that a more complete understanding is not far out of reach. Beyond this, we hope this review serves as an invitation to the study of these systems, and look forward to seeing what other lessons can be learned from this growing family of materials.

\begin{acknowledgements}
We thank S. Bhattacharjee, J. Gaudet, B. Gaulin, A. Hallas, L.  Jaubert, H. Kadowaki, E. Lhotel, P. McClarty,  T. McQueen, S. Petit, and  K. Ross for many useful discussions and comments. We particularly wish to thank P. McClarty for a critical reading of the manuscript. The work at the University of Waterloo was supported by the NSERC of Canada, the Canada Research Chair program (M.J.P.G., Tier 1) and the Canadian Institute for Advanced Research (CIfAR).
\end{acknowledgements}

\bibliography{draft}

\begin{thebibliography}{158}%
\makeatletter
\providecommand \@ifxundefined [1]{%
 \@ifx{#1\undefined}
}%
\providecommand \@ifnum [1]{%
 \ifnum #1\expandafter \@firstoftwo
 \else \expandafter \@secondoftwo
 \fi
}%
\providecommand \@ifx [1]{%
 \ifx #1\expandafter \@firstoftwo
 \else \expandafter \@secondoftwo
 \fi
}%
\providecommand \natexlab [1]{#1}%
\providecommand \enquote  [1]{``#1''}%
\providecommand \bibnamefont  [1]{#1}%
\providecommand \bibfnamefont [1]{#1}%
\providecommand \citenamefont [1]{#1}%
\providecommand \href@noop [0]{\@secondoftwo}%
\providecommand \href [0]{\begingroup \@sanitize@url \@href}%
\providecommand \@href[1]{\@@startlink{#1}\@@href}%
\providecommand \@@href[1]{\endgroup#1\@@endlink}%
\providecommand \@sanitize@url [0]{\catcode `\\12\catcode `\$12\catcode
  `\&12\catcode `\#12\catcode `\^12\catcode `\_12\catcode `\%12\relax}%
\providecommand \@@startlink[1]{}%
\providecommand \@@endlink[0]{}%
\providecommand \url  [0]{\begingroup\@sanitize@url \@url }%
\providecommand \@url [1]{\endgroup\@href {#1}{\urlprefix }}%
\providecommand \urlprefix  [0]{URL }%
\providecommand \Eprint [0]{\href }%
\providecommand \doibase [0]{http://dx.doi.org/}%
\providecommand \selectlanguage [0]{\@gobble}%
\providecommand \bibinfo  [0]{\@secondoftwo}%
\providecommand \bibfield  [0]{\@secondoftwo}%
\providecommand \translation [1]{[#1]}%
\providecommand \BibitemOpen [0]{}%
\providecommand \bibitemStop [0]{}%
\providecommand \bibitemNoStop [0]{.\EOS\space}%
\providecommand \EOS [0]{\spacefactor3000\relax}%
\providecommand \BibitemShut  [1]{\csname bibitem#1\endcsname}%
\let\auto@bib@innerbib\@empty
\bibitem [{\citenamefont {Ramirez}(2001)}]{ramirez2001geometrical}%
  \BibitemOpen
  \bibfield  {author} {\bibinfo {author} {\bibfnamefont {A.~P.}\ \bibnamefont
  {Ramirez}},\ }\href@noop {} {\bibfield  {journal} {\bibinfo  {journal}
  {Handbook of magnetic materials}\ }\textbf {\bibinfo {volume} {13}},\
  \bibinfo {pages} {423} (\bibinfo {year} {2001})}\BibitemShut {NoStop}%
\bibitem [{\citenamefont {Lacroix}\ \emph {et~al.}(2013)\citenamefont
  {Lacroix}, \citenamefont {Mendels},\ and\ \citenamefont
  {Mila}}]{lacroix2013introduction}%
  \BibitemOpen
  \bibfield  {author} {\bibinfo {author} {\bibfnamefont {C.}~\bibnamefont
  {Lacroix}}, \bibinfo {author} {\bibfnamefont {P.}~\bibnamefont {Mendels}}, \
  and\ \bibinfo {author} {\bibfnamefont {F.}~\bibnamefont {Mila}},\ }\href@noop
  {} {\emph {\bibinfo {title} {Introduction to Frustrated Magnetism: Materials,
  Experiments, Theory}}}\ (\bibinfo  {publisher} {Springer},\ \bibinfo {year}
  {2013})\BibitemShut {NoStop}%
\bibitem [{\citenamefont {Savary}\ and\ \citenamefont
  {Balents}(2016)}]{savary2016quantum}%
  \BibitemOpen
  \bibfield  {author} {\bibinfo {author} {\bibfnamefont {L.}~\bibnamefont
  {Savary}}\ and\ \bibinfo {author} {\bibfnamefont {L.}~\bibnamefont
  {Balents}},\ }\href {\doibase 10.1088/0034-4885/80/1/016502} {\bibfield
  {journal} {\bibinfo  {journal} {Reports on Progress in Physics}\ }\textbf
  {\bibinfo {volume} {80}},\ \bibinfo {pages} {016502} (\bibinfo {year}
  {2016})}\BibitemShut {NoStop}%
\bibitem [{\citenamefont {Wen}(2004)}]{wen2004quantum}%
  \BibitemOpen
  \bibfield  {author} {\bibinfo {author} {\bibfnamefont {X.-G.}\ \bibnamefont
  {Wen}},\ }\href@noop {} {\emph {\bibinfo {title} {Quantum field theory of
  many-body systems: from the origin of sound to an origin of light and
  electrons}}}\ (\bibinfo  {publisher} {Oxford University Press on Demand},\
  \bibinfo {year} {2004})\BibitemShut {NoStop}%
\bibitem [{\citenamefont {Rau}\ \emph {et~al.}(2016{\natexlab{a}})\citenamefont
  {Rau}, \citenamefont {Lee},\ and\ \citenamefont {Kee}}]{rau2016spin}%
  \BibitemOpen
  \bibfield  {author} {\bibinfo {author} {\bibfnamefont {J.~G.}\ \bibnamefont
  {Rau}}, \bibinfo {author} {\bibfnamefont {E.~K.-H.}\ \bibnamefont {Lee}}, \
  and\ \bibinfo {author} {\bibfnamefont {H.-Y.}\ \bibnamefont {Kee}},\ }\href
  {\doibase 10.1146/annurev-conmatphys-031115-011319} {\bibfield  {journal}
  {\bibinfo  {journal} {Annual Review of Condensed Matter Physics}\ }\textbf
  {\bibinfo {volume} {7}},\ \bibinfo {pages} {195} (\bibinfo {year}
  {2016}{\natexlab{a}})}\BibitemShut {NoStop}%
\bibitem [{\citenamefont {Winter}\ \emph {et~al.}(2017)\citenamefont {Winter},
  \citenamefont {Tsirlin}, \citenamefont {Daghofer}, \citenamefont {van~den
  Brink}, \citenamefont {Singh}, \citenamefont {Gegenwart},\ and\ \citenamefont
  {Valenti}}]{winter2017models}%
  \BibitemOpen
  \bibfield  {author} {\bibinfo {author} {\bibfnamefont {S.~M.}\ \bibnamefont
  {Winter}}, \bibinfo {author} {\bibfnamefont {A.~A.}\ \bibnamefont {Tsirlin}},
  \bibinfo {author} {\bibfnamefont {M.}~\bibnamefont {Daghofer}}, \bibinfo
  {author} {\bibfnamefont {J.}~\bibnamefont {van~den Brink}}, \bibinfo {author}
  {\bibfnamefont {Y.}~\bibnamefont {Singh}}, \bibinfo {author} {\bibfnamefont
  {P.}~\bibnamefont {Gegenwart}}, \ and\ \bibinfo {author} {\bibfnamefont
  {R.}~\bibnamefont {Valenti}},\ }\href {\doibase 10.1088/1361-648X/aa8cf5}
  {\bibfield  {journal} {\bibinfo  {journal} {Journal of Physics: Condensed
  Matter}\ }\textbf {\bibinfo {volume} {29}},\ \bibinfo {pages} {493002}
  (\bibinfo {year} {2017})}\BibitemShut {NoStop}%
\bibitem [{\citenamefont {Kitaev}(2006)}]{kitaev2006anyons}%
  \BibitemOpen
  \bibfield  {author} {\bibinfo {author} {\bibfnamefont {A.}~\bibnamefont
  {Kitaev}},\ }\href {\doibase 10.1016/j.aop.2005.10.005} {\bibfield  {journal}
  {\bibinfo  {journal} {Annals of Physics}\ }\textbf {\bibinfo {volume}
  {321}},\ \bibinfo {pages} {2} (\bibinfo {year} {2006})}\BibitemShut {NoStop}%
\bibitem [{\citenamefont {Gardner}\ \emph {et~al.}(2010)\citenamefont
  {Gardner}, \citenamefont {Gingras},\ and\ \citenamefont
  {Greedan}}]{gardner2010rmp}%
  \BibitemOpen
  \bibfield  {author} {\bibinfo {author} {\bibfnamefont {J.~S.}\ \bibnamefont
  {Gardner}}, \bibinfo {author} {\bibfnamefont {M.~J.~P.}\ \bibnamefont
  {Gingras}}, \ and\ \bibinfo {author} {\bibfnamefont {J.~E.}\ \bibnamefont
  {Greedan}},\ }\href {\doibase 10.1103/RevModPhys.82.53} {\bibfield  {journal}
  {\bibinfo  {journal} {Rev. Mod. Phys.}\ }\textbf {\bibinfo {volume} {82}},\
  \bibinfo {pages} {53} (\bibinfo {year} {2010})}\BibitemShut {NoStop}%
\bibitem [{\citenamefont {Rau}\ and\ \citenamefont
  {Gingras}(2015)}]{rau2015magnitude}%
  \BibitemOpen
  \bibfield  {author} {\bibinfo {author} {\bibfnamefont {J.~G.}\ \bibnamefont
  {Rau}}\ and\ \bibinfo {author} {\bibfnamefont {M.~J.~P.}\ \bibnamefont
  {Gingras}},\ }\href {\doibase 10.1103/PhysRevB.92.144417} {\bibfield
  {journal} {\bibinfo  {journal} {Phys. Rev. B}\ }\textbf {\bibinfo {volume}
  {92}},\ \bibinfo {pages} {144417} (\bibinfo {year} {2015})}\BibitemShut
  {NoStop}%
\bibitem [{\citenamefont {Gingras}(2011)}]{gingras2011spin}%
  \BibitemOpen
  \bibfield  {author} {\bibinfo {author} {\bibfnamefont {M.~J.~P.}\
  \bibnamefont {Gingras}},\ }in\ \href@noop {} {\emph {\bibinfo {booktitle}
  {Introduction to frustrated magnetism}}}\ (\bibinfo  {publisher} {Springer},\
  \bibinfo {year} {2011})\ pp.\ \bibinfo {pages} {293--329}\BibitemShut
  {NoStop}%
\bibitem [{\citenamefont {Canals}\ and\ \citenamefont
  {Lacroix}(1998)}]{canals1998pyrochlore}%
  \BibitemOpen
  \bibfield  {author} {\bibinfo {author} {\bibfnamefont {B.}~\bibnamefont
  {Canals}}\ and\ \bibinfo {author} {\bibfnamefont {C.}~\bibnamefont
  {Lacroix}},\ }\href {\doibase 10.1103/PhysRevLett.80.2933} {\bibfield
  {journal} {\bibinfo  {journal} {Phys. Rev. Lett.}\ }\textbf {\bibinfo
  {volume} {80}},\ \bibinfo {pages} {2933} (\bibinfo {year}
  {1998})}\BibitemShut {NoStop}%
\bibitem [{\citenamefont {Wybourne}(1965)}]{wybourne1965}%
  \BibitemOpen
  \bibfield  {author} {\bibinfo {author} {\bibfnamefont {B.~G.}\ \bibnamefont
  {Wybourne}},\ }\href@noop {} {\emph {\bibinfo {title} {Spectroscopic
  Properties of Rare Earths}}}\ (\bibinfo  {publisher} {Wiley},\ \bibinfo
  {year} {1965})\BibitemShut {NoStop}%
\bibitem [{\citenamefont {Newman}\ and\ \citenamefont
  {Ng}(2007)}]{newman2007crystal}%
  \BibitemOpen
  \bibfield  {author} {\bibinfo {author} {\bibfnamefont {D.~J.}\ \bibnamefont
  {Newman}}\ and\ \bibinfo {author} {\bibfnamefont {B.}~\bibnamefont {Ng}},\
  }\href@noop {} {\emph {\bibinfo {title} {Crystal field handbook}}}\ (\bibinfo
   {publisher} {Cambridge University Press},\ \bibinfo {year}
  {2007})\BibitemShut {NoStop}%
\bibitem [{Note1()}]{Note1}%
  \BibitemOpen
  \bibinfo {note} {A full accounting of all the sources of crystal field
  effects is complicated, with pure electrostatics being only one contribution.
  We refer the reader to the discussion in Ref.~[\protect \rev@citealp
  {newman2007crystal}] for a more complete discussion.}\BibitemShut {Stop}%
\bibitem [{\citenamefont {Stevens}(1952)}]{stevens1952matrix}%
  \BibitemOpen
  \bibfield  {author} {\bibinfo {author} {\bibfnamefont {K.}~\bibnamefont
  {Stevens}},\ }\href {\doibase 10.1088/0370-1298/65/3/308} {\bibfield
  {journal} {\bibinfo  {journal} {Proceedings of the Physical Society. Section
  A}\ }\textbf {\bibinfo {volume} {65}},\ \bibinfo {pages} {209} (\bibinfo
  {year} {1952})}\BibitemShut {NoStop}%
\bibitem [{\citenamefont {Bradley}\ and\ \citenamefont
  {Cracknell}(2010)}]{bradley2010mathematical}%
  \BibitemOpen
  \bibfield  {author} {\bibinfo {author} {\bibfnamefont {C.}~\bibnamefont
  {Bradley}}\ and\ \bibinfo {author} {\bibfnamefont {A.}~\bibnamefont
  {Cracknell}},\ }\href@noop {} {\emph {\bibinfo {title} {The mathematical
  theory of symmetry in solids: representation theory for point groups and
  space groups}}}\ (\bibinfo  {publisher} {Oxford University Press},\ \bibinfo
  {year} {2010})\BibitemShut {NoStop}%
\bibitem [{\citenamefont {Huang}\ \emph {et~al.}(2014)\citenamefont {Huang},
  \citenamefont {Chen},\ and\ \citenamefont {Hermele}}]{huang2014do}%
  \BibitemOpen
  \bibfield  {author} {\bibinfo {author} {\bibfnamefont {Y.-P.}\ \bibnamefont
  {Huang}}, \bibinfo {author} {\bibfnamefont {G.}~\bibnamefont {Chen}}, \ and\
  \bibinfo {author} {\bibfnamefont {M.}~\bibnamefont {Hermele}},\ }\href
  {\doibase 10.1103/PhysRevLett.112.167203} {\bibfield  {journal} {\bibinfo
  {journal} {Phys. Rev. Lett.}\ }\textbf {\bibinfo {volume} {112}},\ \bibinfo
  {pages} {167203} (\bibinfo {year} {2014})}\BibitemShut {NoStop}%
\bibitem [{Note2()}]{Note2}%
  \BibitemOpen
  \bibinfo {note} {Note that for $D_{3d}$ symmetry, the magnetic dipoles and
  octupoles are not necessarily distinct, with $S^z$ and $S^x$ transforming
  identically.}\BibitemShut {Stop}%
\bibitem [{\citenamefont {Lee}\ \emph {et~al.}(2012)\citenamefont {Lee},
  \citenamefont {Onoda},\ and\ \citenamefont {Balents}}]{lee2012generic}%
  \BibitemOpen
  \bibfield  {author} {\bibinfo {author} {\bibfnamefont {S.}~\bibnamefont
  {Lee}}, \bibinfo {author} {\bibfnamefont {S.}~\bibnamefont {Onoda}}, \ and\
  \bibinfo {author} {\bibfnamefont {L.}~\bibnamefont {Balents}},\ }\href
  {\doibase 10.1103/PhysRevB.86.104412} {\bibfield  {journal} {\bibinfo
  {journal} {Phys. Rev. B}\ }\textbf {\bibinfo {volume} {86}},\ \bibinfo
  {pages} {104412} (\bibinfo {year} {2012})}\BibitemShut {NoStop}%
\bibitem [{\citenamefont {Ross}\ \emph {et~al.}(2011)\citenamefont {Ross},
  \citenamefont {Savary}, \citenamefont {Gaulin},\ and\ \citenamefont
  {Balents}}]{ross2011quantum}%
  \BibitemOpen
  \bibfield  {author} {\bibinfo {author} {\bibfnamefont {K.~A.}\ \bibnamefont
  {Ross}}, \bibinfo {author} {\bibfnamefont {L.}~\bibnamefont {Savary}},
  \bibinfo {author} {\bibfnamefont {B.~D.}\ \bibnamefont {Gaulin}}, \ and\
  \bibinfo {author} {\bibfnamefont {L.}~\bibnamefont {Balents}},\ }\href
  {\doibase 10.1103/PhysRevX.1.021002} {\bibfield  {journal} {\bibinfo
  {journal} {Phys. Rev. X}\ }\textbf {\bibinfo {volume} {1}},\ \bibinfo {pages}
  {021002} (\bibinfo {year} {2011})}\BibitemShut {NoStop}%
\bibitem [{Note3()}]{Note3}%
  \BibitemOpen
  \bibinfo {note} {While this is plainly true for the non-Kramers case, we must
  note that for the dipolar-octupolar doublet, strictly, the statement is that
  one can always find a basis that for the doublet such that $g_{\pm
  }=0$.}\BibitemShut {Stop}%
\bibitem [{\citenamefont {Onoda}\ and\ \citenamefont
  {Tanaka}(2011)}]{onoda2011se}%
  \BibitemOpen
  \bibfield  {author} {\bibinfo {author} {\bibfnamefont {S.}~\bibnamefont
  {Onoda}}\ and\ \bibinfo {author} {\bibfnamefont {Y.}~\bibnamefont {Tanaka}},\
  }\href {\doibase 10.1103/PhysRevB.83.094411} {\bibfield  {journal} {\bibinfo
  {journal} {Phys. Rev. B}\ }\textbf {\bibinfo {volume} {83}},\ \bibinfo
  {pages} {094411} (\bibinfo {year} {2011})}\BibitemShut {NoStop}%
\bibitem [{\citenamefont {Iwahara}\ and\ \citenamefont
  {Chibotaru}(2015)}]{iwahara2015se}%
  \BibitemOpen
  \bibfield  {author} {\bibinfo {author} {\bibfnamefont {N.}~\bibnamefont
  {Iwahara}}\ and\ \bibinfo {author} {\bibfnamefont {L.~F.}\ \bibnamefont
  {Chibotaru}},\ }\href {\doibase 10.1103/PhysRevB.91.174438} {\bibfield
  {journal} {\bibinfo  {journal} {Phys. Rev. B}\ }\textbf {\bibinfo {volume}
  {91}},\ \bibinfo {pages} {174438} (\bibinfo {year} {2015})}\BibitemShut
  {NoStop}%
\bibitem [{\citenamefont {Rau}\ \emph {et~al.}(2016{\natexlab{b}})\citenamefont
  {Rau}, \citenamefont {Petit},\ and\ \citenamefont {Gingras}}]{rau2016vcff}%
  \BibitemOpen
  \bibfield  {author} {\bibinfo {author} {\bibfnamefont {J.~G.}\ \bibnamefont
  {Rau}}, \bibinfo {author} {\bibfnamefont {S.}~\bibnamefont {Petit}}, \ and\
  \bibinfo {author} {\bibfnamefont {M.~J.~P.}\ \bibnamefont {Gingras}},\ }\href
  {\doibase 10.1103/PhysRevB.93.184408} {\bibfield  {journal} {\bibinfo
  {journal} {Phys. Rev. B}\ }\textbf {\bibinfo {volume} {93}},\ \bibinfo
  {pages} {184408} (\bibinfo {year} {2016}{\natexlab{b}})}\BibitemShut
  {NoStop}%
\bibitem [{\citenamefont {Jensen}\ and\ \citenamefont
  {Mackintosh}(1991)}]{jensen1991rare}%
  \BibitemOpen
  \bibfield  {author} {\bibinfo {author} {\bibfnamefont {J.}~\bibnamefont
  {Jensen}}\ and\ \bibinfo {author} {\bibfnamefont {A.~R.}\ \bibnamefont
  {Mackintosh}},\ }\href@noop {} {\emph {\bibinfo {title} {Rare earth
  magnetism}}}\ (\bibinfo  {publisher} {Clarendon Oxford},\ \bibinfo {year}
  {1991})\BibitemShut {NoStop}%
\bibitem [{Note4()}]{Note4}%
  \BibitemOpen
  \bibinfo {note} {Note that, as for the single-ion physics, the spin-only
  moments of {Gd}\protect \textsuperscript {3+} or Eu\protect \textsuperscript
  {2+} are special cases~\cite {gardner2010rmp} that do not readily conform to
  the expectations discussed here.}\BibitemShut {Stop}%
\bibitem [{\citenamefont {Curnoe}(2007)}]{curnoe2007}%
  \BibitemOpen
  \bibfield  {author} {\bibinfo {author} {\bibfnamefont {S.~H.}\ \bibnamefont
  {Curnoe}},\ }\href {\doibase 10.1103/PhysRevB.75.212404} {\bibfield
  {journal} {\bibinfo  {journal} {Phys. Rev. B}\ }\textbf {\bibinfo {volume}
  {75}},\ \bibinfo {pages} {212404} (\bibinfo {year} {2007})}\BibitemShut
  {NoStop}%
\bibitem [{\citenamefont {Yan}\ \emph {et~al.}(2017)\citenamefont {Yan},
  \citenamefont {Benton}, \citenamefont {Jaubert},\ and\ \citenamefont
  {Shannon}}]{yan2017exchange}%
  \BibitemOpen
  \bibfield  {author} {\bibinfo {author} {\bibfnamefont {H.}~\bibnamefont
  {Yan}}, \bibinfo {author} {\bibfnamefont {O.}~\bibnamefont {Benton}},
  \bibinfo {author} {\bibfnamefont {L.}~\bibnamefont {Jaubert}}, \ and\
  \bibinfo {author} {\bibfnamefont {N.}~\bibnamefont {Shannon}},\ }\href
  {\doibase 10.1103/PhysRevB.95.094422} {\bibfield  {journal} {\bibinfo
  {journal} {Phys. Rev. B}\ }\textbf {\bibinfo {volume} {95}},\ \bibinfo
  {pages} {094422} (\bibinfo {year} {2017})}\BibitemShut {NoStop}%
\bibitem [{\citenamefont {Rau}\ and\ \citenamefont
  {Gingras}(2018)}]{rau2018frustration}%
  \BibitemOpen
  \bibfield  {author} {\bibinfo {author} {\bibfnamefont {J.~G.}\ \bibnamefont
  {Rau}}\ and\ \bibinfo {author} {\bibfnamefont {M.~J.}\ \bibnamefont
  {Gingras}},\ }\href@noop {} {\  (\bibinfo {year} {2018})},\ \Eprint
  {http://arxiv.org/abs/1802.03024} {arXiv:1802.03024 [cond-mat.str-el]}
  \BibitemShut {NoStop}%
\bibitem [{\citenamefont {Molavian}\ \emph {et~al.}(2007)\citenamefont
  {Molavian}, \citenamefont {Gingras},\ and\ \citenamefont
  {Canals}}]{molavian2007dyn}%
  \BibitemOpen
  \bibfield  {author} {\bibinfo {author} {\bibfnamefont {H.~R.}\ \bibnamefont
  {Molavian}}, \bibinfo {author} {\bibfnamefont {M.~J.~P.}\ \bibnamefont
  {Gingras}}, \ and\ \bibinfo {author} {\bibfnamefont {B.}~\bibnamefont
  {Canals}},\ }\href {\doibase 10.1103/PhysRevLett.98.157204} {\bibfield
  {journal} {\bibinfo  {journal} {Phys. Rev. Lett.}\ }\textbf {\bibinfo
  {volume} {98}},\ \bibinfo {pages} {157204} (\bibinfo {year}
  {2007})}\BibitemShut {NoStop}%
\bibitem [{\citenamefont {Molavian}\ \emph {et~al.}(2009)\citenamefont
  {Molavian}, \citenamefont {McClarty},\ and\ \citenamefont
  {Gingras}}]{molavian2009towards}%
  \BibitemOpen
  \bibfield  {author} {\bibinfo {author} {\bibfnamefont {H.~R.}\ \bibnamefont
  {Molavian}}, \bibinfo {author} {\bibfnamefont {P.~A.}\ \bibnamefont
  {McClarty}}, \ and\ \bibinfo {author} {\bibfnamefont {M.~J.~P.}\ \bibnamefont
  {Gingras}},\ }\href@noop {} {\  (\bibinfo {year} {2009})},\ \Eprint
  {http://arxiv.org/abs/0912.2957} {arXiv:0912.2957 [cond-mat.stat-mech]}
  \BibitemShut {NoStop}%
\bibitem [{\citenamefont {Savary}\ \emph {et~al.}(2012)\citenamefont {Savary},
  \citenamefont {Ross}, \citenamefont {Gaulin}, \citenamefont {Ruff},\ and\
  \citenamefont {Balents}}]{savary2012obd}%
  \BibitemOpen
  \bibfield  {author} {\bibinfo {author} {\bibfnamefont {L.}~\bibnamefont
  {Savary}}, \bibinfo {author} {\bibfnamefont {K.~A.}\ \bibnamefont {Ross}},
  \bibinfo {author} {\bibfnamefont {B.~D.}\ \bibnamefont {Gaulin}}, \bibinfo
  {author} {\bibfnamefont {J.~P.~C.}\ \bibnamefont {Ruff}}, \ and\ \bibinfo
  {author} {\bibfnamefont {L.}~\bibnamefont {Balents}},\ }\href {\doibase
  10.1103/PhysRevLett.109.167201} {\bibfield  {journal} {\bibinfo  {journal}
  {Phys. Rev. Lett.}\ }\textbf {\bibinfo {volume} {109}},\ \bibinfo {pages}
  {167201} (\bibinfo {year} {2012})}\BibitemShut {NoStop}%
\bibitem [{\citenamefont {Zhitomirsky}\ \emph {et~al.}(2012)\citenamefont
  {Zhitomirsky}, \citenamefont {Gvozdikova}, \citenamefont {Holdsworth},\ and\
  \citenamefont {Moessner}}]{zhitomirsky2012obd}%
  \BibitemOpen
  \bibfield  {author} {\bibinfo {author} {\bibfnamefont {M.~E.}\ \bibnamefont
  {Zhitomirsky}}, \bibinfo {author} {\bibfnamefont {M.~V.}\ \bibnamefont
  {Gvozdikova}}, \bibinfo {author} {\bibfnamefont {P.~C.~W.}\ \bibnamefont
  {Holdsworth}}, \ and\ \bibinfo {author} {\bibfnamefont {R.}~\bibnamefont
  {Moessner}},\ }\href {\doibase 10.1103/PhysRevLett.109.077204} {\bibfield
  {journal} {\bibinfo  {journal} {Phys. Rev. Lett.}\ }\textbf {\bibinfo
  {volume} {109}},\ \bibinfo {pages} {077204} (\bibinfo {year}
  {2012})}\BibitemShut {NoStop}%
\bibitem [{\citenamefont {Wong}\ \emph {et~al.}(2013)\citenamefont {Wong},
  \citenamefont {Hao},\ and\ \citenamefont {Gingras}}]{wong2013obd}%
  \BibitemOpen
  \bibfield  {author} {\bibinfo {author} {\bibfnamefont {A.~W.~C.}\
  \bibnamefont {Wong}}, \bibinfo {author} {\bibfnamefont {Z.}~\bibnamefont
  {Hao}}, \ and\ \bibinfo {author} {\bibfnamefont {M.~J.~P.}\ \bibnamefont
  {Gingras}},\ }\href {\doibase 10.1103/PhysRevB.88.144402} {\bibfield
  {journal} {\bibinfo  {journal} {Phys. Rev. B}\ }\textbf {\bibinfo {volume}
  {88}},\ \bibinfo {pages} {144402} (\bibinfo {year} {2013})}\BibitemShut
  {NoStop}%
\bibitem [{\citenamefont {Gingras}\ and\ \citenamefont
  {McClarty}(2014)}]{gingras2014quantum}%
  \BibitemOpen
  \bibfield  {author} {\bibinfo {author} {\bibfnamefont {M.~J.~P.}\
  \bibnamefont {Gingras}}\ and\ \bibinfo {author} {\bibfnamefont {P.~A.}\
  \bibnamefont {McClarty}},\ }\href {\doibase 10.1088/0034-4885/77/5/056501}
  {\bibfield  {journal} {\bibinfo  {journal} {Reports on Progress in Physics}\
  }\textbf {\bibinfo {volume} {77}},\ \bibinfo {pages} {056501} (\bibinfo
  {year} {2014})}\BibitemShut {NoStop}%
\bibitem [{\citenamefont {Hermele}\ \emph {et~al.}(2004)\citenamefont
  {Hermele}, \citenamefont {Fisher},\ and\ \citenamefont
  {Balents}}]{hermele2004pyro}%
  \BibitemOpen
  \bibfield  {author} {\bibinfo {author} {\bibfnamefont {M.}~\bibnamefont
  {Hermele}}, \bibinfo {author} {\bibfnamefont {M.~P.~A.}\ \bibnamefont
  {Fisher}}, \ and\ \bibinfo {author} {\bibfnamefont {L.}~\bibnamefont
  {Balents}},\ }\href {\doibase 10.1103/PhysRevB.69.064404} {\bibfield
  {journal} {\bibinfo  {journal} {Phys. Rev. B}\ }\textbf {\bibinfo {volume}
  {69}},\ \bibinfo {pages} {064404} (\bibinfo {year} {2004})}\BibitemShut
  {NoStop}%
\bibitem [{\citenamefont {Savary}\ and\ \citenamefont
  {Balents}(2012)}]{savary2012coulombic}%
  \BibitemOpen
  \bibfield  {author} {\bibinfo {author} {\bibfnamefont {L.}~\bibnamefont
  {Savary}}\ and\ \bibinfo {author} {\bibfnamefont {L.}~\bibnamefont
  {Balents}},\ }\href {\doibase 10.1103/PhysRevLett.108.037202} {\bibfield
  {journal} {\bibinfo  {journal} {Phys. Rev. Lett.}\ }\textbf {\bibinfo
  {volume} {108}},\ \bibinfo {pages} {037202} (\bibinfo {year}
  {2012})}\BibitemShut {NoStop}%
\bibitem [{\citenamefont {McClarty}\ \emph {et~al.}(2015)\citenamefont
  {McClarty}, \citenamefont {Sikora}, \citenamefont {Moessner}, \citenamefont
  {Penc}, \citenamefont {Pollmann},\ and\ \citenamefont
  {Shannon}}]{mcclarty2015chain}%
  \BibitemOpen
  \bibfield  {author} {\bibinfo {author} {\bibfnamefont {P.~A.}\ \bibnamefont
  {McClarty}}, \bibinfo {author} {\bibfnamefont {O.}~\bibnamefont {Sikora}},
  \bibinfo {author} {\bibfnamefont {R.}~\bibnamefont {Moessner}}, \bibinfo
  {author} {\bibfnamefont {K.}~\bibnamefont {Penc}}, \bibinfo {author}
  {\bibfnamefont {F.}~\bibnamefont {Pollmann}}, \ and\ \bibinfo {author}
  {\bibfnamefont {N.}~\bibnamefont {Shannon}},\ }\href {\doibase
  10.1103/PhysRevB.92.094418} {\bibfield  {journal} {\bibinfo  {journal} {Phys.
  Rev. B}\ }\textbf {\bibinfo {volume} {92}},\ \bibinfo {pages} {094418}
  (\bibinfo {year} {2015})}\BibitemShut {NoStop}%
\bibitem [{\citenamefont {Moessner}\ and\ \citenamefont
  {Chalker}(1998)}]{moessner1998pyro}%
  \BibitemOpen
  \bibfield  {author} {\bibinfo {author} {\bibfnamefont {R.}~\bibnamefont
  {Moessner}}\ and\ \bibinfo {author} {\bibfnamefont {J.~T.}\ \bibnamefont
  {Chalker}},\ }\href {\doibase 10.1103/PhysRevLett.80.2929} {\bibfield
  {journal} {\bibinfo  {journal} {Phys. Rev. Lett.}\ }\textbf {\bibinfo
  {volume} {80}},\ \bibinfo {pages} {2929} (\bibinfo {year}
  {1998})}\BibitemShut {NoStop}%
\bibitem [{\citenamefont {Henley}(2006)}]{henley2006pyroobd}%
  \BibitemOpen
  \bibfield  {author} {\bibinfo {author} {\bibfnamefont {C.~L.}\ \bibnamefont
  {Henley}},\ }\href {\doibase 10.1103/PhysRevLett.96.047201} {\bibfield
  {journal} {\bibinfo  {journal} {Phys. Rev. Lett.}\ }\textbf {\bibinfo
  {volume} {96}},\ \bibinfo {pages} {047201} (\bibinfo {year}
  {2006})}\BibitemShut {NoStop}%
\bibitem [{\citenamefont {Burnell}\ \emph {et~al.}(2009)\citenamefont
  {Burnell}, \citenamefont {Chakravarty},\ and\ \citenamefont
  {Sondhi}}]{burnell2009}%
  \BibitemOpen
  \bibfield  {author} {\bibinfo {author} {\bibfnamefont {F.~J.}\ \bibnamefont
  {Burnell}}, \bibinfo {author} {\bibfnamefont {S.}~\bibnamefont
  {Chakravarty}}, \ and\ \bibinfo {author} {\bibfnamefont {S.~L.}\ \bibnamefont
  {Sondhi}},\ }\href {\doibase 10.1103/PhysRevB.79.144432} {\bibfield
  {journal} {\bibinfo  {journal} {Phys. Rev. B}\ }\textbf {\bibinfo {volume}
  {79}},\ \bibinfo {pages} {144432} (\bibinfo {year} {2009})}\BibitemShut
  {NoStop}%
\bibitem [{\citenamefont {Benton}\ \emph {et~al.}(2016)\citenamefont {Benton},
  \citenamefont {Jaubert}, \citenamefont {Yan},\ and\ \citenamefont
  {Shannon}}]{benton2016spin}%
  \BibitemOpen
  \bibfield  {author} {\bibinfo {author} {\bibfnamefont {O.}~\bibnamefont
  {Benton}}, \bibinfo {author} {\bibfnamefont {L.~D.}\ \bibnamefont {Jaubert}},
  \bibinfo {author} {\bibfnamefont {H.}~\bibnamefont {Yan}}, \ and\ \bibinfo
  {author} {\bibfnamefont {N.}~\bibnamefont {Shannon}},\ }\href {\doibase
  10.1038/ncomms11572} {\bibfield  {journal} {\bibinfo  {journal} {Nature
  communications}\ }\textbf {\bibinfo {volume} {7}},\ \bibinfo {pages} {11572}
  (\bibinfo {year} {2016})}\BibitemShut {NoStop}%
\bibitem [{\citenamefont {Champion}\ \emph {et~al.}(2003)\citenamefont
  {Champion}, \citenamefont {Harris}, \citenamefont {Holdsworth}, \citenamefont
  {Wills}, \citenamefont {Balakrishnan}, \citenamefont {Bramwell},
  \citenamefont {\ifmmode \check{C}\else \v{C}\fi{}i\ifmmode~\check{z}\else
  \v{z}\fi{}m\'ar}, \citenamefont {Fennell}, \citenamefont {Gardner},
  \citenamefont {Lago}, \citenamefont {McMorrow}, \citenamefont
  {Orend\'a\ifmmode~\check{c}\else \v{c}\fi{}}, \citenamefont
  {Orend\'a\ifmmode~\check{c}\else \v{c}\fi{}ov\'a}, \citenamefont {Paul},
  \citenamefont {Smith}, \citenamefont {Telling},\ and\ \citenamefont
  {Wildes}}]{champion2003obd}%
  \BibitemOpen
  \bibfield  {author} {\bibinfo {author} {\bibfnamefont {J.~D.~M.}\
  \bibnamefont {Champion}}, \bibinfo {author} {\bibfnamefont {M.~J.}\
  \bibnamefont {Harris}}, \bibinfo {author} {\bibfnamefont {P.~C.~W.}\
  \bibnamefont {Holdsworth}}, \bibinfo {author} {\bibfnamefont {A.~S.}\
  \bibnamefont {Wills}}, \bibinfo {author} {\bibfnamefont {G.}~\bibnamefont
  {Balakrishnan}}, \bibinfo {author} {\bibfnamefont {S.~T.}\ \bibnamefont
  {Bramwell}}, \bibinfo {author} {\bibfnamefont {E.}~\bibnamefont {\ifmmode
  \check{C}\else \v{C}\fi{}i\ifmmode~\check{z}\else \v{z}\fi{}m\'ar}}, \bibinfo
  {author} {\bibfnamefont {T.}~\bibnamefont {Fennell}}, \bibinfo {author}
  {\bibfnamefont {J.~S.}\ \bibnamefont {Gardner}}, \bibinfo {author}
  {\bibfnamefont {J.}~\bibnamefont {Lago}}, \bibinfo {author} {\bibfnamefont
  {D.~F.}\ \bibnamefont {McMorrow}}, \bibinfo {author} {\bibfnamefont
  {M.}~\bibnamefont {Orend\'a\ifmmode~\check{c}\else \v{c}\fi{}}}, \bibinfo
  {author} {\bibfnamefont {A.}~\bibnamefont {Orend\'a\ifmmode~\check{c}\else
  \v{c}\fi{}ov\'a}}, \bibinfo {author} {\bibfnamefont {D.~M.}\ \bibnamefont
  {Paul}}, \bibinfo {author} {\bibfnamefont {R.~I.}\ \bibnamefont {Smith}},
  \bibinfo {author} {\bibfnamefont {M.~T.~F.}\ \bibnamefont {Telling}}, \ and\
  \bibinfo {author} {\bibfnamefont {A.}~\bibnamefont {Wildes}},\ }\href
  {\doibase 10.1103/PhysRevB.68.020401} {\bibfield  {journal} {\bibinfo
  {journal} {Phys. Rev. B}\ }\textbf {\bibinfo {volume} {68}},\ \bibinfo
  {pages} {020401} (\bibinfo {year} {2003})}\BibitemShut {NoStop}%
\bibitem [{\citenamefont {Bl\"ote}\ \emph {et~al.}(1969)\citenamefont
  {Bl\"ote}, \citenamefont {Wielinga},\ and\ \citenamefont
  {Huiskamp}}]{huiskamp-1969-physica}%
  \BibitemOpen
  \bibfield  {author} {\bibinfo {author} {\bibfnamefont {H.}~\bibnamefont
  {Bl\"ote}}, \bibinfo {author} {\bibfnamefont {R.}~\bibnamefont {Wielinga}}, \
  and\ \bibinfo {author} {\bibfnamefont {W.}~\bibnamefont {Huiskamp}},\ }\href
  {\doibase 10.1016/0031-8914(69)90187-6} {\bibfield  {journal} {\bibinfo
  {journal} {Physica}\ }\textbf {\bibinfo {volume} {43}},\ \bibinfo {pages}
  {549} (\bibinfo {year} {1969})}\BibitemShut {NoStop}%
\bibitem [{\citenamefont {Champion}\ and\ \citenamefont
  {Holdsworth}(2004)}]{champion2004soft}%
  \BibitemOpen
  \bibfield  {author} {\bibinfo {author} {\bibfnamefont {J.}~\bibnamefont
  {Champion}}\ and\ \bibinfo {author} {\bibfnamefont {P.}~\bibnamefont
  {Holdsworth}},\ }\href {\doibase 10.1088/0953-8984/16/11/013} {\bibfield
  {journal} {\bibinfo  {journal} {Journal of Physics: Condensed Matter}\
  }\textbf {\bibinfo {volume} {16}},\ \bibinfo {pages} {S665} (\bibinfo {year}
  {2004})}\BibitemShut {NoStop}%
\bibitem [{\citenamefont {McClarty}\ \emph {et~al.}(2014)\citenamefont
  {McClarty}, \citenamefont {Stasiak},\ and\ \citenamefont
  {Gingras}}]{mcclarty2014obdxy}%
  \BibitemOpen
  \bibfield  {author} {\bibinfo {author} {\bibfnamefont {P.~A.}\ \bibnamefont
  {McClarty}}, \bibinfo {author} {\bibfnamefont {P.}~\bibnamefont {Stasiak}}, \
  and\ \bibinfo {author} {\bibfnamefont {M.~J.~P.}\ \bibnamefont {Gingras}},\
  }\href {\doibase 10.1103/PhysRevB.89.024425} {\bibfield  {journal} {\bibinfo
  {journal} {Phys. Rev. B}\ }\textbf {\bibinfo {volume} {89}},\ \bibinfo
  {pages} {024425} (\bibinfo {year} {2014})}\BibitemShut {NoStop}%
\bibitem [{\citenamefont {Petit}\ \emph {et~al.}(2014)\citenamefont {Petit},
  \citenamefont {Robert}, \citenamefont {Guitteny}, \citenamefont {Bonville},
  \citenamefont {Decorse}, \citenamefont {Ollivier}, \citenamefont {Mutka},
  \citenamefont {Gingras},\ and\ \citenamefont {Mirebeau}}]{petit2014obd}%
  \BibitemOpen
  \bibfield  {author} {\bibinfo {author} {\bibfnamefont {S.}~\bibnamefont
  {Petit}}, \bibinfo {author} {\bibfnamefont {J.}~\bibnamefont {Robert}},
  \bibinfo {author} {\bibfnamefont {S.}~\bibnamefont {Guitteny}}, \bibinfo
  {author} {\bibfnamefont {P.}~\bibnamefont {Bonville}}, \bibinfo {author}
  {\bibfnamefont {C.}~\bibnamefont {Decorse}}, \bibinfo {author} {\bibfnamefont
  {J.}~\bibnamefont {Ollivier}}, \bibinfo {author} {\bibfnamefont
  {H.}~\bibnamefont {Mutka}}, \bibinfo {author} {\bibfnamefont {M.~J.~P.}\
  \bibnamefont {Gingras}}, \ and\ \bibinfo {author} {\bibfnamefont
  {I.}~\bibnamefont {Mirebeau}},\ }\href {\doibase 10.1103/PhysRevB.90.060410}
  {\bibfield  {journal} {\bibinfo  {journal} {Phys. Rev. B}\ }\textbf {\bibinfo
  {volume} {90}},\ \bibinfo {pages} {060410} (\bibinfo {year}
  {2014})}\BibitemShut {NoStop}%
\bibitem [{\citenamefont {Oitmaa}\ \emph {et~al.}(2013)\citenamefont {Oitmaa},
  \citenamefont {Singh}, \citenamefont {Javanparast}, \citenamefont {Day},
  \citenamefont {Bagheri},\ and\ \citenamefont {Gingras}}]{oitmaa2013obd}%
  \BibitemOpen
  \bibfield  {author} {\bibinfo {author} {\bibfnamefont {J.}~\bibnamefont
  {Oitmaa}}, \bibinfo {author} {\bibfnamefont {R.~R.~P.}\ \bibnamefont
  {Singh}}, \bibinfo {author} {\bibfnamefont {B.}~\bibnamefont {Javanparast}},
  \bibinfo {author} {\bibfnamefont {A.~G.~R.}\ \bibnamefont {Day}}, \bibinfo
  {author} {\bibfnamefont {B.~V.}\ \bibnamefont {Bagheri}}, \ and\ \bibinfo
  {author} {\bibfnamefont {M.~J.~P.}\ \bibnamefont {Gingras}},\ }\href
  {\doibase 10.1103/PhysRevB.88.220404} {\bibfield  {journal} {\bibinfo
  {journal} {Phys. Rev. B}\ }\textbf {\bibinfo {volume} {88}},\ \bibinfo
  {pages} {220404} (\bibinfo {year} {2013})}\BibitemShut {NoStop}%
\bibitem [{\citenamefont {Poole}\ \emph {et~al.}(2007)\citenamefont {Poole},
  \citenamefont {Wills},\ and\ \citenamefont
  {Lelievre-Berna}}]{poole2007magnetic}%
  \BibitemOpen
  \bibfield  {author} {\bibinfo {author} {\bibfnamefont {A.}~\bibnamefont
  {Poole}}, \bibinfo {author} {\bibfnamefont {A.}~\bibnamefont {Wills}}, \ and\
  \bibinfo {author} {\bibfnamefont {E.}~\bibnamefont {Lelievre-Berna}},\ }\href
  {\doibase 10.1088/0953-8984/19/45/452201} {\bibfield  {journal} {\bibinfo
  {journal} {Journal of Physics: Condensed Matter}\ }\textbf {\bibinfo {volume}
  {19}},\ \bibinfo {pages} {452201} (\bibinfo {year} {2007})}\BibitemShut
  {NoStop}%
\bibitem [{\citenamefont {Villain}\ \emph {et~al.}(1980)\citenamefont
  {Villain}, \citenamefont {Bidaux}, \citenamefont {Carton},\ and\
  \citenamefont {Conte}}]{villain1980order}%
  \BibitemOpen
  \bibfield  {author} {\bibinfo {author} {\bibfnamefont {J.}~\bibnamefont
  {Villain}}, \bibinfo {author} {\bibfnamefont {R.}~\bibnamefont {Bidaux}},
  \bibinfo {author} {\bibfnamefont {J.-P.}\ \bibnamefont {Carton}}, \ and\
  \bibinfo {author} {\bibfnamefont {R.}~\bibnamefont {Conte}},\ }\href
  {\doibase 10.1051/jphys:0198000410110126300} {\bibfield  {journal} {\bibinfo
  {journal} {Journal de Physique}\ }\textbf {\bibinfo {volume} {41}},\ \bibinfo
  {pages} {1263} (\bibinfo {year} {1980})}\BibitemShut {NoStop}%
\bibitem [{\citenamefont {Shender}(1982)}]{shender1982antiferromagnetic}%
  \BibitemOpen
  \bibfield  {author} {\bibinfo {author} {\bibfnamefont {E.}~\bibnamefont
  {Shender}},\ }\href
  {http://www.jetp.ac.ru/cgi-bin/e/index/e/56/1/p178?a=list} {\bibfield
  {journal} {\bibinfo  {journal} {Sov. Phys. JETP}\ }\textbf {\bibinfo {volume}
  {56}},\ \bibinfo {pages} {178} (\bibinfo {year} {1982})}\BibitemShut
  {NoStop}%
\bibitem [{\citenamefont {Henley}(1989)}]{henley1989}%
  \BibitemOpen
  \bibfield  {author} {\bibinfo {author} {\bibfnamefont {C.~L.}\ \bibnamefont
  {Henley}},\ }\href {\doibase 10.1103/PhysRevLett.62.2056} {\bibfield
  {journal} {\bibinfo  {journal} {Phys. Rev. Lett.}\ }\textbf {\bibinfo
  {volume} {62}},\ \bibinfo {pages} {2056} (\bibinfo {year}
  {1989})}\BibitemShut {NoStop}%
\bibitem [{\citenamefont {Zhitomirsky}\ \emph {et~al.}(2014)\citenamefont
  {Zhitomirsky}, \citenamefont {Holdsworth},\ and\ \citenamefont
  {Moessner}}]{zhitomirsky2014obd}%
  \BibitemOpen
  \bibfield  {author} {\bibinfo {author} {\bibfnamefont {M.~E.}\ \bibnamefont
  {Zhitomirsky}}, \bibinfo {author} {\bibfnamefont {P.~C.~W.}\ \bibnamefont
  {Holdsworth}}, \ and\ \bibinfo {author} {\bibfnamefont {R.}~\bibnamefont
  {Moessner}},\ }\href {\doibase 10.1103/PhysRevB.89.140403} {\bibfield
  {journal} {\bibinfo  {journal} {Phys. Rev. B}\ }\textbf {\bibinfo {volume}
  {89}},\ \bibinfo {pages} {140403} (\bibinfo {year} {2014})}\BibitemShut
  {NoStop}%
\bibitem [{\citenamefont {Javanparast}\ \emph {et~al.}(2015)\citenamefont
  {Javanparast}, \citenamefont {Day}, \citenamefont {Hao},\ and\ \citenamefont
  {Gingras}}]{javanparast2015}%
  \BibitemOpen
  \bibfield  {author} {\bibinfo {author} {\bibfnamefont {B.}~\bibnamefont
  {Javanparast}}, \bibinfo {author} {\bibfnamefont {A.~G.~R.}\ \bibnamefont
  {Day}}, \bibinfo {author} {\bibfnamefont {Z.}~\bibnamefont {Hao}}, \ and\
  \bibinfo {author} {\bibfnamefont {M.~J.~P.}\ \bibnamefont {Gingras}},\ }\href
  {\doibase 10.1103/PhysRevB.91.174424} {\bibfield  {journal} {\bibinfo
  {journal} {Phys. Rev. B}\ }\textbf {\bibinfo {volume} {91}},\ \bibinfo
  {pages} {174424} (\bibinfo {year} {2015})}\BibitemShut {NoStop}%
\bibitem [{\citenamefont {Andreanov}\ and\ \citenamefont
  {McClarty}(2015)}]{andreanov2015obd}%
  \BibitemOpen
  \bibfield  {author} {\bibinfo {author} {\bibfnamefont {A.}~\bibnamefont
  {Andreanov}}\ and\ \bibinfo {author} {\bibfnamefont {P.~A.}\ \bibnamefont
  {McClarty}},\ }\href {\doibase 10.1103/PhysRevB.91.064401} {\bibfield
  {journal} {\bibinfo  {journal} {Phys. Rev. B}\ }\textbf {\bibinfo {volume}
  {91}},\ \bibinfo {pages} {064401} (\bibinfo {year} {2015})}\BibitemShut
  {NoStop}%
\bibitem [{\citenamefont {Maryasin}\ and\ \citenamefont
  {Zhitomirsky}(2014)}]{maryasin2014obd}%
  \BibitemOpen
  \bibfield  {author} {\bibinfo {author} {\bibfnamefont {V.~S.}\ \bibnamefont
  {Maryasin}}\ and\ \bibinfo {author} {\bibfnamefont {M.~E.}\ \bibnamefont
  {Zhitomirsky}},\ }\href {\doibase 10.1103/PhysRevB.90.094412} {\bibfield
  {journal} {\bibinfo  {journal} {Phys. Rev. B}\ }\textbf {\bibinfo {volume}
  {90}},\ \bibinfo {pages} {094412} (\bibinfo {year} {2014})}\BibitemShut
  {NoStop}%
\bibitem [{\citenamefont {McClarty}\ \emph {et~al.}(2009)\citenamefont
  {McClarty}, \citenamefont {Curnoe},\ and\ \citenamefont
  {Gingras}}]{mcclarty2009energetic}%
  \BibitemOpen
  \bibfield  {author} {\bibinfo {author} {\bibfnamefont {P.~A.}\ \bibnamefont
  {McClarty}}, \bibinfo {author} {\bibfnamefont {S.}~\bibnamefont {Curnoe}}, \
  and\ \bibinfo {author} {\bibfnamefont {M.~J.~P.}\ \bibnamefont {Gingras}},\
  }in\ \href {\doibase 10.1088/1742-6596/145/1/012032} {\emph {\bibinfo
  {booktitle} {Journal of Physics: Conference Series}}},\ Vol.\ \bibinfo
  {volume} {145}\ (\bibinfo {organization} {IOP Publishing},\ \bibinfo {year}
  {2009})\ p.\ \bibinfo {pages} {012032}\BibitemShut {NoStop}%
\bibitem [{Note5()}]{Note5}%
  \BibitemOpen
  \bibinfo {note} {Allowing for a canting of the moments away from the local XY
  plane endows the Landau-Ginzburg free-energy with fourth-order terms~\cite
  {wong2013obd,javanparast2015} such as $\sim m^3 m_z \protect \qopname \relax
  o{cos}(3\phi )$ where $m$ is magnitude of the XY part, and $m_z$ the
  out-of-plane part. Such terms can be removed by solving for the equilibrium
  value of $m_z \propto m^3 \protect \qopname \relax o{cos}(3\phi )$, thus
  obtaining a free-energy of the form given in Eq.~(\ref
  {eq:landau-eto})}\BibitemShut {NoStop}%
\bibitem [{\citenamefont {Ross}\ \emph {et~al.}(2014)\citenamefont {Ross},
  \citenamefont {Qiu}, \citenamefont {Copley}, \citenamefont {Dabkowska},\ and\
  \citenamefont {Gaulin}}]{ross2014obd}%
  \BibitemOpen
  \bibfield  {author} {\bibinfo {author} {\bibfnamefont {K.~A.}\ \bibnamefont
  {Ross}}, \bibinfo {author} {\bibfnamefont {Y.}~\bibnamefont {Qiu}}, \bibinfo
  {author} {\bibfnamefont {J.~R.~D.}\ \bibnamefont {Copley}}, \bibinfo {author}
  {\bibfnamefont {H.~A.}\ \bibnamefont {Dabkowska}}, \ and\ \bibinfo {author}
  {\bibfnamefont {B.~D.}\ \bibnamefont {Gaulin}},\ }\href {\doibase
  10.1103/PhysRevLett.112.057201} {\bibfield  {journal} {\bibinfo  {journal}
  {Phys. Rev. Lett.}\ }\textbf {\bibinfo {volume} {112}},\ \bibinfo {pages}
  {057201} (\bibinfo {year} {2014})}\BibitemShut {NoStop}%
\bibitem [{\citenamefont {Rau}\ \emph {et~al.}(2018)\citenamefont {Rau},
  \citenamefont {McClarty},\ and\ \citenamefont {Moessner}}]{rau2018pseudo}%
  \BibitemOpen
  \bibfield  {author} {\bibinfo {author} {\bibfnamefont {J.~G.}\ \bibnamefont
  {Rau}}, \bibinfo {author} {\bibfnamefont {P.~A.}\ \bibnamefont {McClarty}}, \
  and\ \bibinfo {author} {\bibfnamefont {R.}~\bibnamefont {Moessner}},\
  }\href@noop {} {\  (\bibinfo {year} {2018})},\ \Eprint
  {http://arxiv.org/abs/1805.00947} {arXiv:1805.00947 [cond-mat.str-el]}
  \BibitemShut {NoStop}%
\bibitem [{\citenamefont {Ruff}\ \emph {et~al.}(2008)\citenamefont {Ruff},
  \citenamefont {Clancy}, \citenamefont {Bourque}, \citenamefont {White},
  \citenamefont {Ramazanoglu}, \citenamefont {Gardner}, \citenamefont {Qiu},
  \citenamefont {Copley}, \citenamefont {Johnson}, \citenamefont {Dabkowska},\
  and\ \citenamefont {Gaulin}}]{ruff2008spinwaves}%
  \BibitemOpen
  \bibfield  {author} {\bibinfo {author} {\bibfnamefont {J.~P.~C.}\
  \bibnamefont {Ruff}}, \bibinfo {author} {\bibfnamefont {J.~P.}\ \bibnamefont
  {Clancy}}, \bibinfo {author} {\bibfnamefont {A.}~\bibnamefont {Bourque}},
  \bibinfo {author} {\bibfnamefont {M.~A.}\ \bibnamefont {White}}, \bibinfo
  {author} {\bibfnamefont {M.}~\bibnamefont {Ramazanoglu}}, \bibinfo {author}
  {\bibfnamefont {J.~S.}\ \bibnamefont {Gardner}}, \bibinfo {author}
  {\bibfnamefont {Y.}~\bibnamefont {Qiu}}, \bibinfo {author} {\bibfnamefont
  {J.~R.~D.}\ \bibnamefont {Copley}}, \bibinfo {author} {\bibfnamefont {M.~B.}\
  \bibnamefont {Johnson}}, \bibinfo {author} {\bibfnamefont {H.~A.}\
  \bibnamefont {Dabkowska}}, \ and\ \bibinfo {author} {\bibfnamefont {B.~D.}\
  \bibnamefont {Gaulin}},\ }\href {\doibase 10.1103/PhysRevLett.101.147205}
  {\bibfield  {journal} {\bibinfo  {journal} {Phys. Rev. Lett.}\ }\textbf
  {\bibinfo {volume} {101}},\ \bibinfo {pages} {147205} (\bibinfo {year}
  {2008})}\BibitemShut {NoStop}%
\bibitem [{\citenamefont {Dalmas~de R\'eotier}\ \emph
  {et~al.}(2012)\citenamefont {Dalmas~de R\'eotier}, \citenamefont {Yaouanc},
  \citenamefont {Chapuis}, \citenamefont {Curnoe}, \citenamefont {Grenier},
  \citenamefont {Ressouche}, \citenamefont {Marin}, \citenamefont {Lago},
  \citenamefont {Baines},\ and\ \citenamefont {Giblin}}]{derotier2012eto}%
  \BibitemOpen
  \bibfield  {author} {\bibinfo {author} {\bibfnamefont {P.}~\bibnamefont
  {Dalmas~de R\'eotier}}, \bibinfo {author} {\bibfnamefont {A.}~\bibnamefont
  {Yaouanc}}, \bibinfo {author} {\bibfnamefont {Y.}~\bibnamefont {Chapuis}},
  \bibinfo {author} {\bibfnamefont {S.~H.}\ \bibnamefont {Curnoe}}, \bibinfo
  {author} {\bibfnamefont {B.}~\bibnamefont {Grenier}}, \bibinfo {author}
  {\bibfnamefont {E.}~\bibnamefont {Ressouche}}, \bibinfo {author}
  {\bibfnamefont {C.}~\bibnamefont {Marin}}, \bibinfo {author} {\bibfnamefont
  {J.}~\bibnamefont {Lago}}, \bibinfo {author} {\bibfnamefont {C.}~\bibnamefont
  {Baines}}, \ and\ \bibinfo {author} {\bibfnamefont {S.~R.}\ \bibnamefont
  {Giblin}},\ }\href {\doibase 10.1103/PhysRevB.86.104424} {\bibfield
  {journal} {\bibinfo  {journal} {Phys. Rev. B}\ }\textbf {\bibinfo {volume}
  {86}},\ \bibinfo {pages} {104424} (\bibinfo {year} {2012})}\BibitemShut
  {NoStop}%
\bibitem [{\citenamefont {Sosin}\ \emph {et~al.}(2010)\citenamefont {Sosin},
  \citenamefont {Prozorova}, \citenamefont {Lees}, \citenamefont
  {Balakrishnan},\ and\ \citenamefont {Petrenko}}]{sosin2010eto}%
  \BibitemOpen
  \bibfield  {author} {\bibinfo {author} {\bibfnamefont {S.~S.}\ \bibnamefont
  {Sosin}}, \bibinfo {author} {\bibfnamefont {L.~A.}\ \bibnamefont
  {Prozorova}}, \bibinfo {author} {\bibfnamefont {M.~R.}\ \bibnamefont {Lees}},
  \bibinfo {author} {\bibfnamefont {G.}~\bibnamefont {Balakrishnan}}, \ and\
  \bibinfo {author} {\bibfnamefont {O.~A.}\ \bibnamefont {Petrenko}},\ }\href
  {\doibase 10.1103/PhysRevB.82.094428} {\bibfield  {journal} {\bibinfo
  {journal} {Phys. Rev. B}\ }\textbf {\bibinfo {volume} {82}},\ \bibinfo
  {pages} {094428} (\bibinfo {year} {2010})}\BibitemShut {NoStop}%
\bibitem [{\citenamefont {Lhotel}\ \emph {et~al.}(2017)\citenamefont {Lhotel},
  \citenamefont {Robert}, \citenamefont {Ressouche}, \citenamefont {Damay},
  \citenamefont {Mirebeau}, \citenamefont {Ollivier}, \citenamefont {Mutka},
  \citenamefont {Dalmas~de R\'eotier}, \citenamefont {Yaouanc}, \citenamefont
  {Marin}, \citenamefont {Decorse},\ and\ \citenamefont
  {Petit}}]{lhotel2017gap}%
  \BibitemOpen
  \bibfield  {author} {\bibinfo {author} {\bibfnamefont {E.}~\bibnamefont
  {Lhotel}}, \bibinfo {author} {\bibfnamefont {J.}~\bibnamefont {Robert}},
  \bibinfo {author} {\bibfnamefont {E.}~\bibnamefont {Ressouche}}, \bibinfo
  {author} {\bibfnamefont {F.}~\bibnamefont {Damay}}, \bibinfo {author}
  {\bibfnamefont {I.}~\bibnamefont {Mirebeau}}, \bibinfo {author}
  {\bibfnamefont {J.}~\bibnamefont {Ollivier}}, \bibinfo {author}
  {\bibfnamefont {H.}~\bibnamefont {Mutka}}, \bibinfo {author} {\bibfnamefont
  {P.}~\bibnamefont {Dalmas~de R\'eotier}}, \bibinfo {author} {\bibfnamefont
  {A.}~\bibnamefont {Yaouanc}}, \bibinfo {author} {\bibfnamefont
  {C.}~\bibnamefont {Marin}}, \bibinfo {author} {\bibfnamefont
  {C.}~\bibnamefont {Decorse}}, \ and\ \bibinfo {author} {\bibfnamefont
  {S.}~\bibnamefont {Petit}},\ }\href {\doibase 10.1103/PhysRevB.95.134426}
  {\bibfield  {journal} {\bibinfo  {journal} {Phys. Rev. B}\ }\textbf {\bibinfo
  {volume} {95}},\ \bibinfo {pages} {134426} (\bibinfo {year}
  {2017})}\BibitemShut {NoStop}%
\bibitem [{\citenamefont {Maryasin}\ \emph {et~al.}(2016)\citenamefont
  {Maryasin}, \citenamefont {Zhitomirsky},\ and\ \citenamefont
  {Moessner}}]{maryasin2016clock}%
  \BibitemOpen
  \bibfield  {author} {\bibinfo {author} {\bibfnamefont {V.~S.}\ \bibnamefont
  {Maryasin}}, \bibinfo {author} {\bibfnamefont {M.~E.}\ \bibnamefont
  {Zhitomirsky}}, \ and\ \bibinfo {author} {\bibfnamefont {R.}~\bibnamefont
  {Moessner}},\ }\href {\doibase 10.1103/PhysRevB.93.100406} {\bibfield
  {journal} {\bibinfo  {journal} {Phys. Rev. B}\ }\textbf {\bibinfo {volume}
  {93}},\ \bibinfo {pages} {100406} (\bibinfo {year} {2016})}\BibitemShut
  {NoStop}%
\bibitem [{\citenamefont {Gaudet}\ \emph {et~al.}(2017)\citenamefont {Gaudet},
  \citenamefont {Hallas}, \citenamefont {Thibault}, \citenamefont {Butch},
  \citenamefont {Dabkowska},\ and\ \citenamefont {Gaulin}}]{gaudet2017clock}%
  \BibitemOpen
  \bibfield  {author} {\bibinfo {author} {\bibfnamefont {J.}~\bibnamefont
  {Gaudet}}, \bibinfo {author} {\bibfnamefont {A.~M.}\ \bibnamefont {Hallas}},
  \bibinfo {author} {\bibfnamefont {J.}~\bibnamefont {Thibault}}, \bibinfo
  {author} {\bibfnamefont {N.~P.}\ \bibnamefont {Butch}}, \bibinfo {author}
  {\bibfnamefont {H.~A.}\ \bibnamefont {Dabkowska}}, \ and\ \bibinfo {author}
  {\bibfnamefont {B.~D.}\ \bibnamefont {Gaulin}},\ }\href {\doibase
  10.1103/PhysRevB.95.054407} {\bibfield  {journal} {\bibinfo  {journal} {Phys.
  Rev. B}\ }\textbf {\bibinfo {volume} {95}},\ \bibinfo {pages} {054407}
  (\bibinfo {year} {2017})}\BibitemShut {NoStop}%
\bibitem [{\citenamefont {Andrade}\ \emph {et~al.}(2018)\citenamefont
  {Andrade}, \citenamefont {Hoyos}, \citenamefont {Rachel},\ and\ \citenamefont
  {Vojta}}]{andrade2018clusterglass}%
  \BibitemOpen
  \bibfield  {author} {\bibinfo {author} {\bibfnamefont {E.~C.}\ \bibnamefont
  {Andrade}}, \bibinfo {author} {\bibfnamefont {J.~A.}\ \bibnamefont {Hoyos}},
  \bibinfo {author} {\bibfnamefont {S.}~\bibnamefont {Rachel}}, \ and\ \bibinfo
  {author} {\bibfnamefont {M.}~\bibnamefont {Vojta}},\ }\href {\doibase
  10.1103/PhysRevLett.120.097204} {\bibfield  {journal} {\bibinfo  {journal}
  {Phys. Rev. Lett.}\ }\textbf {\bibinfo {volume} {120}},\ \bibinfo {pages}
  {097204} (\bibinfo {year} {2018})}\BibitemShut {NoStop}%
\bibitem [{\citenamefont {Niven}\ \emph {et~al.}(2014)\citenamefont {Niven},
  \citenamefont {Johnson}, \citenamefont {Bourque}, \citenamefont {Murray},
  \citenamefont {James}, \citenamefont {Da̧bkowska}, \citenamefont {Gaulin},\
  and\ \citenamefont {White}}]{niven2014magnetic}%
  \BibitemOpen
  \bibfield  {author} {\bibinfo {author} {\bibfnamefont {J.~F.}\ \bibnamefont
  {Niven}}, \bibinfo {author} {\bibfnamefont {M.~B.}\ \bibnamefont {Johnson}},
  \bibinfo {author} {\bibfnamefont {A.}~\bibnamefont {Bourque}}, \bibinfo
  {author} {\bibfnamefont {P.~J.}\ \bibnamefont {Murray}}, \bibinfo {author}
  {\bibfnamefont {D.~D.}\ \bibnamefont {James}}, \bibinfo {author}
  {\bibfnamefont {H.~A.}\ \bibnamefont {Da̧bkowska}}, \bibinfo {author}
  {\bibfnamefont {B.~D.}\ \bibnamefont {Gaulin}}, \ and\ \bibinfo {author}
  {\bibfnamefont {M.~A.}\ \bibnamefont {White}},\ }\href {\doibase
  10.1098/rspa.2014.0387} {\bibfield  {journal} {\bibinfo  {journal} {Proc. R.
  Soc. A}\ }\textbf {\bibinfo {volume} {470}},\ \bibinfo {pages} {20140387}
  (\bibinfo {year} {2014})}\BibitemShut {NoStop}%
\bibitem [{\citenamefont {Gaudet}\ \emph
  {et~al.}(2016{\natexlab{a}})\citenamefont {Gaudet}, \citenamefont {Hallas},
  \citenamefont {Maharaj}, \citenamefont {Buhariwalla}, \citenamefont
  {Kermarrec}, \citenamefont {Butch}, \citenamefont {Munsie}, \citenamefont
  {Dabkowska}, \citenamefont {Luke},\ and\ \citenamefont
  {Gaulin}}]{gaudet2016dilution}%
  \BibitemOpen
  \bibfield  {author} {\bibinfo {author} {\bibfnamefont {J.}~\bibnamefont
  {Gaudet}}, \bibinfo {author} {\bibfnamefont {A.~M.}\ \bibnamefont {Hallas}},
  \bibinfo {author} {\bibfnamefont {D.~D.}\ \bibnamefont {Maharaj}}, \bibinfo
  {author} {\bibfnamefont {C.~R.~C.}\ \bibnamefont {Buhariwalla}}, \bibinfo
  {author} {\bibfnamefont {E.}~\bibnamefont {Kermarrec}}, \bibinfo {author}
  {\bibfnamefont {N.~P.}\ \bibnamefont {Butch}}, \bibinfo {author}
  {\bibfnamefont {T.~J.~S.}\ \bibnamefont {Munsie}}, \bibinfo {author}
  {\bibfnamefont {H.~A.}\ \bibnamefont {Dabkowska}}, \bibinfo {author}
  {\bibfnamefont {G.~M.}\ \bibnamefont {Luke}}, \ and\ \bibinfo {author}
  {\bibfnamefont {B.~D.}\ \bibnamefont {Gaulin}},\ }\href {\doibase
  10.1103/PhysRevB.94.060407} {\bibfield  {journal} {\bibinfo  {journal} {Phys.
  Rev. B}\ }\textbf {\bibinfo {volume} {94}},\ \bibinfo {pages} {060407}
  (\bibinfo {year} {2016}{\natexlab{a}})}\BibitemShut {NoStop}%
\bibitem [{\citenamefont {Dun}\ \emph {et~al.}(2015)\citenamefont {Dun},
  \citenamefont {Li}, \citenamefont {Freitas}, \citenamefont {Arrighi},
  \citenamefont {Dela~Cruz}, \citenamefont {Lee}, \citenamefont {Choi},
  \citenamefont {Cao}, \citenamefont {Silverstein}, \citenamefont {Wiebe},
  \citenamefont {Cheng},\ and\ \citenamefont {Zhou}}]{dun2015ergeo}%
  \BibitemOpen
  \bibfield  {author} {\bibinfo {author} {\bibfnamefont {Z.~L.}\ \bibnamefont
  {Dun}}, \bibinfo {author} {\bibfnamefont {X.}~\bibnamefont {Li}}, \bibinfo
  {author} {\bibfnamefont {R.~S.}\ \bibnamefont {Freitas}}, \bibinfo {author}
  {\bibfnamefont {E.}~\bibnamefont {Arrighi}}, \bibinfo {author} {\bibfnamefont
  {C.~R.}\ \bibnamefont {Dela~Cruz}}, \bibinfo {author} {\bibfnamefont
  {M.}~\bibnamefont {Lee}}, \bibinfo {author} {\bibfnamefont {E.~S.}\
  \bibnamefont {Choi}}, \bibinfo {author} {\bibfnamefont {H.~B.}\ \bibnamefont
  {Cao}}, \bibinfo {author} {\bibfnamefont {H.~J.}\ \bibnamefont
  {Silverstein}}, \bibinfo {author} {\bibfnamefont {C.~R.}\ \bibnamefont
  {Wiebe}}, \bibinfo {author} {\bibfnamefont {J.~G.}\ \bibnamefont {Cheng}}, \
  and\ \bibinfo {author} {\bibfnamefont {H.~D.}\ \bibnamefont {Zhou}},\ }\href
  {\doibase 10.1103/PhysRevB.92.140407} {\bibfield  {journal} {\bibinfo
  {journal} {Phys. Rev. B}\ }\textbf {\bibinfo {volume} {92}},\ \bibinfo
  {pages} {140407} (\bibinfo {year} {2015})}\BibitemShut {NoStop}%
\bibitem [{\citenamefont {Hallas}\ \emph
  {et~al.}(2016{\natexlab{a}})\citenamefont {Hallas}, \citenamefont {Gaudet},
  \citenamefont {Wilson}, \citenamefont {Munsie}, \citenamefont {Aczel},
  \citenamefont {Stone}, \citenamefont {Freitas}, \citenamefont
  {Arevalo-Lopez}, \citenamefont {Attfield}, \citenamefont {Tachibana},
  \citenamefont {Wiebe}, \citenamefont {Luke},\ and\ \citenamefont
  {Gaulin}}]{hallas2016ybgeo}%
  \BibitemOpen
  \bibfield  {author} {\bibinfo {author} {\bibfnamefont {A.~M.}\ \bibnamefont
  {Hallas}}, \bibinfo {author} {\bibfnamefont {J.}~\bibnamefont {Gaudet}},
  \bibinfo {author} {\bibfnamefont {M.~N.}\ \bibnamefont {Wilson}}, \bibinfo
  {author} {\bibfnamefont {T.~J.}\ \bibnamefont {Munsie}}, \bibinfo {author}
  {\bibfnamefont {A.~A.}\ \bibnamefont {Aczel}}, \bibinfo {author}
  {\bibfnamefont {M.~B.}\ \bibnamefont {Stone}}, \bibinfo {author}
  {\bibfnamefont {R.~S.}\ \bibnamefont {Freitas}}, \bibinfo {author}
  {\bibfnamefont {A.~M.}\ \bibnamefont {Arevalo-Lopez}}, \bibinfo {author}
  {\bibfnamefont {J.~P.}\ \bibnamefont {Attfield}}, \bibinfo {author}
  {\bibfnamefont {M.}~\bibnamefont {Tachibana}}, \bibinfo {author}
  {\bibfnamefont {C.~R.}\ \bibnamefont {Wiebe}}, \bibinfo {author}
  {\bibfnamefont {G.~M.}\ \bibnamefont {Luke}}, \ and\ \bibinfo {author}
  {\bibfnamefont {B.~D.}\ \bibnamefont {Gaulin}},\ }\href {\doibase
  10.1103/PhysRevB.93.104405} {\bibfield  {journal} {\bibinfo  {journal} {Phys.
  Rev. B}\ }\textbf {\bibinfo {volume} {93}},\ \bibinfo {pages} {104405}
  (\bibinfo {year} {2016}{\natexlab{a}})}\BibitemShut {NoStop}%
\bibitem [{\citenamefont {Yasui}\ \emph {et~al.}(2003)\citenamefont {Yasui},
  \citenamefont {Soda}, \citenamefont {Iikubo}, \citenamefont {Ito},
  \citenamefont {Sato}, \citenamefont {Hamaguchi}, \citenamefont {Matsushita},
  \citenamefont {Wada}, \citenamefont {Takeuchi}, \citenamefont {Aso} \emph
  {et~al.}}]{yasui2003ferromagnetic}%
  \BibitemOpen
  \bibfield  {author} {\bibinfo {author} {\bibfnamefont {Y.}~\bibnamefont
  {Yasui}}, \bibinfo {author} {\bibfnamefont {M.}~\bibnamefont {Soda}},
  \bibinfo {author} {\bibfnamefont {S.}~\bibnamefont {Iikubo}}, \bibinfo
  {author} {\bibfnamefont {M.}~\bibnamefont {Ito}}, \bibinfo {author}
  {\bibfnamefont {M.}~\bibnamefont {Sato}}, \bibinfo {author} {\bibfnamefont
  {N.}~\bibnamefont {Hamaguchi}}, \bibinfo {author} {\bibfnamefont
  {T.}~\bibnamefont {Matsushita}}, \bibinfo {author} {\bibfnamefont
  {N.}~\bibnamefont {Wada}}, \bibinfo {author} {\bibfnamefont {T.}~\bibnamefont
  {Takeuchi}}, \bibinfo {author} {\bibfnamefont {N.}~\bibnamefont {Aso}},
  \emph {et~al.},\ }\href {\doibase 10.1143/jpsj.72.3014} {\bibfield  {journal}
  {\bibinfo  {journal} {Journal of the Physical Society of Japan}\ }\textbf
  {\bibinfo {volume} {72}},\ \bibinfo {pages} {3014} (\bibinfo {year}
  {2003})}\BibitemShut {NoStop}%
\bibitem [{\citenamefont {Chang}\ \emph {et~al.}(2012)\citenamefont {Chang},
  \citenamefont {Onoda}, \citenamefont {Su}, \citenamefont {Kao}, \citenamefont
  {Tsuei}, \citenamefont {Yasui}, \citenamefont {Kakurai},\ and\ \citenamefont
  {Lees}}]{chang2012higgs}%
  \BibitemOpen
  \bibfield  {author} {\bibinfo {author} {\bibfnamefont {L.-J.}\ \bibnamefont
  {Chang}}, \bibinfo {author} {\bibfnamefont {S.}~\bibnamefont {Onoda}},
  \bibinfo {author} {\bibfnamefont {Y.}~\bibnamefont {Su}}, \bibinfo {author}
  {\bibfnamefont {Y.-J.}\ \bibnamefont {Kao}}, \bibinfo {author} {\bibfnamefont
  {K.-D.}\ \bibnamefont {Tsuei}}, \bibinfo {author} {\bibfnamefont
  {Y.}~\bibnamefont {Yasui}}, \bibinfo {author} {\bibfnamefont
  {K.}~\bibnamefont {Kakurai}}, \ and\ \bibinfo {author} {\bibfnamefont
  {M.~R.}\ \bibnamefont {Lees}},\ }\href {\doibase 10.1038/ncomms1989}
  {\bibfield  {journal} {\bibinfo  {journal} {Nature communications}\ }\textbf
  {\bibinfo {volume} {3}},\ \bibinfo {pages} {992} (\bibinfo {year}
  {2012})}\BibitemShut {NoStop}%
\bibitem [{\citenamefont {Yaouanc}\ \emph {et~al.}(2016)\citenamefont
  {Yaouanc}, \citenamefont {de~R{\'e}otier}, \citenamefont {Keller},
  \citenamefont {Roessli},\ and\ \citenamefont {Forget}}]{yaouanc2016novel}%
  \BibitemOpen
  \bibfield  {author} {\bibinfo {author} {\bibfnamefont {A.}~\bibnamefont
  {Yaouanc}}, \bibinfo {author} {\bibfnamefont {P.~D.}\ \bibnamefont
  {de~R{\'e}otier}}, \bibinfo {author} {\bibfnamefont {L.}~\bibnamefont
  {Keller}}, \bibinfo {author} {\bibfnamefont {B.}~\bibnamefont {Roessli}}, \
  and\ \bibinfo {author} {\bibfnamefont {A.}~\bibnamefont {Forget}},\ }\href
  {\doibase 10.1088/0953-8984/28/42/426002} {\bibfield  {journal} {\bibinfo
  {journal} {Journal of Physics: Condensed Matter}\ }\textbf {\bibinfo {volume}
  {28}},\ \bibinfo {pages} {426002} (\bibinfo {year} {2016})}\BibitemShut
  {NoStop}%
\bibitem [{\citenamefont {Gaudet}\ \emph
  {et~al.}(2016{\natexlab{b}})\citenamefont {Gaudet}, \citenamefont {Ross},
  \citenamefont {Kermarrec}, \citenamefont {Butch}, \citenamefont {Ehlers},
  \citenamefont {Dabkowska},\ and\ \citenamefont {Gaulin}}]{gaudet2016gapless}%
  \BibitemOpen
  \bibfield  {author} {\bibinfo {author} {\bibfnamefont {J.}~\bibnamefont
  {Gaudet}}, \bibinfo {author} {\bibfnamefont {K.~A.}\ \bibnamefont {Ross}},
  \bibinfo {author} {\bibfnamefont {E.}~\bibnamefont {Kermarrec}}, \bibinfo
  {author} {\bibfnamefont {N.~P.}\ \bibnamefont {Butch}}, \bibinfo {author}
  {\bibfnamefont {G.}~\bibnamefont {Ehlers}}, \bibinfo {author} {\bibfnamefont
  {H.~A.}\ \bibnamefont {Dabkowska}}, \ and\ \bibinfo {author} {\bibfnamefont
  {B.~D.}\ \bibnamefont {Gaulin}},\ }\href {\doibase
  10.1103/PhysRevB.93.064406} {\bibfield  {journal} {\bibinfo  {journal} {Phys.
  Rev. B}\ }\textbf {\bibinfo {volume} {93}},\ \bibinfo {pages} {064406}
  (\bibinfo {year} {2016}{\natexlab{b}})}\BibitemShut {NoStop}%
\bibitem [{\citenamefont {Kermarrec}\ \emph {et~al.}(2017)\citenamefont
  {Kermarrec}, \citenamefont {Gaudet}, \citenamefont {Fritsch}, \citenamefont
  {Khasanov}, \citenamefont {Guguchia}, \citenamefont {Ritter}, \citenamefont
  {Ross}, \citenamefont {Dabkowska},\ and\ \citenamefont
  {Gaulin}}]{kermarrec2017ground}%
  \BibitemOpen
  \bibfield  {author} {\bibinfo {author} {\bibfnamefont {E.}~\bibnamefont
  {Kermarrec}}, \bibinfo {author} {\bibfnamefont {J.}~\bibnamefont {Gaudet}},
  \bibinfo {author} {\bibfnamefont {K.}~\bibnamefont {Fritsch}}, \bibinfo
  {author} {\bibfnamefont {R.}~\bibnamefont {Khasanov}}, \bibinfo {author}
  {\bibfnamefont {Z.}~\bibnamefont {Guguchia}}, \bibinfo {author}
  {\bibfnamefont {C.}~\bibnamefont {Ritter}}, \bibinfo {author} {\bibfnamefont
  {K.}~\bibnamefont {Ross}}, \bibinfo {author} {\bibfnamefont {H.}~\bibnamefont
  {Dabkowska}}, \ and\ \bibinfo {author} {\bibfnamefont {B.}~\bibnamefont
  {Gaulin}},\ }\href {\doibase 10.1038/ncomms14810} {\bibfield  {journal}
  {\bibinfo  {journal} {Nature communications}\ }\textbf {\bibinfo {volume}
  {8}},\ \bibinfo {pages} {14810} (\bibinfo {year} {2017})}\BibitemShut
  {NoStop}%
\bibitem [{\citenamefont {Pe\c{c}anha-Antonio}\ \emph
  {et~al.}(2017)\citenamefont {Pe\c{c}anha-Antonio}, \citenamefont {Feng},
  \citenamefont {Su}, \citenamefont {Pomjakushin}, \citenamefont {Demmel},
  \citenamefont {Chang}, \citenamefont {Aldus}, \citenamefont {Xiao},
  \citenamefont {Lees},\ and\ \citenamefont {Br\"uckel}}]{viv2017excitations}%
  \BibitemOpen
  \bibfield  {author} {\bibinfo {author} {\bibfnamefont {V.}~\bibnamefont
  {Pe\c{c}anha-Antonio}}, \bibinfo {author} {\bibfnamefont {E.}~\bibnamefont
  {Feng}}, \bibinfo {author} {\bibfnamefont {Y.}~\bibnamefont {Su}}, \bibinfo
  {author} {\bibfnamefont {V.}~\bibnamefont {Pomjakushin}}, \bibinfo {author}
  {\bibfnamefont {F.}~\bibnamefont {Demmel}}, \bibinfo {author} {\bibfnamefont
  {L.-J.}\ \bibnamefont {Chang}}, \bibinfo {author} {\bibfnamefont {R.~J.}\
  \bibnamefont {Aldus}}, \bibinfo {author} {\bibfnamefont {Y.}~\bibnamefont
  {Xiao}}, \bibinfo {author} {\bibfnamefont {M.~R.}\ \bibnamefont {Lees}}, \
  and\ \bibinfo {author} {\bibfnamefont {T.}~\bibnamefont {Br\"uckel}},\ }\href
  {\doibase 10.1103/PhysRevB.96.214415} {\bibfield  {journal} {\bibinfo
  {journal} {Phys. Rev. B}\ }\textbf {\bibinfo {volume} {96}},\ \bibinfo
  {pages} {214415} (\bibinfo {year} {2017})}\BibitemShut {NoStop}%
\bibitem [{\citenamefont {Gaudet}\ \emph {et~al.}(2015)\citenamefont {Gaudet},
  \citenamefont {Maharaj}, \citenamefont {Sala}, \citenamefont {Kermarrec},
  \citenamefont {Ross}, \citenamefont {Dabkowska}, \citenamefont {Kolesnikov},
  \citenamefont {Granroth},\ and\ \citenamefont {Gaulin}}]{gaudet2015cef}%
  \BibitemOpen
  \bibfield  {author} {\bibinfo {author} {\bibfnamefont {J.}~\bibnamefont
  {Gaudet}}, \bibinfo {author} {\bibfnamefont {D.~D.}\ \bibnamefont {Maharaj}},
  \bibinfo {author} {\bibfnamefont {G.}~\bibnamefont {Sala}}, \bibinfo {author}
  {\bibfnamefont {E.}~\bibnamefont {Kermarrec}}, \bibinfo {author}
  {\bibfnamefont {K.~A.}\ \bibnamefont {Ross}}, \bibinfo {author}
  {\bibfnamefont {H.~A.}\ \bibnamefont {Dabkowska}}, \bibinfo {author}
  {\bibfnamefont {A.~I.}\ \bibnamefont {Kolesnikov}}, \bibinfo {author}
  {\bibfnamefont {G.~E.}\ \bibnamefont {Granroth}}, \ and\ \bibinfo {author}
  {\bibfnamefont {B.~D.}\ \bibnamefont {Gaulin}},\ }\href {\doibase
  10.1103/PhysRevB.92.134420} {\bibfield  {journal} {\bibinfo  {journal} {Phys.
  Rev. B}\ }\textbf {\bibinfo {volume} {92}},\ \bibinfo {pages} {134420}
  (\bibinfo {year} {2015})}\BibitemShut {NoStop}%
\bibitem [{\citenamefont {Robert}\ \emph {et~al.}(2015)\citenamefont {Robert},
  \citenamefont {Lhotel}, \citenamefont {Remenyi}, \citenamefont {Sahling},
  \citenamefont {Mirebeau}, \citenamefont {Decorse}, \citenamefont {Canals},\
  and\ \citenamefont {Petit}}]{petit2015dynamics}%
  \BibitemOpen
  \bibfield  {author} {\bibinfo {author} {\bibfnamefont {J.}~\bibnamefont
  {Robert}}, \bibinfo {author} {\bibfnamefont {E.}~\bibnamefont {Lhotel}},
  \bibinfo {author} {\bibfnamefont {G.}~\bibnamefont {Remenyi}}, \bibinfo
  {author} {\bibfnamefont {S.}~\bibnamefont {Sahling}}, \bibinfo {author}
  {\bibfnamefont {I.}~\bibnamefont {Mirebeau}}, \bibinfo {author}
  {\bibfnamefont {C.}~\bibnamefont {Decorse}}, \bibinfo {author} {\bibfnamefont
  {B.}~\bibnamefont {Canals}}, \ and\ \bibinfo {author} {\bibfnamefont
  {S.}~\bibnamefont {Petit}},\ }\href {\doibase 10.1103/PhysRevB.92.064425}
  {\bibfield  {journal} {\bibinfo  {journal} {Phys. Rev. B}\ }\textbf {\bibinfo
  {volume} {92}},\ \bibinfo {pages} {064425} (\bibinfo {year}
  {2015})}\BibitemShut {NoStop}%
\bibitem [{\citenamefont {Thompson}\ \emph {et~al.}(2017)\citenamefont
  {Thompson}, \citenamefont {McClarty}, \citenamefont {Prabhakaran},
  \citenamefont {Cabrera}, \citenamefont {Guidi},\ and\ \citenamefont
  {Coldea}}]{thompson2017}%
  \BibitemOpen
  \bibfield  {author} {\bibinfo {author} {\bibfnamefont {J.~D.}\ \bibnamefont
  {Thompson}}, \bibinfo {author} {\bibfnamefont {P.~A.}\ \bibnamefont
  {McClarty}}, \bibinfo {author} {\bibfnamefont {D.}~\bibnamefont
  {Prabhakaran}}, \bibinfo {author} {\bibfnamefont {I.}~\bibnamefont
  {Cabrera}}, \bibinfo {author} {\bibfnamefont {T.}~\bibnamefont {Guidi}}, \
  and\ \bibinfo {author} {\bibfnamefont {R.}~\bibnamefont {Coldea}},\ }\href
  {\doibase 10.1103/PhysRevLett.119.057203} {\bibfield  {journal} {\bibinfo
  {journal} {Phys. Rev. Lett.}\ }\textbf {\bibinfo {volume} {119}},\ \bibinfo
  {pages} {057203} (\bibinfo {year} {2017})}\BibitemShut {NoStop}%
\bibitem [{Note6()}]{Note6}%
  \BibitemOpen
  \bibinfo {note} {Some aspects of this transition are contentious, given that
  the magnetic Bragg peaks are difficult to resolve atop the structural Bragg
  peaks.}\BibitemShut {Stop}%
\bibitem [{\citenamefont {Hallas}\ \emph
  {et~al.}(2016{\natexlab{b}})\citenamefont {Hallas}, \citenamefont {Gaudet},
  \citenamefont {Butch}, \citenamefont {Tachibana}, \citenamefont {Freitas},
  \citenamefont {Luke}, \citenamefont {Wiebe},\ and\ \citenamefont
  {Gaulin}}]{hallas2016universal}%
  \BibitemOpen
  \bibfield  {author} {\bibinfo {author} {\bibfnamefont {A.~M.}\ \bibnamefont
  {Hallas}}, \bibinfo {author} {\bibfnamefont {J.}~\bibnamefont {Gaudet}},
  \bibinfo {author} {\bibfnamefont {N.~P.}\ \bibnamefont {Butch}}, \bibinfo
  {author} {\bibfnamefont {M.}~\bibnamefont {Tachibana}}, \bibinfo {author}
  {\bibfnamefont {R.~S.}\ \bibnamefont {Freitas}}, \bibinfo {author}
  {\bibfnamefont {G.~M.}\ \bibnamefont {Luke}}, \bibinfo {author}
  {\bibfnamefont {C.~R.}\ \bibnamefont {Wiebe}}, \ and\ \bibinfo {author}
  {\bibfnamefont {B.~D.}\ \bibnamefont {Gaulin}},\ }\href {\doibase
  10.1103/PhysRevB.93.100403} {\bibfield  {journal} {\bibinfo  {journal} {Phys.
  Rev. B}\ }\textbf {\bibinfo {volume} {93}},\ \bibinfo {pages} {100403}
  (\bibinfo {year} {2016}{\natexlab{b}})}\BibitemShut {NoStop}%
\bibitem [{\citenamefont {Pan}\ \emph {et~al.}(2014)\citenamefont {Pan},
  \citenamefont {Kim}, \citenamefont {Ghosh}, \citenamefont {Morris},
  \citenamefont {Ross}, \citenamefont {Kermarrec}, \citenamefont {Gaulin},
  \citenamefont {Koohpayeh}, \citenamefont {Tchernyshyov},\ and\ \citenamefont
  {Armitage}}]{pan2014low}%
  \BibitemOpen
  \bibfield  {author} {\bibinfo {author} {\bibfnamefont {L.}~\bibnamefont
  {Pan}}, \bibinfo {author} {\bibfnamefont {S.~K.}\ \bibnamefont {Kim}},
  \bibinfo {author} {\bibfnamefont {A.}~\bibnamefont {Ghosh}}, \bibinfo
  {author} {\bibfnamefont {C.~M.}\ \bibnamefont {Morris}}, \bibinfo {author}
  {\bibfnamefont {K.~A.}\ \bibnamefont {Ross}}, \bibinfo {author}
  {\bibfnamefont {E.}~\bibnamefont {Kermarrec}}, \bibinfo {author}
  {\bibfnamefont {B.~D.}\ \bibnamefont {Gaulin}}, \bibinfo {author}
  {\bibfnamefont {S.}~\bibnamefont {Koohpayeh}}, \bibinfo {author}
  {\bibfnamefont {O.}~\bibnamefont {Tchernyshyov}}, \ and\ \bibinfo {author}
  {\bibfnamefont {N.}~\bibnamefont {Armitage}},\ }\href {\doibase
  10.1038/ncomms5970} {\bibfield  {journal} {\bibinfo  {journal} {Nature
  communications}\ }\textbf {\bibinfo {volume} {5}},\ \bibinfo {pages} {4970}
  (\bibinfo {year} {2014})}\BibitemShut {NoStop}%
\bibitem [{\citenamefont {Wang}\ and\ \citenamefont
  {Senthil}(2016)}]{wang2016qsl}%
  \BibitemOpen
  \bibfield  {author} {\bibinfo {author} {\bibfnamefont {C.}~\bibnamefont
  {Wang}}\ and\ \bibinfo {author} {\bibfnamefont {T.}~\bibnamefont {Senthil}},\
  }\href {\doibase 10.1103/PhysRevX.6.011034} {\bibfield  {journal} {\bibinfo
  {journal} {Phys. Rev. X}\ }\textbf {\bibinfo {volume} {6}},\ \bibinfo {pages}
  {011034} (\bibinfo {year} {2016})}\BibitemShut {NoStop}%
\bibitem [{\citenamefont {Dun}\ \emph {et~al.}(2014)\citenamefont {Dun},
  \citenamefont {Lee}, \citenamefont {Choi}, \citenamefont {Hallas},
  \citenamefont {Wiebe}, \citenamefont {Gardner}, \citenamefont {Arrighi},
  \citenamefont {Freitas}, \citenamefont {Arevalo-Lopez}, \citenamefont
  {Attfield}, \citenamefont {Zhou},\ and\ \citenamefont
  {Cheng}}]{hallas2014chem}%
  \BibitemOpen
  \bibfield  {author} {\bibinfo {author} {\bibfnamefont {Z.~L.}\ \bibnamefont
  {Dun}}, \bibinfo {author} {\bibfnamefont {M.}~\bibnamefont {Lee}}, \bibinfo
  {author} {\bibfnamefont {E.~S.}\ \bibnamefont {Choi}}, \bibinfo {author}
  {\bibfnamefont {A.~M.}\ \bibnamefont {Hallas}}, \bibinfo {author}
  {\bibfnamefont {C.~R.}\ \bibnamefont {Wiebe}}, \bibinfo {author}
  {\bibfnamefont {J.~S.}\ \bibnamefont {Gardner}}, \bibinfo {author}
  {\bibfnamefont {E.}~\bibnamefont {Arrighi}}, \bibinfo {author} {\bibfnamefont
  {R.~S.}\ \bibnamefont {Freitas}}, \bibinfo {author} {\bibfnamefont {A.~M.}\
  \bibnamefont {Arevalo-Lopez}}, \bibinfo {author} {\bibfnamefont {J.~P.}\
  \bibnamefont {Attfield}}, \bibinfo {author} {\bibfnamefont {H.~D.}\
  \bibnamefont {Zhou}}, \ and\ \bibinfo {author} {\bibfnamefont {J.~G.}\
  \bibnamefont {Cheng}},\ }\href {\doibase 10.1103/PhysRevB.89.064401}
  {\bibfield  {journal} {\bibinfo  {journal} {Phys. Rev. B}\ }\textbf {\bibinfo
  {volume} {89}},\ \bibinfo {pages} {064401} (\bibinfo {year}
  {2014})}\BibitemShut {NoStop}%
\bibitem [{\citenamefont {Jaubert}\ \emph {et~al.}(2015)\citenamefont
  {Jaubert}, \citenamefont {Benton}, \citenamefont {Rau}, \citenamefont
  {Oitmaa}, \citenamefont {Singh}, \citenamefont {Shannon},\ and\ \citenamefont
  {Gingras}}]{jaubert2015multiphase}%
  \BibitemOpen
  \bibfield  {author} {\bibinfo {author} {\bibfnamefont {L.~D.~C.}\
  \bibnamefont {Jaubert}}, \bibinfo {author} {\bibfnamefont {O.}~\bibnamefont
  {Benton}}, \bibinfo {author} {\bibfnamefont {J.~G.}\ \bibnamefont {Rau}},
  \bibinfo {author} {\bibfnamefont {J.}~\bibnamefont {Oitmaa}}, \bibinfo
  {author} {\bibfnamefont {R.~R.~P.}\ \bibnamefont {Singh}}, \bibinfo {author}
  {\bibfnamefont {N.}~\bibnamefont {Shannon}}, \ and\ \bibinfo {author}
  {\bibfnamefont {M.~J.~P.}\ \bibnamefont {Gingras}},\ }\href {\doibase
  10.1103/PhysRevLett.115.267208} {\bibfield  {journal} {\bibinfo  {journal}
  {Phys. Rev. Lett.}\ }\textbf {\bibinfo {volume} {115}},\ \bibinfo {pages}
  {267208} (\bibinfo {year} {2015})}\BibitemShut {NoStop}%
\bibitem [{\citenamefont {Thompson}\ \emph {et~al.}(2011)\citenamefont
  {Thompson}, \citenamefont {McClarty}, \citenamefont {R\o{}nnow},
  \citenamefont {Regnault}, \citenamefont {Sorge},\ and\ \citenamefont
  {Gingras}}]{ytorods}%
  \BibitemOpen
  \bibfield  {author} {\bibinfo {author} {\bibfnamefont {J.~D.}\ \bibnamefont
  {Thompson}}, \bibinfo {author} {\bibfnamefont {P.~A.}\ \bibnamefont
  {McClarty}}, \bibinfo {author} {\bibfnamefont {H.~M.}\ \bibnamefont
  {R\o{}nnow}}, \bibinfo {author} {\bibfnamefont {L.~P.}\ \bibnamefont
  {Regnault}}, \bibinfo {author} {\bibfnamefont {A.}~\bibnamefont {Sorge}}, \
  and\ \bibinfo {author} {\bibfnamefont {M.~J.~P.}\ \bibnamefont {Gingras}},\
  }\href {\doibase 10.1103/PhysRevLett.106.187202} {\bibfield  {journal}
  {\bibinfo  {journal} {Phys. Rev. Lett.}\ }\textbf {\bibinfo {volume} {106}},\
  \bibinfo {pages} {187202} (\bibinfo {year} {2011})}\BibitemShut {NoStop}%
\bibitem [{\citenamefont {Elhajal}\ \emph {et~al.}(2005)\citenamefont
  {Elhajal}, \citenamefont {Canals}, \citenamefont {Sunyer},\ and\
  \citenamefont {Lacroix}}]{elhajal2005}%
  \BibitemOpen
  \bibfield  {author} {\bibinfo {author} {\bibfnamefont {M.}~\bibnamefont
  {Elhajal}}, \bibinfo {author} {\bibfnamefont {B.}~\bibnamefont {Canals}},
  \bibinfo {author} {\bibfnamefont {R.}~\bibnamefont {Sunyer}}, \ and\ \bibinfo
  {author} {\bibfnamefont {C.}~\bibnamefont {Lacroix}},\ }\href {\doibase
  10.1103/PhysRevB.71.094420} {\bibfield  {journal} {\bibinfo  {journal} {Phys.
  Rev. B}\ }\textbf {\bibinfo {volume} {71}},\ \bibinfo {pages} {094420}
  (\bibinfo {year} {2005})}\BibitemShut {NoStop}%
\bibitem [{\citenamefont {Canals}\ \emph {et~al.}(2008)\citenamefont {Canals},
  \citenamefont {Elhajal},\ and\ \citenamefont {Lacroix}}]{canals2008}%
  \BibitemOpen
  \bibfield  {author} {\bibinfo {author} {\bibfnamefont {B.}~\bibnamefont
  {Canals}}, \bibinfo {author} {\bibfnamefont {M.}~\bibnamefont {Elhajal}}, \
  and\ \bibinfo {author} {\bibfnamefont {C.}~\bibnamefont {Lacroix}},\ }\href
  {\doibase 10.1103/PhysRevB.78.214431} {\bibfield  {journal} {\bibinfo
  {journal} {Phys. Rev. B}\ }\textbf {\bibinfo {volume} {78}},\ \bibinfo
  {pages} {214431} (\bibinfo {year} {2008})}\BibitemShut {NoStop}%
\bibitem [{\citenamefont {Chern}(2010)}]{chern2010pyrochlore}%
  \BibitemOpen
  \bibfield  {author} {\bibinfo {author} {\bibfnamefont {G.-W.}\ \bibnamefont
  {Chern}},\ }\href@noop {} {\  (\bibinfo {year} {2010})},\ \Eprint
  {http://arxiv.org/abs/1008.3038} {arXiv:1008.3038 [cond-mat.str-el]}
  \BibitemShut {NoStop}%
\bibitem [{\citenamefont {Arpino}\ \emph {et~al.}(2017)\citenamefont {Arpino},
  \citenamefont {Trump}, \citenamefont {Scheie}, \citenamefont {McQueen},\ and\
  \citenamefont {Koohpayeh}}]{arpino2017}%
  \BibitemOpen
  \bibfield  {author} {\bibinfo {author} {\bibfnamefont {K.~E.}\ \bibnamefont
  {Arpino}}, \bibinfo {author} {\bibfnamefont {B.~A.}\ \bibnamefont {Trump}},
  \bibinfo {author} {\bibfnamefont {A.~O.}\ \bibnamefont {Scheie}}, \bibinfo
  {author} {\bibfnamefont {T.~M.}\ \bibnamefont {McQueen}}, \ and\ \bibinfo
  {author} {\bibfnamefont {S.~M.}\ \bibnamefont {Koohpayeh}},\ }\href {\doibase
  10.1103/PhysRevB.95.094407} {\bibfield  {journal} {\bibinfo  {journal} {Phys.
  Rev. B}\ }\textbf {\bibinfo {volume} {95}},\ \bibinfo {pages} {094407}
  (\bibinfo {year} {2017})}\BibitemShut {NoStop}%
\bibitem [{\citenamefont {Cai}\ \emph {et~al.}(2016)\citenamefont {Cai},
  \citenamefont {Cui}, \citenamefont {Li}, \citenamefont {Dun}, \citenamefont
  {Ma}, \citenamefont {dela Cruz}, \citenamefont {Jiao}, \citenamefont {Liao},
  \citenamefont {Sun}, \citenamefont {Li}, \citenamefont {Zhou}, \citenamefont
  {Goodenough}, \citenamefont {Zhou},\ and\ \citenamefont
  {Cheng}}]{cai2016platinum}%
  \BibitemOpen
  \bibfield  {author} {\bibinfo {author} {\bibfnamefont {Y.~Q.}\ \bibnamefont
  {Cai}}, \bibinfo {author} {\bibfnamefont {Q.}~\bibnamefont {Cui}}, \bibinfo
  {author} {\bibfnamefont {X.}~\bibnamefont {Li}}, \bibinfo {author}
  {\bibfnamefont {Z.~L.}\ \bibnamefont {Dun}}, \bibinfo {author} {\bibfnamefont
  {J.}~\bibnamefont {Ma}}, \bibinfo {author} {\bibfnamefont {C.}~\bibnamefont
  {dela Cruz}}, \bibinfo {author} {\bibfnamefont {Y.~Y.}\ \bibnamefont {Jiao}},
  \bibinfo {author} {\bibfnamefont {J.}~\bibnamefont {Liao}}, \bibinfo {author}
  {\bibfnamefont {P.~J.}\ \bibnamefont {Sun}}, \bibinfo {author} {\bibfnamefont
  {Y.~Q.}\ \bibnamefont {Li}}, \bibinfo {author} {\bibfnamefont {J.~S.}\
  \bibnamefont {Zhou}}, \bibinfo {author} {\bibfnamefont {J.~B.}\ \bibnamefont
  {Goodenough}}, \bibinfo {author} {\bibfnamefont {H.~D.}\ \bibnamefont
  {Zhou}}, \ and\ \bibinfo {author} {\bibfnamefont {J.-G.}\ \bibnamefont
  {Cheng}},\ }\href {\doibase 10.1103/PhysRevB.93.014443} {\bibfield  {journal}
  {\bibinfo  {journal} {Phys. Rev. B}\ }\textbf {\bibinfo {volume} {93}},\
  \bibinfo {pages} {014443} (\bibinfo {year} {2016})}\BibitemShut {NoStop}%
\bibitem [{\citenamefont {Gingras}\ \emph {et~al.}(2000)\citenamefont
  {Gingras}, \citenamefont {den Hertog}, \citenamefont {Faucher}, \citenamefont
  {Gardner}, \citenamefont {Dunsiger}, \citenamefont {Chang}, \citenamefont
  {Gaulin}, \citenamefont {Raju},\ and\ \citenamefont
  {Greedan}}]{gingras2000iontto}%
  \BibitemOpen
  \bibfield  {author} {\bibinfo {author} {\bibfnamefont {M.~J.~P.}\
  \bibnamefont {Gingras}}, \bibinfo {author} {\bibfnamefont {B.~C.}\
  \bibnamefont {den Hertog}}, \bibinfo {author} {\bibfnamefont
  {M.}~\bibnamefont {Faucher}}, \bibinfo {author} {\bibfnamefont {J.~S.}\
  \bibnamefont {Gardner}}, \bibinfo {author} {\bibfnamefont {S.~R.}\
  \bibnamefont {Dunsiger}}, \bibinfo {author} {\bibfnamefont {L.~J.}\
  \bibnamefont {Chang}}, \bibinfo {author} {\bibfnamefont {B.~D.}\ \bibnamefont
  {Gaulin}}, \bibinfo {author} {\bibfnamefont {N.~P.}\ \bibnamefont {Raju}}, \
  and\ \bibinfo {author} {\bibfnamefont {J.~E.}\ \bibnamefont {Greedan}},\
  }\href {\doibase 10.1103/PhysRevB.62.6496} {\bibfield  {journal} {\bibinfo
  {journal} {Phys. Rev. B}\ }\textbf {\bibinfo {volume} {62}},\ \bibinfo
  {pages} {6496} (\bibinfo {year} {2000})}\BibitemShut {NoStop}%
\bibitem [{\citenamefont {Mirebeau}\ \emph {et~al.}(2007)\citenamefont
  {Mirebeau}, \citenamefont {Bonville},\ and\ \citenamefont
  {Hennion}}]{mirebeau2007ttotsno}%
  \BibitemOpen
  \bibfield  {author} {\bibinfo {author} {\bibfnamefont {I.}~\bibnamefont
  {Mirebeau}}, \bibinfo {author} {\bibfnamefont {P.}~\bibnamefont {Bonville}},
  \ and\ \bibinfo {author} {\bibfnamefont {M.}~\bibnamefont {Hennion}},\ }\href
  {\doibase 10.1103/PhysRevB.76.184436} {\bibfield  {journal} {\bibinfo
  {journal} {Phys. Rev. B}\ }\textbf {\bibinfo {volume} {76}},\ \bibinfo
  {pages} {184436} (\bibinfo {year} {2007})}\BibitemShut {NoStop}%
\bibitem [{\citenamefont {Zhang}\ \emph {et~al.}(2014)\citenamefont {Zhang},
  \citenamefont {Fritsch}, \citenamefont {Hao}, \citenamefont {Bagheri},
  \citenamefont {Gingras}, \citenamefont {Granroth}, \citenamefont
  {Jiramongkolchai}, \citenamefont {Cava},\ and\ \citenamefont
  {Gaulin}}]{zhang2014ttocef}%
  \BibitemOpen
  \bibfield  {author} {\bibinfo {author} {\bibfnamefont {J.}~\bibnamefont
  {Zhang}}, \bibinfo {author} {\bibfnamefont {K.}~\bibnamefont {Fritsch}},
  \bibinfo {author} {\bibfnamefont {Z.}~\bibnamefont {Hao}}, \bibinfo {author}
  {\bibfnamefont {B.~V.}\ \bibnamefont {Bagheri}}, \bibinfo {author}
  {\bibfnamefont {M.~J.~P.}\ \bibnamefont {Gingras}}, \bibinfo {author}
  {\bibfnamefont {G.~E.}\ \bibnamefont {Granroth}}, \bibinfo {author}
  {\bibfnamefont {P.}~\bibnamefont {Jiramongkolchai}}, \bibinfo {author}
  {\bibfnamefont {R.~J.}\ \bibnamefont {Cava}}, \ and\ \bibinfo {author}
  {\bibfnamefont {B.~D.}\ \bibnamefont {Gaulin}},\ }\href {\doibase
  10.1103/PhysRevB.89.134410} {\bibfield  {journal} {\bibinfo  {journal} {Phys.
  Rev. B}\ }\textbf {\bibinfo {volume} {89}},\ \bibinfo {pages} {134410}
  (\bibinfo {year} {2014})}\BibitemShut {NoStop}%
\bibitem [{\citenamefont {Princep}\ \emph {et~al.}(2015)\citenamefont
  {Princep}, \citenamefont {Walker}, \citenamefont {Adroja}, \citenamefont
  {Prabhakaran},\ and\ \citenamefont {Boothroyd}}]{princep2015ttocef}%
  \BibitemOpen
  \bibfield  {author} {\bibinfo {author} {\bibfnamefont {A.~J.}\ \bibnamefont
  {Princep}}, \bibinfo {author} {\bibfnamefont {H.~C.}\ \bibnamefont {Walker}},
  \bibinfo {author} {\bibfnamefont {D.~T.}\ \bibnamefont {Adroja}}, \bibinfo
  {author} {\bibfnamefont {D.}~\bibnamefont {Prabhakaran}}, \ and\ \bibinfo
  {author} {\bibfnamefont {A.~T.}\ \bibnamefont {Boothroyd}},\ }\href {\doibase
  10.1103/PhysRevB.91.224430} {\bibfield  {journal} {\bibinfo  {journal} {Phys.
  Rev. B}\ }\textbf {\bibinfo {volume} {91}},\ \bibinfo {pages} {224430}
  (\bibinfo {year} {2015})}\BibitemShut {NoStop}%
\bibitem [{\citenamefont {Gardner}\ \emph {et~al.}(2004)\citenamefont
  {Gardner}, \citenamefont {Ehlers}, \citenamefont {Bramwell},\ and\
  \citenamefont {Gaulin}}]{gardner2004spin}%
  \BibitemOpen
  \bibfield  {author} {\bibinfo {author} {\bibfnamefont {J.~S.}\ \bibnamefont
  {Gardner}}, \bibinfo {author} {\bibfnamefont {G.}~\bibnamefont {Ehlers}},
  \bibinfo {author} {\bibfnamefont {S.~T.}\ \bibnamefont {Bramwell}}, \ and\
  \bibinfo {author} {\bibfnamefont {B.~D.}\ \bibnamefont {Gaulin}},\ }\href
  {\doibase 10.1088/0953-8984/16/11/011} {\bibfield  {journal} {\bibinfo
  {journal} {Journal of Physics: Condensed Matter}\ }\textbf {\bibinfo {volume}
  {16}},\ \bibinfo {pages} {S643} (\bibinfo {year} {2004})}\BibitemShut
  {NoStop}%
\bibitem [{\citenamefont {Guitteny}\ \emph {et~al.}(2013)\citenamefont
  {Guitteny}, \citenamefont {Robert}, \citenamefont {Bonville}, \citenamefont
  {Ollivier}, \citenamefont {Decorse}, \citenamefont {Steffens}, \citenamefont
  {Boehm}, \citenamefont {Mutka}, \citenamefont {Mirebeau},\ and\ \citenamefont
  {Petit}}]{guitteny2013ttoexcitations}%
  \BibitemOpen
  \bibfield  {author} {\bibinfo {author} {\bibfnamefont {S.}~\bibnamefont
  {Guitteny}}, \bibinfo {author} {\bibfnamefont {J.}~\bibnamefont {Robert}},
  \bibinfo {author} {\bibfnamefont {P.}~\bibnamefont {Bonville}}, \bibinfo
  {author} {\bibfnamefont {J.}~\bibnamefont {Ollivier}}, \bibinfo {author}
  {\bibfnamefont {C.}~\bibnamefont {Decorse}}, \bibinfo {author} {\bibfnamefont
  {P.}~\bibnamefont {Steffens}}, \bibinfo {author} {\bibfnamefont
  {M.}~\bibnamefont {Boehm}}, \bibinfo {author} {\bibfnamefont
  {H.}~\bibnamefont {Mutka}}, \bibinfo {author} {\bibfnamefont
  {I.}~\bibnamefont {Mirebeau}}, \ and\ \bibinfo {author} {\bibfnamefont
  {S.}~\bibnamefont {Petit}},\ }\href {\doibase 10.1103/PhysRevLett.111.087201}
  {\bibfield  {journal} {\bibinfo  {journal} {Phys. Rev. Lett.}\ }\textbf
  {\bibinfo {volume} {111}},\ \bibinfo {pages} {087201} (\bibinfo {year}
  {2013})}\BibitemShut {NoStop}%
\bibitem [{\citenamefont {Fennell}\ \emph {et~al.}(2014)\citenamefont
  {Fennell}, \citenamefont {Kenzelmann}, \citenamefont {Roessli}, \citenamefont
  {Mutka}, \citenamefont {Ollivier}, \citenamefont {Ruminy}, \citenamefont
  {Stuhr}, \citenamefont {Zaharko}, \citenamefont {Bovo}, \citenamefont
  {Cervellino}, \citenamefont {Haas},\ and\ \citenamefont
  {Cava}}]{fennell2014magnetoelastic2}%
  \BibitemOpen
  \bibfield  {author} {\bibinfo {author} {\bibfnamefont {T.}~\bibnamefont
  {Fennell}}, \bibinfo {author} {\bibfnamefont {M.}~\bibnamefont {Kenzelmann}},
  \bibinfo {author} {\bibfnamefont {B.}~\bibnamefont {Roessli}}, \bibinfo
  {author} {\bibfnamefont {H.}~\bibnamefont {Mutka}}, \bibinfo {author}
  {\bibfnamefont {J.}~\bibnamefont {Ollivier}}, \bibinfo {author}
  {\bibfnamefont {M.}~\bibnamefont {Ruminy}}, \bibinfo {author} {\bibfnamefont
  {U.}~\bibnamefont {Stuhr}}, \bibinfo {author} {\bibfnamefont
  {O.}~\bibnamefont {Zaharko}}, \bibinfo {author} {\bibfnamefont
  {L.}~\bibnamefont {Bovo}}, \bibinfo {author} {\bibfnamefont {A.}~\bibnamefont
  {Cervellino}}, \bibinfo {author} {\bibfnamefont {M.~K.}\ \bibnamefont
  {Haas}}, \ and\ \bibinfo {author} {\bibfnamefont {R.~J.}\ \bibnamefont
  {Cava}},\ }\href {\doibase 10.1103/PhysRevLett.112.017203} {\bibfield
  {journal} {\bibinfo  {journal} {Phys. Rev. Lett.}\ }\textbf {\bibinfo
  {volume} {112}},\ \bibinfo {pages} {017203} (\bibinfo {year}
  {2014})}\BibitemShut {NoStop}%
\bibitem [{\citenamefont {Kao}\ \emph {et~al.}(2003)\citenamefont {Kao},
  \citenamefont {Enjalran}, \citenamefont {Del~Maestro}, \citenamefont
  {Molavian},\ and\ \citenamefont {Gingras}}]{kao2003understanding}%
  \BibitemOpen
  \bibfield  {author} {\bibinfo {author} {\bibfnamefont {Y.-J.}\ \bibnamefont
  {Kao}}, \bibinfo {author} {\bibfnamefont {M.}~\bibnamefont {Enjalran}},
  \bibinfo {author} {\bibfnamefont {A.}~\bibnamefont {Del~Maestro}}, \bibinfo
  {author} {\bibfnamefont {H.~R.}\ \bibnamefont {Molavian}}, \ and\ \bibinfo
  {author} {\bibfnamefont {M.~J.~P.}\ \bibnamefont {Gingras}},\ }\href
  {\doibase 10.1103/PhysRevB.68.172407} {\bibfield  {journal} {\bibinfo
  {journal} {Phys. Rev. B}\ }\textbf {\bibinfo {volume} {68}},\ \bibinfo
  {pages} {172407} (\bibinfo {year} {2003})}\BibitemShut {NoStop}%
\bibitem [{\citenamefont {Enjalran}\ and\ \citenamefont
  {Gingras}(2004)}]{enjalran2004theory}%
  \BibitemOpen
  \bibfield  {author} {\bibinfo {author} {\bibfnamefont {M.}~\bibnamefont
  {Enjalran}}\ and\ \bibinfo {author} {\bibfnamefont {M.~J.~P.}\ \bibnamefont
  {Gingras}},\ }\href {\doibase 10.1103/PhysRevB.70.174426} {\bibfield
  {journal} {\bibinfo  {journal} {Phys. Rev. B}\ }\textbf {\bibinfo {volume}
  {70}},\ \bibinfo {pages} {174426} (\bibinfo {year} {2004})}\BibitemShut
  {NoStop}%
\bibitem [{\citenamefont {Petit}\ \emph {et~al.}(2012)\citenamefont {Petit},
  \citenamefont {Bonville}, \citenamefont {Robert}, \citenamefont {Decorse},\
  and\ \citenamefont {Mirebeau}}]{petit2012tto}%
  \BibitemOpen
  \bibfield  {author} {\bibinfo {author} {\bibfnamefont {S.}~\bibnamefont
  {Petit}}, \bibinfo {author} {\bibfnamefont {P.}~\bibnamefont {Bonville}},
  \bibinfo {author} {\bibfnamefont {J.}~\bibnamefont {Robert}}, \bibinfo
  {author} {\bibfnamefont {C.}~\bibnamefont {Decorse}}, \ and\ \bibinfo
  {author} {\bibfnamefont {I.}~\bibnamefont {Mirebeau}},\ }\href {\doibase
  10.1103/PhysRevB.86.174403} {\bibfield  {journal} {\bibinfo  {journal} {Phys.
  Rev. B}\ }\textbf {\bibinfo {volume} {86}},\ \bibinfo {pages} {174403}
  (\bibinfo {year} {2012})}\BibitemShut {NoStop}%
\bibitem [{\citenamefont {Takatsu}\ \emph {et~al.}(2016)\citenamefont
  {Takatsu}, \citenamefont {Onoda}, \citenamefont {Kittaka}, \citenamefont
  {Kasahara}, \citenamefont {Kono}, \citenamefont {Sakakibara}, \citenamefont
  {Kato}, \citenamefont {F\aa{}k}, \citenamefont {Ollivier}, \citenamefont
  {Lynn}, \citenamefont {Taniguchi}, \citenamefont {Wakita},\ and\
  \citenamefont {Kadowaki}}]{takatsu2016quadrupole}%
  \BibitemOpen
  \bibfield  {author} {\bibinfo {author} {\bibfnamefont {H.}~\bibnamefont
  {Takatsu}}, \bibinfo {author} {\bibfnamefont {S.}~\bibnamefont {Onoda}},
  \bibinfo {author} {\bibfnamefont {S.}~\bibnamefont {Kittaka}}, \bibinfo
  {author} {\bibfnamefont {A.}~\bibnamefont {Kasahara}}, \bibinfo {author}
  {\bibfnamefont {Y.}~\bibnamefont {Kono}}, \bibinfo {author} {\bibfnamefont
  {T.}~\bibnamefont {Sakakibara}}, \bibinfo {author} {\bibfnamefont
  {Y.}~\bibnamefont {Kato}}, \bibinfo {author} {\bibfnamefont {B.}~\bibnamefont
  {F\aa{}k}}, \bibinfo {author} {\bibfnamefont {J.}~\bibnamefont {Ollivier}},
  \bibinfo {author} {\bibfnamefont {J.~W.}\ \bibnamefont {Lynn}}, \bibinfo
  {author} {\bibfnamefont {T.}~\bibnamefont {Taniguchi}}, \bibinfo {author}
  {\bibfnamefont {M.}~\bibnamefont {Wakita}}, \ and\ \bibinfo {author}
  {\bibfnamefont {H.}~\bibnamefont {Kadowaki}},\ }\href {\doibase
  10.1103/PhysRevLett.116.217201} {\bibfield  {journal} {\bibinfo  {journal}
  {Phys. Rev. Lett.}\ }\textbf {\bibinfo {volume} {116}},\ \bibinfo {pages}
  {217201} (\bibinfo {year} {2016})}\BibitemShut {NoStop}%
\bibitem [{\citenamefont {Kermarrec}\ \emph {et~al.}(2015)\citenamefont
  {Kermarrec}, \citenamefont {Maharaj}, \citenamefont {Gaudet}, \citenamefont
  {Fritsch}, \citenamefont {Pomaranski}, \citenamefont {Kycia}, \citenamefont
  {Qiu}, \citenamefont {Copley}, \citenamefont {Couchman}, \citenamefont
  {Morningstar}, \citenamefont {Dabkowska},\ and\ \citenamefont
  {Gaulin}}]{kermarrec2015hhh3}%
  \BibitemOpen
  \bibfield  {author} {\bibinfo {author} {\bibfnamefont {E.}~\bibnamefont
  {Kermarrec}}, \bibinfo {author} {\bibfnamefont {D.~D.}\ \bibnamefont
  {Maharaj}}, \bibinfo {author} {\bibfnamefont {J.}~\bibnamefont {Gaudet}},
  \bibinfo {author} {\bibfnamefont {K.}~\bibnamefont {Fritsch}}, \bibinfo
  {author} {\bibfnamefont {D.}~\bibnamefont {Pomaranski}}, \bibinfo {author}
  {\bibfnamefont {J.~B.}\ \bibnamefont {Kycia}}, \bibinfo {author}
  {\bibfnamefont {Y.}~\bibnamefont {Qiu}}, \bibinfo {author} {\bibfnamefont
  {J.~R.~D.}\ \bibnamefont {Copley}}, \bibinfo {author} {\bibfnamefont
  {M.~M.~P.}\ \bibnamefont {Couchman}}, \bibinfo {author} {\bibfnamefont
  {A.~O.~R.}\ \bibnamefont {Morningstar}}, \bibinfo {author} {\bibfnamefont
  {H.~A.}\ \bibnamefont {Dabkowska}}, \ and\ \bibinfo {author} {\bibfnamefont
  {B.~D.}\ \bibnamefont {Gaulin}},\ }\href {\doibase
  10.1103/PhysRevB.92.245114} {\bibfield  {journal} {\bibinfo  {journal} {Phys.
  Rev. B}\ }\textbf {\bibinfo {volume} {92}},\ \bibinfo {pages} {245114}
  (\bibinfo {year} {2015})}\BibitemShut {NoStop}%
\bibitem [{\citenamefont {Gardner}\ \emph {et~al.}(1999)\citenamefont
  {Gardner}, \citenamefont {Dunsiger}, \citenamefont {Gaulin}, \citenamefont
  {Gingras}, \citenamefont {Greedan}, \citenamefont {Kiefl}, \citenamefont
  {Lumsden}, \citenamefont {MacFarlane}, \citenamefont {Raju}, \citenamefont
  {Sonier}, \citenamefont {Swainson},\ and\ \citenamefont
  {Tun}}]{gardner1999first}%
  \BibitemOpen
  \bibfield  {author} {\bibinfo {author} {\bibfnamefont {J.~S.}\ \bibnamefont
  {Gardner}}, \bibinfo {author} {\bibfnamefont {S.~R.}\ \bibnamefont
  {Dunsiger}}, \bibinfo {author} {\bibfnamefont {B.~D.}\ \bibnamefont
  {Gaulin}}, \bibinfo {author} {\bibfnamefont {M.~J.~P.}\ \bibnamefont
  {Gingras}}, \bibinfo {author} {\bibfnamefont {J.~E.}\ \bibnamefont
  {Greedan}}, \bibinfo {author} {\bibfnamefont {R.~F.}\ \bibnamefont {Kiefl}},
  \bibinfo {author} {\bibfnamefont {M.~D.}\ \bibnamefont {Lumsden}}, \bibinfo
  {author} {\bibfnamefont {W.~A.}\ \bibnamefont {MacFarlane}}, \bibinfo
  {author} {\bibfnamefont {N.~P.}\ \bibnamefont {Raju}}, \bibinfo {author}
  {\bibfnamefont {J.~E.}\ \bibnamefont {Sonier}}, \bibinfo {author}
  {\bibfnamefont {I.}~\bibnamefont {Swainson}}, \ and\ \bibinfo {author}
  {\bibfnamefont {Z.}~\bibnamefont {Tun}},\ }\href {\doibase
  10.1103/PhysRevLett.82.1012} {\bibfield  {journal} {\bibinfo  {journal}
  {Phys. Rev. Lett.}\ }\textbf {\bibinfo {volume} {82}},\ \bibinfo {pages}
  {1012} (\bibinfo {year} {1999})}\BibitemShut {NoStop}%
\bibitem [{\citenamefont {Gardner}\ \emph {et~al.}(2001)\citenamefont
  {Gardner}, \citenamefont {Gaulin}, \citenamefont {Berlinsky}, \citenamefont
  {Waldron}, \citenamefont {Dunsiger}, \citenamefont {Raju},\ and\
  \citenamefont {Greedan}}]{gardner2001tto}%
  \BibitemOpen
  \bibfield  {author} {\bibinfo {author} {\bibfnamefont {J.~S.}\ \bibnamefont
  {Gardner}}, \bibinfo {author} {\bibfnamefont {B.~D.}\ \bibnamefont {Gaulin}},
  \bibinfo {author} {\bibfnamefont {A.~J.}\ \bibnamefont {Berlinsky}}, \bibinfo
  {author} {\bibfnamefont {P.}~\bibnamefont {Waldron}}, \bibinfo {author}
  {\bibfnamefont {S.~R.}\ \bibnamefont {Dunsiger}}, \bibinfo {author}
  {\bibfnamefont {N.~P.}\ \bibnamefont {Raju}}, \ and\ \bibinfo {author}
  {\bibfnamefont {J.~E.}\ \bibnamefont {Greedan}},\ }\href {\doibase
  10.1103/PhysRevB.64.224416} {\bibfield  {journal} {\bibinfo  {journal} {Phys.
  Rev. B}\ }\textbf {\bibinfo {volume} {64}},\ \bibinfo {pages} {224416}
  (\bibinfo {year} {2001})}\BibitemShut {NoStop}%
\bibitem [{\citenamefont {Fritsch}\ \emph {et~al.}(2013)\citenamefont
  {Fritsch}, \citenamefont {Ross}, \citenamefont {Qiu}, \citenamefont {Copley},
  \citenamefont {Guidi}, \citenamefont {Bewley}, \citenamefont {Dabkowska},\
  and\ \citenamefont {Gaulin}}]{fritsch2013hhh}%
  \BibitemOpen
  \bibfield  {author} {\bibinfo {author} {\bibfnamefont {K.}~\bibnamefont
  {Fritsch}}, \bibinfo {author} {\bibfnamefont {K.~A.}\ \bibnamefont {Ross}},
  \bibinfo {author} {\bibfnamefont {Y.}~\bibnamefont {Qiu}}, \bibinfo {author}
  {\bibfnamefont {J.~R.~D.}\ \bibnamefont {Copley}}, \bibinfo {author}
  {\bibfnamefont {T.}~\bibnamefont {Guidi}}, \bibinfo {author} {\bibfnamefont
  {R.~I.}\ \bibnamefont {Bewley}}, \bibinfo {author} {\bibfnamefont {H.~A.}\
  \bibnamefont {Dabkowska}}, \ and\ \bibinfo {author} {\bibfnamefont {B.~D.}\
  \bibnamefont {Gaulin}},\ }\href {\doibase 10.1103/PhysRevB.87.094410}
  {\bibfield  {journal} {\bibinfo  {journal} {Phys. Rev. B}\ }\textbf {\bibinfo
  {volume} {87}},\ \bibinfo {pages} {094410} (\bibinfo {year}
  {2013})}\BibitemShut {NoStop}%
\bibitem [{\citenamefont {Fritsch}\ \emph {et~al.}(2014)\citenamefont
  {Fritsch}, \citenamefont {Kermarrec}, \citenamefont {Ross}, \citenamefont
  {Qiu}, \citenamefont {Copley}, \citenamefont {Pomaranski}, \citenamefont
  {Kycia}, \citenamefont {Dabkowska},\ and\ \citenamefont
  {Gaulin}}]{fritsch2014hhh2}%
  \BibitemOpen
  \bibfield  {author} {\bibinfo {author} {\bibfnamefont {K.}~\bibnamefont
  {Fritsch}}, \bibinfo {author} {\bibfnamefont {E.}~\bibnamefont {Kermarrec}},
  \bibinfo {author} {\bibfnamefont {K.~A.}\ \bibnamefont {Ross}}, \bibinfo
  {author} {\bibfnamefont {Y.}~\bibnamefont {Qiu}}, \bibinfo {author}
  {\bibfnamefont {J.~R.~D.}\ \bibnamefont {Copley}}, \bibinfo {author}
  {\bibfnamefont {D.}~\bibnamefont {Pomaranski}}, \bibinfo {author}
  {\bibfnamefont {J.~B.}\ \bibnamefont {Kycia}}, \bibinfo {author}
  {\bibfnamefont {H.~A.}\ \bibnamefont {Dabkowska}}, \ and\ \bibinfo {author}
  {\bibfnamefont {B.~D.}\ \bibnamefont {Gaulin}},\ }\href {\doibase
  10.1103/PhysRevB.90.014429} {\bibfield  {journal} {\bibinfo  {journal} {Phys.
  Rev. B}\ }\textbf {\bibinfo {volume} {90}},\ \bibinfo {pages} {014429}
  (\bibinfo {year} {2014})}\BibitemShut {NoStop}%
\bibitem [{\citenamefont {Fennell}\ \emph {et~al.}(2012)\citenamefont
  {Fennell}, \citenamefont {Kenzelmann}, \citenamefont {Roessli}, \citenamefont
  {Haas},\ and\ \citenamefont {Cava}}]{fennell2012powerlaw}%
  \BibitemOpen
  \bibfield  {author} {\bibinfo {author} {\bibfnamefont {T.}~\bibnamefont
  {Fennell}}, \bibinfo {author} {\bibfnamefont {M.}~\bibnamefont {Kenzelmann}},
  \bibinfo {author} {\bibfnamefont {B.}~\bibnamefont {Roessli}}, \bibinfo
  {author} {\bibfnamefont {M.~K.}\ \bibnamefont {Haas}}, \ and\ \bibinfo
  {author} {\bibfnamefont {R.~J.}\ \bibnamefont {Cava}},\ }\href {\doibase
  10.1103/PhysRevLett.109.017201} {\bibfield  {journal} {\bibinfo  {journal}
  {Phys. Rev. Lett.}\ }\textbf {\bibinfo {volume} {109}},\ \bibinfo {pages}
  {017201} (\bibinfo {year} {2012})}\BibitemShut {NoStop}%
\bibitem [{\citenamefont {Kanada}\ \emph {et~al.}(1999)\citenamefont {Kanada},
  \citenamefont {Yasui}, \citenamefont {Ito}, \citenamefont {Harashina},
  \citenamefont {Sato}, \citenamefont {Okumura},\ and\ \citenamefont
  {Kakurai}}]{kanada1999neutron}%
  \BibitemOpen
  \bibfield  {author} {\bibinfo {author} {\bibfnamefont {M.}~\bibnamefont
  {Kanada}}, \bibinfo {author} {\bibfnamefont {Y.}~\bibnamefont {Yasui}},
  \bibinfo {author} {\bibfnamefont {M.}~\bibnamefont {Ito}}, \bibinfo {author}
  {\bibfnamefont {H.}~\bibnamefont {Harashina}}, \bibinfo {author}
  {\bibfnamefont {M.}~\bibnamefont {Sato}}, \bibinfo {author} {\bibfnamefont
  {H.}~\bibnamefont {Okumura}}, \ and\ \bibinfo {author} {\bibfnamefont
  {K.}~\bibnamefont {Kakurai}},\ }\href {\doibase 10.1143/JPSJ.68.3802}
  {\bibfield  {journal} {\bibinfo  {journal} {Journal of the Physical Society
  of Japan}\ }\textbf {\bibinfo {volume} {68}},\ \bibinfo {pages} {3802}
  (\bibinfo {year} {1999})}\BibitemShut {NoStop}%
\bibitem [{\citenamefont {Gardner}\ \emph {et~al.}(2003)\citenamefont
  {Gardner}, \citenamefont {Keren}, \citenamefont {Ehlers}, \citenamefont
  {Stock}, \citenamefont {Segal}, \citenamefont {Roper}, \citenamefont
  {F\aa{}k}, \citenamefont {Stone}, \citenamefont {Hammar}, \citenamefont
  {Reich},\ and\ \citenamefont {Gaulin}}]{gardner2003dynamictto}%
  \BibitemOpen
  \bibfield  {author} {\bibinfo {author} {\bibfnamefont {J.~S.}\ \bibnamefont
  {Gardner}}, \bibinfo {author} {\bibfnamefont {A.}~\bibnamefont {Keren}},
  \bibinfo {author} {\bibfnamefont {G.}~\bibnamefont {Ehlers}}, \bibinfo
  {author} {\bibfnamefont {C.}~\bibnamefont {Stock}}, \bibinfo {author}
  {\bibfnamefont {E.}~\bibnamefont {Segal}}, \bibinfo {author} {\bibfnamefont
  {J.~M.}\ \bibnamefont {Roper}}, \bibinfo {author} {\bibfnamefont
  {B.}~\bibnamefont {F\aa{}k}}, \bibinfo {author} {\bibfnamefont {M.~B.}\
  \bibnamefont {Stone}}, \bibinfo {author} {\bibfnamefont {P.~R.}\ \bibnamefont
  {Hammar}}, \bibinfo {author} {\bibfnamefont {D.~H.}\ \bibnamefont {Reich}}, \
  and\ \bibinfo {author} {\bibfnamefont {B.~D.}\ \bibnamefont {Gaulin}},\
  }\href {\doibase 10.1103/PhysRevB.68.180401} {\bibfield  {journal} {\bibinfo
  {journal} {Phys. Rev. B}\ }\textbf {\bibinfo {volume} {68}},\ \bibinfo
  {pages} {180401} (\bibinfo {year} {2003})}\BibitemShut {NoStop}%
\bibitem [{\citenamefont {Takatsu}\ \emph {et~al.}(2012)\citenamefont
  {Takatsu}, \citenamefont {Kadowaki}, \citenamefont {Sato}, \citenamefont
  {Lynn}, \citenamefont {Tabata}, \citenamefont {Yamazaki},\ and\ \citenamefont
  {Matsuhira}}]{takatsu2012ins}%
  \BibitemOpen
  \bibfield  {author} {\bibinfo {author} {\bibfnamefont {H.}~\bibnamefont
  {Takatsu}}, \bibinfo {author} {\bibfnamefont {H.}~\bibnamefont {Kadowaki}},
  \bibinfo {author} {\bibfnamefont {T.~J.}\ \bibnamefont {Sato}}, \bibinfo
  {author} {\bibfnamefont {J.~W.}\ \bibnamefont {Lynn}}, \bibinfo {author}
  {\bibfnamefont {Y.}~\bibnamefont {Tabata}}, \bibinfo {author} {\bibfnamefont
  {T.}~\bibnamefont {Yamazaki}}, \ and\ \bibinfo {author} {\bibfnamefont
  {K.}~\bibnamefont {Matsuhira}},\ }\href {\doibase
  10.1088/0953-8984/24/5/052201} {\bibfield  {journal} {\bibinfo  {journal}
  {Journal of Physics: Condensed Matter}\ }\textbf {\bibinfo {volume} {24}},\
  \bibinfo {pages} {052201} (\bibinfo {year} {2012})}\BibitemShut {NoStop}%
\bibitem [{\citenamefont {Kadowaki}\ \emph {et~al.}(2015)\citenamefont
  {Kadowaki}, \citenamefont {Takatsu}, \citenamefont {Taniguchi}, \citenamefont
  {F{\aa}k},\ and\ \citenamefont {Ollivier}}]{kadowaki2015composite}%
  \BibitemOpen
  \bibfield  {author} {\bibinfo {author} {\bibfnamefont {H.}~\bibnamefont
  {Kadowaki}}, \bibinfo {author} {\bibfnamefont {H.}~\bibnamefont {Takatsu}},
  \bibinfo {author} {\bibfnamefont {T.}~\bibnamefont {Taniguchi}}, \bibinfo
  {author} {\bibfnamefont {B.}~\bibnamefont {F{\aa}k}}, \ and\ \bibinfo
  {author} {\bibfnamefont {J.}~\bibnamefont {Ollivier}},\ }in\ \href {\doibase
  10.1142/S2010324715400032} {\emph {\bibinfo {booktitle} {Spin}}},\
  Vol.~\bibinfo {volume} {5}\ (\bibinfo {organization} {World Scientific},\
  \bibinfo {year} {2015})\ p.\ \bibinfo {pages} {1540003}\BibitemShut {NoStop}%
\bibitem [{\citenamefont {Hamaguchi}\ \emph {et~al.}(2004)\citenamefont
  {Hamaguchi}, \citenamefont {Matsushita}, \citenamefont {Wada}, \citenamefont
  {Yasui},\ and\ \citenamefont {Sato}}]{hamaguchi2004tto}%
  \BibitemOpen
  \bibfield  {author} {\bibinfo {author} {\bibfnamefont {N.}~\bibnamefont
  {Hamaguchi}}, \bibinfo {author} {\bibfnamefont {T.}~\bibnamefont
  {Matsushita}}, \bibinfo {author} {\bibfnamefont {N.}~\bibnamefont {Wada}},
  \bibinfo {author} {\bibfnamefont {Y.}~\bibnamefont {Yasui}}, \ and\ \bibinfo
  {author} {\bibfnamefont {M.}~\bibnamefont {Sato}},\ }\href {\doibase
  10.1103/PhysRevB.69.132413} {\bibfield  {journal} {\bibinfo  {journal} {Phys.
  Rev. B}\ }\textbf {\bibinfo {volume} {69}},\ \bibinfo {pages} {132413}
  (\bibinfo {year} {2004})}\BibitemShut {NoStop}%
\bibitem [{\citenamefont {Wakita}\ \emph {et~al.}(2016)\citenamefont {Wakita},
  \citenamefont {Taniguchi}, \citenamefont {Edamoto}, \citenamefont {Takatsu},\
  and\ \citenamefont {Kadowaki}}]{wakita2016quantum}%
  \BibitemOpen
  \bibfield  {author} {\bibinfo {author} {\bibfnamefont {M.}~\bibnamefont
  {Wakita}}, \bibinfo {author} {\bibfnamefont {T.}~\bibnamefont {Taniguchi}},
  \bibinfo {author} {\bibfnamefont {H.}~\bibnamefont {Edamoto}}, \bibinfo
  {author} {\bibfnamefont {H.}~\bibnamefont {Takatsu}}, \ and\ \bibinfo
  {author} {\bibfnamefont {H.}~\bibnamefont {Kadowaki}},\ }in\ \href {\doibase
  10.1088/1742-6596/683/1/012023} {\emph {\bibinfo {booktitle} {Journal of
  Physics: Conference Series}}},\ Vol.\ \bibinfo {volume} {683}\ (\bibinfo
  {organization} {IOP Publishing},\ \bibinfo {year} {2016})\ p.\ \bibinfo
  {pages} {012023}\BibitemShut {NoStop}%
\bibitem [{\citenamefont {Taniguchi}\ \emph {et~al.}(2013)\citenamefont
  {Taniguchi}, \citenamefont {Kadowaki}, \citenamefont {Takatsu}, \citenamefont
  {F\aa{}k}, \citenamefont {Ollivier}, \citenamefont {Yamazaki}, \citenamefont
  {Sato}, \citenamefont {Yoshizawa}, \citenamefont {Shimura}, \citenamefont
  {Sakakibara}, \citenamefont {Hong}, \citenamefont {Goto}, \citenamefont
  {Yaraskavitch},\ and\ \citenamefont {Kycia}}]{taniguichi2002longrangeorder}%
  \BibitemOpen
  \bibfield  {author} {\bibinfo {author} {\bibfnamefont {T.}~\bibnamefont
  {Taniguchi}}, \bibinfo {author} {\bibfnamefont {H.}~\bibnamefont {Kadowaki}},
  \bibinfo {author} {\bibfnamefont {H.}~\bibnamefont {Takatsu}}, \bibinfo
  {author} {\bibfnamefont {B.}~\bibnamefont {F\aa{}k}}, \bibinfo {author}
  {\bibfnamefont {J.}~\bibnamefont {Ollivier}}, \bibinfo {author}
  {\bibfnamefont {T.}~\bibnamefont {Yamazaki}}, \bibinfo {author}
  {\bibfnamefont {T.~J.}\ \bibnamefont {Sato}}, \bibinfo {author}
  {\bibfnamefont {H.}~\bibnamefont {Yoshizawa}}, \bibinfo {author}
  {\bibfnamefont {Y.}~\bibnamefont {Shimura}}, \bibinfo {author} {\bibfnamefont
  {T.}~\bibnamefont {Sakakibara}}, \bibinfo {author} {\bibfnamefont
  {T.}~\bibnamefont {Hong}}, \bibinfo {author} {\bibfnamefont {K.}~\bibnamefont
  {Goto}}, \bibinfo {author} {\bibfnamefont {L.~R.}\ \bibnamefont
  {Yaraskavitch}}, \ and\ \bibinfo {author} {\bibfnamefont {J.~B.}\
  \bibnamefont {Kycia}},\ }\href {\doibase 10.1103/PhysRevB.87.060408}
  {\bibfield  {journal} {\bibinfo  {journal} {Phys. Rev. B}\ }\textbf {\bibinfo
  {volume} {87}},\ \bibinfo {pages} {060408} (\bibinfo {year}
  {2013})}\BibitemShut {NoStop}%
\bibitem [{\citenamefont {Kadowaki}\ \emph {et~al.}(2018)\citenamefont
  {Kadowaki}, \citenamefont {Wakita}, \citenamefont {F{\aa}k}, \citenamefont
  {Ollivier}, \citenamefont {Ohira-Kawamura}, \citenamefont {Nakajima},
  \citenamefont {Takatsu},\ and\ \citenamefont
  {Tamai}}]{kadowaki2018continuum}%
  \BibitemOpen
  \bibfield  {author} {\bibinfo {author} {\bibfnamefont {H.}~\bibnamefont
  {Kadowaki}}, \bibinfo {author} {\bibfnamefont {M.}~\bibnamefont {Wakita}},
  \bibinfo {author} {\bibfnamefont {B.}~\bibnamefont {F{\aa}k}}, \bibinfo
  {author} {\bibfnamefont {J.}~\bibnamefont {Ollivier}}, \bibinfo {author}
  {\bibfnamefont {S.}~\bibnamefont {Ohira-Kawamura}}, \bibinfo {author}
  {\bibfnamefont {K.}~\bibnamefont {Nakajima}}, \bibinfo {author}
  {\bibfnamefont {H.}~\bibnamefont {Takatsu}}, \ and\ \bibinfo {author}
  {\bibfnamefont {M.}~\bibnamefont {Tamai}},\ }\href@noop {} {\  (\bibinfo
  {year} {2018})},\ \Eprint {http://arxiv.org/abs/1803.00688} {arXiv:1803.00688
  [cond-mat.str-el]} \BibitemShut {NoStop}%
\bibitem [{\citenamefont {Mirebeau}\ \emph {et~al.}(2005)\citenamefont
  {Mirebeau}, \citenamefont {Apetrei}, \citenamefont {Rodr\'{\i}guez-Carvajal},
  \citenamefont {Bonville}, \citenamefont {Forget}, \citenamefont {Colson},
  \citenamefont {Glazkov}, \citenamefont {Sanchez}, \citenamefont {Isnard},\
  and\ \citenamefont {Suard}}]{mirebeau2005ordered}%
  \BibitemOpen
  \bibfield  {author} {\bibinfo {author} {\bibfnamefont {I.}~\bibnamefont
  {Mirebeau}}, \bibinfo {author} {\bibfnamefont {A.}~\bibnamefont {Apetrei}},
  \bibinfo {author} {\bibfnamefont {J.}~\bibnamefont
  {Rodr\'{\i}guez-Carvajal}}, \bibinfo {author} {\bibfnamefont
  {P.}~\bibnamefont {Bonville}}, \bibinfo {author} {\bibfnamefont
  {A.}~\bibnamefont {Forget}}, \bibinfo {author} {\bibfnamefont
  {D.}~\bibnamefont {Colson}}, \bibinfo {author} {\bibfnamefont
  {V.}~\bibnamefont {Glazkov}}, \bibinfo {author} {\bibfnamefont {J.~P.}\
  \bibnamefont {Sanchez}}, \bibinfo {author} {\bibfnamefont {O.}~\bibnamefont
  {Isnard}}, \ and\ \bibinfo {author} {\bibfnamefont {E.}~\bibnamefont
  {Suard}},\ }\href {\doibase 10.1103/PhysRevLett.94.246402} {\bibfield
  {journal} {\bibinfo  {journal} {Phys. Rev. Lett.}\ }\textbf {\bibinfo
  {volume} {94}},\ \bibinfo {pages} {246402} (\bibinfo {year}
  {2005})}\BibitemShut {NoStop}%
\bibitem [{\citenamefont {Sibille}\ \emph
  {et~al.}(2017{\natexlab{a}})\citenamefont {Sibille}, \citenamefont {Lhotel},
  \citenamefont {Hatnean}, \citenamefont {Nilsen}, \citenamefont {Ehlers},
  \citenamefont {Cervellino}, \citenamefont {Ressouche}, \citenamefont
  {Frontzek}, \citenamefont {Zaharko}, \citenamefont {Pomjakushin} \emph
  {et~al.}}]{sibille2017coulomb}%
  \BibitemOpen
  \bibfield  {author} {\bibinfo {author} {\bibfnamefont {R.}~\bibnamefont
  {Sibille}}, \bibinfo {author} {\bibfnamefont {E.}~\bibnamefont {Lhotel}},
  \bibinfo {author} {\bibfnamefont {M.~C.}\ \bibnamefont {Hatnean}}, \bibinfo
  {author} {\bibfnamefont {G.~J.}\ \bibnamefont {Nilsen}}, \bibinfo {author}
  {\bibfnamefont {G.}~\bibnamefont {Ehlers}}, \bibinfo {author} {\bibfnamefont
  {A.}~\bibnamefont {Cervellino}}, \bibinfo {author} {\bibfnamefont
  {E.}~\bibnamefont {Ressouche}}, \bibinfo {author} {\bibfnamefont
  {M.}~\bibnamefont {Frontzek}}, \bibinfo {author} {\bibfnamefont
  {O.}~\bibnamefont {Zaharko}}, \bibinfo {author} {\bibfnamefont
  {V.}~\bibnamefont {Pomjakushin}},  \emph {et~al.},\ }\href {\doibase
  10.1038/s41467-017-00905-w} {\bibfield  {journal} {\bibinfo  {journal}
  {Nature communications}\ }\textbf {\bibinfo {volume} {8}},\ \bibinfo {pages}
  {892} (\bibinfo {year} {2017}{\natexlab{a}})}\BibitemShut {NoStop}%
\bibitem [{\citenamefont {Ruff}\ \emph {et~al.}(2010)\citenamefont {Ruff},
  \citenamefont {Islam}, \citenamefont {Clancy}, \citenamefont {Ross},
  \citenamefont {Nojiri}, \citenamefont {Matsuda}, \citenamefont {Dabkowska},
  \citenamefont {Dabkowski},\ and\ \citenamefont
  {Gaulin}}]{ruff2010magnetoelastic}%
  \BibitemOpen
  \bibfield  {author} {\bibinfo {author} {\bibfnamefont {J.~P.~C.}\
  \bibnamefont {Ruff}}, \bibinfo {author} {\bibfnamefont {Z.}~\bibnamefont
  {Islam}}, \bibinfo {author} {\bibfnamefont {J.~P.}\ \bibnamefont {Clancy}},
  \bibinfo {author} {\bibfnamefont {K.~A.}\ \bibnamefont {Ross}}, \bibinfo
  {author} {\bibfnamefont {H.}~\bibnamefont {Nojiri}}, \bibinfo {author}
  {\bibfnamefont {Y.~H.}\ \bibnamefont {Matsuda}}, \bibinfo {author}
  {\bibfnamefont {H.~A.}\ \bibnamefont {Dabkowska}}, \bibinfo {author}
  {\bibfnamefont {A.~D.}\ \bibnamefont {Dabkowski}}, \ and\ \bibinfo {author}
  {\bibfnamefont {B.~D.}\ \bibnamefont {Gaulin}},\ }\href {\doibase
  10.1103/PhysRevLett.105.077203} {\bibfield  {journal} {\bibinfo  {journal}
  {Phys. Rev. Lett.}\ }\textbf {\bibinfo {volume} {105}},\ \bibinfo {pages}
  {077203} (\bibinfo {year} {2010})}\BibitemShut {NoStop}%
\bibitem [{\citenamefont {Constable}\ \emph {et~al.}(2017)\citenamefont
  {Constable}, \citenamefont {Ballou}, \citenamefont {Robert}, \citenamefont
  {Decorse}, \citenamefont {Brubach}, \citenamefont {Roy}, \citenamefont
  {Lhotel}, \citenamefont {Del-Rey}, \citenamefont {Simonet}, \citenamefont
  {Petit},\ and\ \citenamefont {deBrion}}]{constable2017vibronic}%
  \BibitemOpen
  \bibfield  {author} {\bibinfo {author} {\bibfnamefont {E.}~\bibnamefont
  {Constable}}, \bibinfo {author} {\bibfnamefont {R.}~\bibnamefont {Ballou}},
  \bibinfo {author} {\bibfnamefont {J.}~\bibnamefont {Robert}}, \bibinfo
  {author} {\bibfnamefont {C.}~\bibnamefont {Decorse}}, \bibinfo {author}
  {\bibfnamefont {J.-B.}\ \bibnamefont {Brubach}}, \bibinfo {author}
  {\bibfnamefont {P.}~\bibnamefont {Roy}}, \bibinfo {author} {\bibfnamefont
  {E.}~\bibnamefont {Lhotel}}, \bibinfo {author} {\bibfnamefont
  {L.}~\bibnamefont {Del-Rey}}, \bibinfo {author} {\bibfnamefont
  {V.}~\bibnamefont {Simonet}}, \bibinfo {author} {\bibfnamefont
  {S.}~\bibnamefont {Petit}}, \ and\ \bibinfo {author} {\bibfnamefont
  {S.}~\bibnamefont {deBrion}},\ }\href {\doibase 10.1103/PhysRevB.95.020415}
  {\bibfield  {journal} {\bibinfo  {journal} {Phys. Rev. B}\ }\textbf {\bibinfo
  {volume} {95}},\ \bibinfo {pages} {020415} (\bibinfo {year}
  {2017})}\BibitemShut {NoStop}%
\bibitem [{\citenamefont {Belov}\ \emph {et~al.}(1983)\citenamefont {Belov},
  \citenamefont {Kataev}, \citenamefont {Levitin}, \citenamefont {Nikitin},\
  and\ \citenamefont {Sokolov}}]{belov1983giant}%
  \BibitemOpen
  \bibfield  {author} {\bibinfo {author} {\bibfnamefont {K.~P.}\ \bibnamefont
  {Belov}}, \bibinfo {author} {\bibfnamefont {G.~I.}\ \bibnamefont {Kataev}},
  \bibinfo {author} {\bibfnamefont {R.~Z.}\ \bibnamefont {Levitin}}, \bibinfo
  {author} {\bibfnamefont {S.~A.}\ \bibnamefont {Nikitin}}, \ and\ \bibinfo
  {author} {\bibfnamefont {V.~I.}\ \bibnamefont {Sokolov}},\ }\href@noop {}
  {\bibfield  {journal} {\bibinfo  {journal} {Physics-Uspekhi}\ }\textbf
  {\bibinfo {volume} {26}},\ \bibinfo {pages} {518} (\bibinfo {year}
  {1983})}\BibitemShut {NoStop}%
\bibitem [{\citenamefont {Mamsurova}\ \emph {et~al.}(1986)\citenamefont
  {Mamsurova}, \citenamefont {Pigal'Skii},\ and\ \citenamefont
  {Pukhov}}]{mamsurova1986low}%
  \BibitemOpen
  \bibfield  {author} {\bibinfo {author} {\bibfnamefont {L.~G.}\ \bibnamefont
  {Mamsurova}}, \bibinfo {author} {\bibfnamefont {K.~S.}\ \bibnamefont
  {Pigal'Skii}}, \ and\ \bibinfo {author} {\bibfnamefont {K.~K.}\ \bibnamefont
  {Pukhov}},\ }\href@noop {} {\bibfield  {journal} {\bibinfo  {journal} {JETP
  Lett}\ }\textbf {\bibinfo {volume} {43}} (\bibinfo {year}
  {1986})}\BibitemShut {NoStop}%
\bibitem [{\citenamefont {Ruminy}\ \emph {et~al.}(2016)\citenamefont {Ruminy},
  \citenamefont {Bovo}, \citenamefont {Pomjakushina}, \citenamefont {Haas},
  \citenamefont {Stuhr}, \citenamefont {Cervellino}, \citenamefont {Cava},
  \citenamefont {Kenzelmann},\ and\ \citenamefont
  {Fennell}}]{ruminy2016magnetoelastic1}%
  \BibitemOpen
  \bibfield  {author} {\bibinfo {author} {\bibfnamefont {M.}~\bibnamefont
  {Ruminy}}, \bibinfo {author} {\bibfnamefont {L.}~\bibnamefont {Bovo}},
  \bibinfo {author} {\bibfnamefont {E.}~\bibnamefont {Pomjakushina}}, \bibinfo
  {author} {\bibfnamefont {M.~K.}\ \bibnamefont {Haas}}, \bibinfo {author}
  {\bibfnamefont {U.}~\bibnamefont {Stuhr}}, \bibinfo {author} {\bibfnamefont
  {A.}~\bibnamefont {Cervellino}}, \bibinfo {author} {\bibfnamefont {R.~J.}\
  \bibnamefont {Cava}}, \bibinfo {author} {\bibfnamefont {M.}~\bibnamefont
  {Kenzelmann}}, \ and\ \bibinfo {author} {\bibfnamefont {T.}~\bibnamefont
  {Fennell}},\ }\href {\doibase 10.1103/PhysRevB.93.144407} {\bibfield
  {journal} {\bibinfo  {journal} {Phys. Rev. B}\ }\textbf {\bibinfo {volume}
  {93}},\ \bibinfo {pages} {144407} (\bibinfo {year} {2016})}\BibitemShut
  {NoStop}%
\bibitem [{\citenamefont {Sazonov}\ \emph {et~al.}(2012)\citenamefont
  {Sazonov}, \citenamefont {Gukasov}, \citenamefont {Mirebeau},\ and\
  \citenamefont {Bonville}}]{ttodoublelayer}%
  \BibitemOpen
  \bibfield  {author} {\bibinfo {author} {\bibfnamefont {A.~P.}\ \bibnamefont
  {Sazonov}}, \bibinfo {author} {\bibfnamefont {A.}~\bibnamefont {Gukasov}},
  \bibinfo {author} {\bibfnamefont {I.}~\bibnamefont {Mirebeau}}, \ and\
  \bibinfo {author} {\bibfnamefont {P.}~\bibnamefont {Bonville}},\ }\href
  {\doibase 10.1103/PhysRevB.85.214420} {\bibfield  {journal} {\bibinfo
  {journal} {Phys. Rev. B}\ }\textbf {\bibinfo {volume} {85}},\ \bibinfo
  {pages} {214420} (\bibinfo {year} {2012})}\BibitemShut {NoStop}%
\bibitem [{\citenamefont {Legl}\ \emph {et~al.}(2012)\citenamefont {Legl},
  \citenamefont {Krey}, \citenamefont {Dunsiger}, \citenamefont {Dabkowska},
  \citenamefont {Rodriguez}, \citenamefont {Luke},\ and\ \citenamefont
  {Pfleiderer}}]{ttofield2}%
  \BibitemOpen
  \bibfield  {author} {\bibinfo {author} {\bibfnamefont {S.}~\bibnamefont
  {Legl}}, \bibinfo {author} {\bibfnamefont {C.}~\bibnamefont {Krey}}, \bibinfo
  {author} {\bibfnamefont {S.~R.}\ \bibnamefont {Dunsiger}}, \bibinfo {author}
  {\bibfnamefont {H.~A.}\ \bibnamefont {Dabkowska}}, \bibinfo {author}
  {\bibfnamefont {J.~A.}\ \bibnamefont {Rodriguez}}, \bibinfo {author}
  {\bibfnamefont {G.~M.}\ \bibnamefont {Luke}}, \ and\ \bibinfo {author}
  {\bibfnamefont {C.}~\bibnamefont {Pfleiderer}},\ }\href {\doibase
  10.1103/PhysRevLett.109.047201} {\bibfield  {journal} {\bibinfo  {journal}
  {Phys. Rev. Lett.}\ }\textbf {\bibinfo {volume} {109}},\ \bibinfo {pages}
  {047201} (\bibinfo {year} {2012})}\BibitemShut {NoStop}%
\bibitem [{\citenamefont {Yin}\ \emph {et~al.}(2013)\citenamefont {Yin},
  \citenamefont {Xia}, \citenamefont {Takano}, \citenamefont {Sullivan},
  \citenamefont {Li},\ and\ \citenamefont {Sun}}]{yin2013field}%
  \BibitemOpen
  \bibfield  {author} {\bibinfo {author} {\bibfnamefont {L.}~\bibnamefont
  {Yin}}, \bibinfo {author} {\bibfnamefont {J.~S.}\ \bibnamefont {Xia}},
  \bibinfo {author} {\bibfnamefont {Y.}~\bibnamefont {Takano}}, \bibinfo
  {author} {\bibfnamefont {N.~S.}\ \bibnamefont {Sullivan}}, \bibinfo {author}
  {\bibfnamefont {Q.~J.}\ \bibnamefont {Li}}, \ and\ \bibinfo {author}
  {\bibfnamefont {X.~F.}\ \bibnamefont {Sun}},\ }\href {\doibase
  10.1103/PhysRevLett.110.137201} {\bibfield  {journal} {\bibinfo  {journal}
  {Phys. Rev. Lett.}\ }\textbf {\bibinfo {volume} {110}},\ \bibinfo {pages}
  {137201} (\bibinfo {year} {2013})}\BibitemShut {NoStop}%
\bibitem [{\citenamefont {Hirschberger}\ \emph {et~al.}(2015)\citenamefont
  {Hirschberger}, \citenamefont {Krizan}, \citenamefont {Cava},\ and\
  \citenamefont {Ong}}]{hirschberger2015large}%
  \BibitemOpen
  \bibfield  {author} {\bibinfo {author} {\bibfnamefont {M.}~\bibnamefont
  {Hirschberger}}, \bibinfo {author} {\bibfnamefont {J.~W.}\ \bibnamefont
  {Krizan}}, \bibinfo {author} {\bibfnamefont {R.~J.}\ \bibnamefont {Cava}}, \
  and\ \bibinfo {author} {\bibfnamefont {N.~P.}\ \bibnamefont {Ong}},\ }\href
  {\doibase 10.1126/science.1257340} {\bibfield  {journal} {\bibinfo  {journal}
  {Science}\ }\textbf {\bibinfo {volume} {348}},\ \bibinfo {pages} {106}
  (\bibinfo {year} {2015})}\BibitemShut {NoStop}%
\bibitem [{\citenamefont {Savary}\ and\ \citenamefont
  {Balents}(2017)}]{savary2017disorder}%
  \BibitemOpen
  \bibfield  {author} {\bibinfo {author} {\bibfnamefont {L.}~\bibnamefont
  {Savary}}\ and\ \bibinfo {author} {\bibfnamefont {L.}~\bibnamefont
  {Balents}},\ }\href {\doibase 10.1103/PhysRevLett.118.087203} {\bibfield
  {journal} {\bibinfo  {journal} {Phys. Rev. Lett.}\ }\textbf {\bibinfo
  {volume} {118}},\ \bibinfo {pages} {087203} (\bibinfo {year}
  {2017})}\BibitemShut {NoStop}%
\bibitem [{\citenamefont {Benton}(2017)}]{benton2017quantum}%
  \BibitemOpen
  \bibfield  {author} {\bibinfo {author} {\bibfnamefont {O.}~\bibnamefont
  {Benton}},\ }\href@noop {} {\  (\bibinfo {year} {2017})},\ \Eprint
  {http://arxiv.org/abs/1706.09238} {arXiv:1706.09238 [cond-mat.str-el]}
  \BibitemShut {NoStop}%
\bibitem [{\citenamefont {Kimura}\ \emph {et~al.}(2013)\citenamefont {Kimura},
  \citenamefont {Nakatsuji}, \citenamefont {Wen}, \citenamefont {Broholm},
  \citenamefont {Stone}, \citenamefont {Nishibori},\ and\ \citenamefont
  {Sawa}}]{kimura2013quantum}%
  \BibitemOpen
  \bibfield  {author} {\bibinfo {author} {\bibfnamefont {K.}~\bibnamefont
  {Kimura}}, \bibinfo {author} {\bibfnamefont {S.}~\bibnamefont {Nakatsuji}},
  \bibinfo {author} {\bibfnamefont {J.~J.}\ \bibnamefont {Wen}}, \bibinfo
  {author} {\bibfnamefont {C.}~\bibnamefont {Broholm}}, \bibinfo {author}
  {\bibfnamefont {M.~B.}\ \bibnamefont {Stone}}, \bibinfo {author}
  {\bibfnamefont {E.}~\bibnamefont {Nishibori}}, \ and\ \bibinfo {author}
  {\bibfnamefont {H.}~\bibnamefont {Sawa}},\ }\href {\doibase
  10.1038/ncomms2914} {\bibfield  {journal} {\bibinfo  {journal} {Nature
  communications}\ }\textbf {\bibinfo {volume} {4}},\ \bibinfo {pages} {1934}
  (\bibinfo {year} {2013})}\BibitemShut {NoStop}%
\bibitem [{\citenamefont {Petit}\ \emph {et~al.}(2016)\citenamefont {Petit},
  \citenamefont {Lhotel}, \citenamefont {Guitteny}, \citenamefont {Florea},
  \citenamefont {Robert}, \citenamefont {Bonville}, \citenamefont {Mirebeau},
  \citenamefont {Ollivier}, \citenamefont {Mutka}, \citenamefont {Ressouche},
  \citenamefont {Decorse}, \citenamefont {Ciomaga~Hatnean},\ and\ \citenamefont
  {Balakrishnan}}]{petit2016przro1}%
  \BibitemOpen
  \bibfield  {author} {\bibinfo {author} {\bibfnamefont {S.}~\bibnamefont
  {Petit}}, \bibinfo {author} {\bibfnamefont {E.}~\bibnamefont {Lhotel}},
  \bibinfo {author} {\bibfnamefont {S.}~\bibnamefont {Guitteny}}, \bibinfo
  {author} {\bibfnamefont {O.}~\bibnamefont {Florea}}, \bibinfo {author}
  {\bibfnamefont {J.}~\bibnamefont {Robert}}, \bibinfo {author} {\bibfnamefont
  {P.}~\bibnamefont {Bonville}}, \bibinfo {author} {\bibfnamefont
  {I.}~\bibnamefont {Mirebeau}}, \bibinfo {author} {\bibfnamefont
  {J.}~\bibnamefont {Ollivier}}, \bibinfo {author} {\bibfnamefont
  {H.}~\bibnamefont {Mutka}}, \bibinfo {author} {\bibfnamefont
  {E.}~\bibnamefont {Ressouche}}, \bibinfo {author} {\bibfnamefont
  {C.}~\bibnamefont {Decorse}}, \bibinfo {author} {\bibfnamefont
  {M.}~\bibnamefont {Ciomaga~Hatnean}}, \ and\ \bibinfo {author} {\bibfnamefont
  {G.}~\bibnamefont {Balakrishnan}},\ }\href {\doibase
  10.1103/PhysRevB.94.165153} {\bibfield  {journal} {\bibinfo  {journal} {Phys.
  Rev. B}\ }\textbf {\bibinfo {volume} {94}},\ \bibinfo {pages} {165153}
  (\bibinfo {year} {2016})}\BibitemShut {NoStop}%
\bibitem [{\citenamefont {Bonville}\ \emph {et~al.}(2016)\citenamefont
  {Bonville}, \citenamefont {Guitteny}, \citenamefont {Gukasov}, \citenamefont
  {Mirebeau}, \citenamefont {Petit}, \citenamefont {Decorse}, \citenamefont
  {Hatnean},\ and\ \citenamefont {Balakrishnan}}]{bonville2016przro2}%
  \BibitemOpen
  \bibfield  {author} {\bibinfo {author} {\bibfnamefont {P.}~\bibnamefont
  {Bonville}}, \bibinfo {author} {\bibfnamefont {S.}~\bibnamefont {Guitteny}},
  \bibinfo {author} {\bibfnamefont {A.}~\bibnamefont {Gukasov}}, \bibinfo
  {author} {\bibfnamefont {I.}~\bibnamefont {Mirebeau}}, \bibinfo {author}
  {\bibfnamefont {S.}~\bibnamefont {Petit}}, \bibinfo {author} {\bibfnamefont
  {C.}~\bibnamefont {Decorse}}, \bibinfo {author} {\bibfnamefont {M.~C.}\
  \bibnamefont {Hatnean}}, \ and\ \bibinfo {author} {\bibfnamefont
  {G.}~\bibnamefont {Balakrishnan}},\ }\href {\doibase
  10.1103/PhysRevB.94.134428} {\bibfield  {journal} {\bibinfo  {journal} {Phys.
  Rev. B}\ }\textbf {\bibinfo {volume} {94}},\ \bibinfo {pages} {134428}
  (\bibinfo {year} {2016})}\BibitemShut {NoStop}%
\bibitem [{\citenamefont {Martin}\ \emph {et~al.}(2017)\citenamefont {Martin},
  \citenamefont {Bonville}, \citenamefont {Lhotel}, \citenamefont {Guitteny},
  \citenamefont {Wildes}, \citenamefont {Decorse}, \citenamefont
  {Ciomaga~Hatnean}, \citenamefont {Balakrishnan}, \citenamefont {Mirebeau},\
  and\ \citenamefont {Petit}}]{martin2017disorder}%
  \BibitemOpen
  \bibfield  {author} {\bibinfo {author} {\bibfnamefont {N.}~\bibnamefont
  {Martin}}, \bibinfo {author} {\bibfnamefont {P.}~\bibnamefont {Bonville}},
  \bibinfo {author} {\bibfnamefont {E.}~\bibnamefont {Lhotel}}, \bibinfo
  {author} {\bibfnamefont {S.}~\bibnamefont {Guitteny}}, \bibinfo {author}
  {\bibfnamefont {A.}~\bibnamefont {Wildes}}, \bibinfo {author} {\bibfnamefont
  {C.}~\bibnamefont {Decorse}}, \bibinfo {author} {\bibfnamefont
  {M.}~\bibnamefont {Ciomaga~Hatnean}}, \bibinfo {author} {\bibfnamefont
  {G.}~\bibnamefont {Balakrishnan}}, \bibinfo {author} {\bibfnamefont
  {I.}~\bibnamefont {Mirebeau}}, \ and\ \bibinfo {author} {\bibfnamefont
  {S.}~\bibnamefont {Petit}},\ }\href {\doibase 10.1103/PhysRevX.7.041028}
  {\bibfield  {journal} {\bibinfo  {journal} {Phys. Rev. X}\ }\textbf {\bibinfo
  {volume} {7}},\ \bibinfo {pages} {041028} (\bibinfo {year}
  {2017})}\BibitemShut {NoStop}%
\bibitem [{\citenamefont {Wen}\ \emph {et~al.}(2017)\citenamefont {Wen},
  \citenamefont {Koohpayeh}, \citenamefont {Ross}, \citenamefont {Trump},
  \citenamefont {McQueen}, \citenamefont {Kimura}, \citenamefont {Nakatsuji},
  \citenamefont {Qiu}, \citenamefont {Pajerowski}, \citenamefont {Copley},\
  and\ \citenamefont {Broholm}}]{wen2017disorder}%
  \BibitemOpen
  \bibfield  {author} {\bibinfo {author} {\bibfnamefont {J.-J.}\ \bibnamefont
  {Wen}}, \bibinfo {author} {\bibfnamefont {S.~M.}\ \bibnamefont {Koohpayeh}},
  \bibinfo {author} {\bibfnamefont {K.~A.}\ \bibnamefont {Ross}}, \bibinfo
  {author} {\bibfnamefont {B.~A.}\ \bibnamefont {Trump}}, \bibinfo {author}
  {\bibfnamefont {T.~M.}\ \bibnamefont {McQueen}}, \bibinfo {author}
  {\bibfnamefont {K.}~\bibnamefont {Kimura}}, \bibinfo {author} {\bibfnamefont
  {S.}~\bibnamefont {Nakatsuji}}, \bibinfo {author} {\bibfnamefont
  {Y.}~\bibnamefont {Qiu}}, \bibinfo {author} {\bibfnamefont {D.~M.}\
  \bibnamefont {Pajerowski}}, \bibinfo {author} {\bibfnamefont {J.~R.~D.}\
  \bibnamefont {Copley}}, \ and\ \bibinfo {author} {\bibfnamefont {C.~L.}\
  \bibnamefont {Broholm}},\ }\href {\doibase 10.1103/PhysRevLett.118.107206}
  {\bibfield  {journal} {\bibinfo  {journal} {Phys. Rev. Lett.}\ }\textbf
  {\bibinfo {volume} {118}},\ \bibinfo {pages} {107206} (\bibinfo {year}
  {2017})}\BibitemShut {NoStop}%
\bibitem [{\citenamefont {Koohpayeh}\ \emph {et~al.}(2014)\citenamefont
  {Koohpayeh}, \citenamefont {Wen}, \citenamefont {Trump}, \citenamefont
  {Broholm},\ and\ \citenamefont {McQueen}}]{koohpayeh2014synthesis}%
  \BibitemOpen
  \bibfield  {author} {\bibinfo {author} {\bibfnamefont {S.~M.}\ \bibnamefont
  {Koohpayeh}}, \bibinfo {author} {\bibfnamefont {J.-J.}\ \bibnamefont {Wen}},
  \bibinfo {author} {\bibfnamefont {B.~A.}\ \bibnamefont {Trump}}, \bibinfo
  {author} {\bibfnamefont {C.~L.}\ \bibnamefont {Broholm}}, \ and\ \bibinfo
  {author} {\bibfnamefont {T.~M.}\ \bibnamefont {McQueen}},\ }\href {\doibase
  10.1016/j.jcrysgro.2014.06.037} {\bibfield  {journal} {\bibinfo  {journal}
  {Journal of Crystal Growth}\ }\textbf {\bibinfo {volume} {402}},\ \bibinfo
  {pages} {291} (\bibinfo {year} {2014})}\BibitemShut {NoStop}%
\bibitem [{\citenamefont {Tabira}\ \emph {et~al.}(1999)\citenamefont {Tabira},
  \citenamefont {Withers}, \citenamefont {Thompson},\ and\ \citenamefont
  {Schmid}}]{tabira1999structured}%
  \BibitemOpen
  \bibfield  {author} {\bibinfo {author} {\bibfnamefont {Y.}~\bibnamefont
  {Tabira}}, \bibinfo {author} {\bibfnamefont {R.}~\bibnamefont {Withers}},
  \bibinfo {author} {\bibfnamefont {J.}~\bibnamefont {Thompson}}, \ and\
  \bibinfo {author} {\bibfnamefont {S.}~\bibnamefont {Schmid}},\ }\href
  {\doibase 10.1006/jssc.1998.8054} {\bibfield  {journal} {\bibinfo  {journal}
  {Journal of Solid State Chemistry}\ }\textbf {\bibinfo {volume} {142}},\
  \bibinfo {pages} {393} (\bibinfo {year} {1999})}\BibitemShut {NoStop}%
\bibitem [{\citenamefont {Tabira}\ \emph {et~al.}(2001)\citenamefont {Tabira},
  \citenamefont {Withers}, \citenamefont {Yamada},\ and\ \citenamefont
  {Ishizawa}}]{tabira2001annular}%
  \BibitemOpen
  \bibfield  {author} {\bibinfo {author} {\bibfnamefont {Y.}~\bibnamefont
  {Tabira}}, \bibinfo {author} {\bibfnamefont {R.~L.}\ \bibnamefont {Withers}},
  \bibinfo {author} {\bibfnamefont {T.}~\bibnamefont {Yamada}}, \ and\ \bibinfo
  {author} {\bibfnamefont {N.}~\bibnamefont {Ishizawa}},\ }\href {\doibase
  10.1524/zkri.216.2.92.20338} {\bibfield  {journal} {\bibinfo  {journal} {Z.
  Kristallogr.}\ }\textbf {\bibinfo {volume} {216}},\ \bibinfo {pages} {92}
  (\bibinfo {year} {2001})}\BibitemShut {NoStop}%
\bibitem [{\citenamefont {Shoemaker}\ \emph {et~al.}(2011)\citenamefont
  {Shoemaker}, \citenamefont {Seshadri}, \citenamefont {Tachibana},\ and\
  \citenamefont {Hector}}]{shoemaker2011biruo}%
  \BibitemOpen
  \bibfield  {author} {\bibinfo {author} {\bibfnamefont {D.~P.}\ \bibnamefont
  {Shoemaker}}, \bibinfo {author} {\bibfnamefont {R.}~\bibnamefont {Seshadri}},
  \bibinfo {author} {\bibfnamefont {M.}~\bibnamefont {Tachibana}}, \ and\
  \bibinfo {author} {\bibfnamefont {A.~L.}\ \bibnamefont {Hector}},\ }\href
  {\doibase 10.1103/PhysRevB.84.064117} {\bibfield  {journal} {\bibinfo
  {journal} {Phys. Rev. B}\ }\textbf {\bibinfo {volume} {84}},\ \bibinfo
  {pages} {064117} (\bibinfo {year} {2011})}\BibitemShut {NoStop}%
\bibitem [{\citenamefont {Tokiwa}\ \emph {et~al.}(2018)\citenamefont {Tokiwa},
  \citenamefont {Yamashita}, \citenamefont {Terazawa}, \citenamefont {Kimura},
  \citenamefont {Kasahara}, \citenamefont {Onishi}, \citenamefont {Kato},
  \citenamefont {Halim}, \citenamefont {Gegenwart}, \citenamefont {Shibauchi}
  \emph {et~al.}}]{tokiwa2018discovery}%
  \BibitemOpen
  \bibfield  {author} {\bibinfo {author} {\bibfnamefont {Y.}~\bibnamefont
  {Tokiwa}}, \bibinfo {author} {\bibfnamefont {T.}~\bibnamefont {Yamashita}},
  \bibinfo {author} {\bibfnamefont {D.}~\bibnamefont {Terazawa}}, \bibinfo
  {author} {\bibfnamefont {K.}~\bibnamefont {Kimura}}, \bibinfo {author}
  {\bibfnamefont {Y.}~\bibnamefont {Kasahara}}, \bibinfo {author}
  {\bibfnamefont {T.}~\bibnamefont {Onishi}}, \bibinfo {author} {\bibfnamefont
  {Y.}~\bibnamefont {Kato}}, \bibinfo {author} {\bibfnamefont {M.}~\bibnamefont
  {Halim}}, \bibinfo {author} {\bibfnamefont {P.}~\bibnamefont {Gegenwart}},
  \bibinfo {author} {\bibfnamefont {T.}~\bibnamefont {Shibauchi}},  \emph
  {et~al.},\ }\href@noop {} {\  (\bibinfo {year} {2018})},\ \Eprint
  {http://arxiv.org/abs/1803.05557} {arXiv:1803.05557 [cond-mat.str-el]}
  \BibitemShut {NoStop}%
\bibitem [{\citenamefont {Anand}\ \emph {et~al.}(2016)\citenamefont {Anand},
  \citenamefont {Opherden}, \citenamefont {Xu}, \citenamefont {Adroja},
  \citenamefont {Islam}, \citenamefont {Herrmannsd\"orfer}, \citenamefont
  {Hornung}, \citenamefont {Sch\"onemann}, \citenamefont {Uhlarz},
  \citenamefont {Walker}, \citenamefont {Casati},\ and\ \citenamefont
  {Lake}}]{anand2016prhfo1}%
  \BibitemOpen
  \bibfield  {author} {\bibinfo {author} {\bibfnamefont {V.~K.}\ \bibnamefont
  {Anand}}, \bibinfo {author} {\bibfnamefont {L.}~\bibnamefont {Opherden}},
  \bibinfo {author} {\bibfnamefont {J.}~\bibnamefont {Xu}}, \bibinfo {author}
  {\bibfnamefont {D.~T.}\ \bibnamefont {Adroja}}, \bibinfo {author}
  {\bibfnamefont {A.~T. M.~N.}\ \bibnamefont {Islam}}, \bibinfo {author}
  {\bibfnamefont {T.}~\bibnamefont {Herrmannsd\"orfer}}, \bibinfo {author}
  {\bibfnamefont {J.}~\bibnamefont {Hornung}}, \bibinfo {author} {\bibfnamefont
  {R.}~\bibnamefont {Sch\"onemann}}, \bibinfo {author} {\bibfnamefont
  {M.}~\bibnamefont {Uhlarz}}, \bibinfo {author} {\bibfnamefont {H.~C.}\
  \bibnamefont {Walker}}, \bibinfo {author} {\bibfnamefont {N.}~\bibnamefont
  {Casati}}, \ and\ \bibinfo {author} {\bibfnamefont {B.}~\bibnamefont
  {Lake}},\ }\href {\doibase 10.1103/PhysRevB.94.144415} {\bibfield  {journal}
  {\bibinfo  {journal} {Phys. Rev. B}\ }\textbf {\bibinfo {volume} {94}},\
  \bibinfo {pages} {144415} (\bibinfo {year} {2016})}\BibitemShut {NoStop}%
\bibitem [{\citenamefont {Sibille}\ \emph {et~al.}(2016)\citenamefont
  {Sibille}, \citenamefont {Lhotel}, \citenamefont {Hatnean}, \citenamefont
  {Balakrishnan}, \citenamefont {F\aa{}k}, \citenamefont {Gauthier},
  \citenamefont {Fennell},\ and\ \citenamefont
  {Kenzelmann}}]{sibille2016prhfo}%
  \BibitemOpen
  \bibfield  {author} {\bibinfo {author} {\bibfnamefont {R.}~\bibnamefont
  {Sibille}}, \bibinfo {author} {\bibfnamefont {E.}~\bibnamefont {Lhotel}},
  \bibinfo {author} {\bibfnamefont {M.~C.}\ \bibnamefont {Hatnean}}, \bibinfo
  {author} {\bibfnamefont {G.}~\bibnamefont {Balakrishnan}}, \bibinfo {author}
  {\bibfnamefont {B.}~\bibnamefont {F\aa{}k}}, \bibinfo {author} {\bibfnamefont
  {N.}~\bibnamefont {Gauthier}}, \bibinfo {author} {\bibfnamefont
  {T.}~\bibnamefont {Fennell}}, \ and\ \bibinfo {author} {\bibfnamefont
  {M.}~\bibnamefont {Kenzelmann}},\ }\href {\doibase
  10.1103/PhysRevB.94.024436} {\bibfield  {journal} {\bibinfo  {journal} {Phys.
  Rev. B}\ }\textbf {\bibinfo {volume} {94}},\ \bibinfo {pages} {024436}
  (\bibinfo {year} {2016})}\BibitemShut {NoStop}%
\bibitem [{\citenamefont {Sibille}\ \emph
  {et~al.}(2017{\natexlab{b}})\citenamefont {Sibille}, \citenamefont
  {Gauthier}, \citenamefont {Yan}, \citenamefont {Hatnean}, \citenamefont
  {Ollivier}, \citenamefont {Winn}, \citenamefont {Balakrishnan}, \citenamefont
  {Kenzelmann}, \citenamefont {Shannon},\ and\ \citenamefont
  {Fennell}}]{sibille2017experimental}%
  \BibitemOpen
  \bibfield  {author} {\bibinfo {author} {\bibfnamefont {R.}~\bibnamefont
  {Sibille}}, \bibinfo {author} {\bibfnamefont {N.}~\bibnamefont {Gauthier}},
  \bibinfo {author} {\bibfnamefont {H.}~\bibnamefont {Yan}}, \bibinfo {author}
  {\bibfnamefont {M.~C.}\ \bibnamefont {Hatnean}}, \bibinfo {author}
  {\bibfnamefont {J.}~\bibnamefont {Ollivier}}, \bibinfo {author}
  {\bibfnamefont {B.}~\bibnamefont {Winn}}, \bibinfo {author} {\bibfnamefont
  {G.}~\bibnamefont {Balakrishnan}}, \bibinfo {author} {\bibfnamefont
  {M.}~\bibnamefont {Kenzelmann}}, \bibinfo {author} {\bibfnamefont
  {N.}~\bibnamefont {Shannon}}, \ and\ \bibinfo {author} {\bibfnamefont
  {T.}~\bibnamefont {Fennell}},\ }\href@noop {} {\  (\bibinfo {year}
  {2017}{\natexlab{b}})},\ \Eprint {http://arxiv.org/abs/1706.03604}
  {arXiv:1706.03604 [cond-mat.str-el]} \BibitemShut {NoStop}%
\bibitem [{\citenamefont {Lau}\ \emph {et~al.}(2005)\citenamefont {Lau},
  \citenamefont {Freitas}, \citenamefont {Ueland}, \citenamefont {Schiffer},\
  and\ \citenamefont {Cava}}]{lau2005}%
  \BibitemOpen
  \bibfield  {author} {\bibinfo {author} {\bibfnamefont {G.~C.}\ \bibnamefont
  {Lau}}, \bibinfo {author} {\bibfnamefont {R.~S.}\ \bibnamefont {Freitas}},
  \bibinfo {author} {\bibfnamefont {B.~G.}\ \bibnamefont {Ueland}}, \bibinfo
  {author} {\bibfnamefont {P.}~\bibnamefont {Schiffer}}, \ and\ \bibinfo
  {author} {\bibfnamefont {R.~J.}\ \bibnamefont {Cava}},\ }\href {\doibase
  10.1103/PhysRevB.72.054411} {\bibfield  {journal} {\bibinfo  {journal} {Phys.
  Rev. B}\ }\textbf {\bibinfo {volume} {72}},\ \bibinfo {pages} {054411}
  (\bibinfo {year} {2005})}\BibitemShut {NoStop}%
\bibitem [{\citenamefont {Lago}\ \emph {et~al.}(2010)\citenamefont {Lago},
  \citenamefont {\ifmmode \check{Z}\else
  \v{Z}\fi{}ivkovi\ifmmode~\acute{c}\else \'{c}\fi{}}, \citenamefont {Malkin},
  \citenamefont {Rodriguez~Fernandez}, \citenamefont {Ghigna}, \citenamefont
  {Dalmas~de R\'eotier}, \citenamefont {Yaouanc},\ and\ \citenamefont
  {Rojo}}]{erspinel1}%
  \BibitemOpen
  \bibfield  {author} {\bibinfo {author} {\bibfnamefont {J.}~\bibnamefont
  {Lago}}, \bibinfo {author} {\bibfnamefont {I.}~\bibnamefont {\ifmmode
  \check{Z}\else \v{Z}\fi{}ivkovi\ifmmode~\acute{c}\else \'{c}\fi{}}}, \bibinfo
  {author} {\bibfnamefont {B.~Z.}\ \bibnamefont {Malkin}}, \bibinfo {author}
  {\bibfnamefont {J.}~\bibnamefont {Rodriguez~Fernandez}}, \bibinfo {author}
  {\bibfnamefont {P.}~\bibnamefont {Ghigna}}, \bibinfo {author} {\bibfnamefont
  {P.}~\bibnamefont {Dalmas~de R\'eotier}}, \bibinfo {author} {\bibfnamefont
  {A.}~\bibnamefont {Yaouanc}}, \ and\ \bibinfo {author} {\bibfnamefont
  {T.}~\bibnamefont {Rojo}},\ }\href {\doibase 10.1103/PhysRevLett.104.247203}
  {\bibfield  {journal} {\bibinfo  {journal} {Phys. Rev. Lett.}\ }\textbf
  {\bibinfo {volume} {104}},\ \bibinfo {pages} {247203} (\bibinfo {year}
  {2010})}\BibitemShut {NoStop}%
\bibitem [{\citenamefont {Gao}\ \emph {et~al.}(2017)\citenamefont {Gao},
  \citenamefont {Zaharko}, \citenamefont {Tsurkan}, \citenamefont {Prodan},
  \citenamefont {Riordan}, \citenamefont {Lago}, \citenamefont {Wildes},
  \citenamefont {Koza}, \citenamefont {Ritter}, \citenamefont {Fouquet} \emph
  {et~al.}}]{erspinel2}%
  \BibitemOpen
  \bibfield  {author} {\bibinfo {author} {\bibfnamefont {S.}~\bibnamefont
  {Gao}}, \bibinfo {author} {\bibfnamefont {O.}~\bibnamefont {Zaharko}},
  \bibinfo {author} {\bibfnamefont {V.}~\bibnamefont {Tsurkan}}, \bibinfo
  {author} {\bibfnamefont {L.}~\bibnamefont {Prodan}}, \bibinfo {author}
  {\bibfnamefont {E.}~\bibnamefont {Riordan}}, \bibinfo {author} {\bibfnamefont
  {J.}~\bibnamefont {Lago}}, \bibinfo {author} {\bibfnamefont {A.}~\bibnamefont
  {Wildes}}, \bibinfo {author} {\bibfnamefont {M.~M.}\ \bibnamefont {Koza}},
  \bibinfo {author} {\bibfnamefont {C.}~\bibnamefont {Ritter}}, \bibinfo
  {author} {\bibfnamefont {P.}~\bibnamefont {Fouquet}},  \emph {et~al.},\
  }\href@noop {} {\  (\bibinfo {year} {2017})},\ \Eprint
  {http://arxiv.org/abs/1705.10737} {arXiv:1705.10737 [cond-mat.str-el]}
  \BibitemShut {NoStop}%
\bibitem [{\citenamefont {Reig-i Plessis}\ \emph {et~al.}(2017)\citenamefont
  {Reig-i Plessis}, \citenamefont {Geldern}, \citenamefont {Aczel},\ and\
  \citenamefont {MacDougall}}]{erspinel3}%
  \BibitemOpen
  \bibfield  {author} {\bibinfo {author} {\bibfnamefont {D.}~\bibnamefont
  {Reig-i Plessis}}, \bibinfo {author} {\bibfnamefont {S.~V.}\ \bibnamefont
  {Geldern}}, \bibinfo {author} {\bibfnamefont {A.~A.}\ \bibnamefont {Aczel}},
  \ and\ \bibinfo {author} {\bibfnamefont {G.~J.}\ \bibnamefont {MacDougall}},\
  }\href@noop {} {\  (\bibinfo {year} {2017})},\ \Eprint
  {http://arxiv.org/abs/1703.04267} {arXiv:1703.04267 [cond-mat.dis-nn]}
  \BibitemShut {NoStop}%
\bibitem [{\citenamefont {Higo}\ \emph {et~al.}(2017)\citenamefont {Higo},
  \citenamefont {Iritani}, \citenamefont {Halim}, \citenamefont {Higemoto},
  \citenamefont {Ito}, \citenamefont {Kuga}, \citenamefont {Kimura},\ and\
  \citenamefont {Nakatsuji}}]{higo2017spinels1}%
  \BibitemOpen
  \bibfield  {author} {\bibinfo {author} {\bibfnamefont {T.}~\bibnamefont
  {Higo}}, \bibinfo {author} {\bibfnamefont {K.}~\bibnamefont {Iritani}},
  \bibinfo {author} {\bibfnamefont {M.}~\bibnamefont {Halim}}, \bibinfo
  {author} {\bibfnamefont {W.}~\bibnamefont {Higemoto}}, \bibinfo {author}
  {\bibfnamefont {T.~U.}\ \bibnamefont {Ito}}, \bibinfo {author} {\bibfnamefont
  {K.}~\bibnamefont {Kuga}}, \bibinfo {author} {\bibfnamefont {K.}~\bibnamefont
  {Kimura}}, \ and\ \bibinfo {author} {\bibfnamefont {S.}~\bibnamefont
  {Nakatsuji}},\ }\href {\doibase 10.1103/PhysRevB.95.174443} {\bibfield
  {journal} {\bibinfo  {journal} {Phys. Rev. B}\ }\textbf {\bibinfo {volume}
  {95}},\ \bibinfo {pages} {174443} (\bibinfo {year} {2017})}\BibitemShut
  {NoStop}%
\bibitem [{\citenamefont {Dalmas~de R\'eotier}\ \emph
  {et~al.}(2017)\citenamefont {Dalmas~de R\'eotier}, \citenamefont {Marin},
  \citenamefont {Yaouanc}, \citenamefont {Ritter}, \citenamefont {Maisuradze},
  \citenamefont {Roessli}, \citenamefont {Bertin}, \citenamefont {Baker},\ and\
  \citenamefont {Amato}}]{ybspinel2}%
  \BibitemOpen
  \bibfield  {author} {\bibinfo {author} {\bibfnamefont {P.}~\bibnamefont
  {Dalmas~de R\'eotier}}, \bibinfo {author} {\bibfnamefont {C.}~\bibnamefont
  {Marin}}, \bibinfo {author} {\bibfnamefont {A.}~\bibnamefont {Yaouanc}},
  \bibinfo {author} {\bibfnamefont {C.}~\bibnamefont {Ritter}}, \bibinfo
  {author} {\bibfnamefont {A.}~\bibnamefont {Maisuradze}}, \bibinfo {author}
  {\bibfnamefont {B.}~\bibnamefont {Roessli}}, \bibinfo {author} {\bibfnamefont
  {A.}~\bibnamefont {Bertin}}, \bibinfo {author} {\bibfnamefont {P.~J.}\
  \bibnamefont {Baker}}, \ and\ \bibinfo {author} {\bibfnamefont
  {A.}~\bibnamefont {Amato}},\ }\href {\doibase 10.1103/PhysRevB.96.134403}
  {\bibfield  {journal} {\bibinfo  {journal} {Phys. Rev. B}\ }\textbf {\bibinfo
  {volume} {96}},\ \bibinfo {pages} {134403} (\bibinfo {year}
  {2017})}\BibitemShut {NoStop}%
\bibitem [{\citenamefont {Krizan}\ and\ \citenamefont
  {Cava}(2014)}]{fluoridesynthesis1}%
  \BibitemOpen
  \bibfield  {author} {\bibinfo {author} {\bibfnamefont {J.~W.}\ \bibnamefont
  {Krizan}}\ and\ \bibinfo {author} {\bibfnamefont {R.~J.}\ \bibnamefont
  {Cava}},\ }\href {\doibase 10.1103/PhysRevB.89.214401} {\bibfield  {journal}
  {\bibinfo  {journal} {Phys. Rev. B}\ }\textbf {\bibinfo {volume} {89}},\
  \bibinfo {pages} {214401} (\bibinfo {year} {2014})}\BibitemShut {NoStop}%
\bibitem [{\citenamefont {Krizan}\ and\ \citenamefont
  {Cava}(2015)}]{fluoridesynthesis4}%
  \BibitemOpen
  \bibfield  {author} {\bibinfo {author} {\bibfnamefont {J.~W.}\ \bibnamefont
  {Krizan}}\ and\ \bibinfo {author} {\bibfnamefont {R.~J.}\ \bibnamefont
  {Cava}},\ }\href {\doibase 10.1103/PhysRevB.92.014406} {\bibfield  {journal}
  {\bibinfo  {journal} {Phys. Rev. B}\ }\textbf {\bibinfo {volume} {92}},\
  \bibinfo {pages} {014406} (\bibinfo {year} {2015})}\BibitemShut {NoStop}%
\bibitem [{\citenamefont {Sanders}\ \emph {et~al.}(2016)\citenamefont
  {Sanders}, \citenamefont {Krizan}, \citenamefont {Plumb}, \citenamefont
  {McQueen},\ and\ \citenamefont {Cava}}]{fluoridesynthesis3}%
  \BibitemOpen
  \bibfield  {author} {\bibinfo {author} {\bibfnamefont {M.~B.}\ \bibnamefont
  {Sanders}}, \bibinfo {author} {\bibfnamefont {J.~W.}\ \bibnamefont {Krizan}},
  \bibinfo {author} {\bibfnamefont {K.~W.}\ \bibnamefont {Plumb}}, \bibinfo
  {author} {\bibfnamefont {T.~M.}\ \bibnamefont {McQueen}}, \ and\ \bibinfo
  {author} {\bibfnamefont {R.~J.}\ \bibnamefont {Cava}},\ }\href {\doibase
  10.1088/1361-648X/29/4/045801} {\bibfield  {journal} {\bibinfo  {journal}
  {Journal of Physics: Condensed Matter}\ }\textbf {\bibinfo {volume} {29}},\
  \bibinfo {pages} {045801} (\bibinfo {year} {2016})}\BibitemShut {NoStop}%
\bibitem [{\citenamefont {Ross}\ \emph {et~al.}(2016)\citenamefont {Ross},
  \citenamefont {Krizan}, \citenamefont {Rodriguez-Rivera}, \citenamefont
  {Cava},\ and\ \citenamefont {Broholm}}]{fluoridexy1}%
  \BibitemOpen
  \bibfield  {author} {\bibinfo {author} {\bibfnamefont {K.~A.}\ \bibnamefont
  {Ross}}, \bibinfo {author} {\bibfnamefont {J.~W.}\ \bibnamefont {Krizan}},
  \bibinfo {author} {\bibfnamefont {J.~A.}\ \bibnamefont {Rodriguez-Rivera}},
  \bibinfo {author} {\bibfnamefont {R.~J.}\ \bibnamefont {Cava}}, \ and\
  \bibinfo {author} {\bibfnamefont {C.~L.}\ \bibnamefont {Broholm}},\ }\href
  {\doibase 10.1103/PhysRevB.93.014433} {\bibfield  {journal} {\bibinfo
  {journal} {Phys. Rev. B}\ }\textbf {\bibinfo {volume} {93}},\ \bibinfo
  {pages} {014433} (\bibinfo {year} {2016})}\BibitemShut {NoStop}%
\bibitem [{\citenamefont {Ross}\ \emph {et~al.}(2017)\citenamefont {Ross},
  \citenamefont {Brown}, \citenamefont {Cava}, \citenamefont {Krizan},
  \citenamefont {Nagler}, \citenamefont {Rodriguez-Rivera},\ and\ \citenamefont
  {Stone}}]{ross2017fluoride}%
  \BibitemOpen
  \bibfield  {author} {\bibinfo {author} {\bibfnamefont {K.~A.}\ \bibnamefont
  {Ross}}, \bibinfo {author} {\bibfnamefont {J.~M.}\ \bibnamefont {Brown}},
  \bibinfo {author} {\bibfnamefont {R.~J.}\ \bibnamefont {Cava}}, \bibinfo
  {author} {\bibfnamefont {J.~W.}\ \bibnamefont {Krizan}}, \bibinfo {author}
  {\bibfnamefont {S.~E.}\ \bibnamefont {Nagler}}, \bibinfo {author}
  {\bibfnamefont {J.~A.}\ \bibnamefont {Rodriguez-Rivera}}, \ and\ \bibinfo
  {author} {\bibfnamefont {M.~B.}\ \bibnamefont {Stone}},\ }\href {\doibase
  10.1103/PhysRevB.95.144414} {\bibfield  {journal} {\bibinfo  {journal} {Phys.
  Rev. B}\ }\textbf {\bibinfo {volume} {95}},\ \bibinfo {pages} {144414}
  (\bibinfo {year} {2017})}\BibitemShut {NoStop}%
\bibitem [{\citenamefont {Plumb}\ \emph {et~al.}(2017)\citenamefont {Plumb},
  \citenamefont {Changlani}, \citenamefont {Scheie}, \citenamefont {Zhang},
  \citenamefont {Kriza}, \citenamefont {Rodriguez-Rivera}, \citenamefont {Qiu},
  \citenamefont {Winn}, \citenamefont {Cava},\ and\ \citenamefont
  {Broholm}}]{plumb2017continuum}%
  \BibitemOpen
  \bibfield  {author} {\bibinfo {author} {\bibfnamefont {K.~W.}\ \bibnamefont
  {Plumb}}, \bibinfo {author} {\bibfnamefont {H.~J.}\ \bibnamefont
  {Changlani}}, \bibinfo {author} {\bibfnamefont {A.}~\bibnamefont {Scheie}},
  \bibinfo {author} {\bibfnamefont {S.}~\bibnamefont {Zhang}}, \bibinfo
  {author} {\bibfnamefont {J.~W.}\ \bibnamefont {Kriza}}, \bibinfo {author}
  {\bibfnamefont {J.~A.}\ \bibnamefont {Rodriguez-Rivera}}, \bibinfo {author}
  {\bibfnamefont {Y.}~\bibnamefont {Qiu}}, \bibinfo {author} {\bibfnamefont
  {B.}~\bibnamefont {Winn}}, \bibinfo {author} {\bibfnamefont {R.~J.}\
  \bibnamefont {Cava}}, \ and\ \bibinfo {author} {\bibfnamefont {C.~L.}\
  \bibnamefont {Broholm}},\ }\href@noop {} {\  (\bibinfo {year} {2017})},\
  \Eprint {http://arxiv.org/abs/1711.07509} {arXiv:1711.07509
  [cond-mat.str-el]} \BibitemShut {NoStop}%
\bibitem [{\citenamefont {Conlon}\ and\ \citenamefont
  {Chalker}(2009)}]{conlon2009}%
  \BibitemOpen
  \bibfield  {author} {\bibinfo {author} {\bibfnamefont {P.~H.}\ \bibnamefont
  {Conlon}}\ and\ \bibinfo {author} {\bibfnamefont {J.~T.}\ \bibnamefont
  {Chalker}},\ }\href {\doibase 10.1103/PhysRevLett.102.237206} {\bibfield
  {journal} {\bibinfo  {journal} {Phys. Rev. Lett.}\ }\textbf {\bibinfo
  {volume} {102}},\ \bibinfo {pages} {237206} (\bibinfo {year}
  {2009})}\BibitemShut {NoStop}%
\bibitem [{\citenamefont {Anderson}(1956)}]{anderson}%
  \BibitemOpen
  \bibfield  {author} {\bibinfo {author} {\bibfnamefont {P.~W.}\ \bibnamefont
  {Anderson}},\ }\href {\doibase 10.1103/PhysRev.102.1008} {\bibfield
  {journal} {\bibinfo  {journal} {Phys. Rev.}\ }\textbf {\bibinfo {volume}
  {102}},\ \bibinfo {pages} {1008} (\bibinfo {year} {1956})}\BibitemShut
  {NoStop}%
\bibitem [{\citenamefont {Witczak-Krempa}\ \emph {et~al.}(2014)\citenamefont
  {Witczak-Krempa}, \citenamefont {Chen}, \citenamefont {Kim},\ and\
  \citenamefont {Balents}}]{witczak2014correlated}%
  \BibitemOpen
  \bibfield  {author} {\bibinfo {author} {\bibfnamefont {W.}~\bibnamefont
  {Witczak-Krempa}}, \bibinfo {author} {\bibfnamefont {G.}~\bibnamefont
  {Chen}}, \bibinfo {author} {\bibfnamefont {Y.~B.}\ \bibnamefont {Kim}}, \
  and\ \bibinfo {author} {\bibfnamefont {L.}~\bibnamefont {Balents}},\ }\href
  {\doibase 10.1146/annurev-conmatphys-020911-125138} {\bibfield  {journal}
  {\bibinfo  {journal} {Annu. Rev. Condens. Matter Phys.}\ }\textbf {\bibinfo
  {volume} {5}},\ \bibinfo {pages} {57} (\bibinfo {year} {2014})}\BibitemShut
  {NoStop}%
\end{thebibliography}%

\end{document}